\documentclass[%
a4paper,
reprint,
superscriptaddress,
nofootinbib,
longbibliography,
amsmath,amssymb,
aps, prd,
]{revtex4-2}

\usepackage[dvipsnames,svgnames,x11names]{xcolor}
\usepackage{graphicx}
\usepackage{setspace}
\usepackage[caption=false]{subfig}
\usepackage{placeins}
\usepackage{dcolumn}
\usepackage{multirow}
\usepackage{enumitem}
\usepackage{calc}
\usepackage{bm}
\usepackage[colorlinks]{hyperref}
\usepackage[shortcuts]{extdash}
\usepackage{physics} 
\usepackage[free-standing-units]{siunitx} 
\usepackage[capitalise,nameinlink]{cleveref} 

\makeatletter
    \g@addto@macro\bfseries{\boldmath}
\makeatother

\makeatletter
\def\@sect@ltx#1#2#3#4#5#6[#7]#8{%
    \@ifnum{#2>\c@secnumdepth}{%
        \def\H@svsec{\phantomsection}%
        \let\@svsec\@empty
    }{%
        \H@refstepcounter{#1}%
        \def\H@svsec{\phantomsection}%
        \protected@edef\@svsec{{#1}}%
        \@ifundefined{@#1cntformat}{%
            \prepdef\@svsec\@seccntformat
        }{%
            \expandafter\prepdef
            \expandafter\@svsec
            \csname @#1cntformat\endcsname
        }%
    }%
    \@tempskipa #5\relax
    \@ifdim{\@tempskipa>\z@}{%
        \begingroup
        \interlinepenalty \@M
        #6{%
            \@ifundefined{@hangfrom@#1}{\@hang@from}{\csname @hangfrom@#1\endcsname}%
            {\hskip#3\relax\H@svsec}{\@svsec}{#8}%
        }%
        \@@par
        \endgroup
        \@ifundefined{#1mark}{\@gobble}{\csname #1mark\endcsname}{#7}%
        \addcontentsline{toc}{#1}{
            \@ifnum{#2>\c@secnumdepth}{\protect\numberline{}}{\protect\numberline{\csname the#1\endcsname}}#7
        }%
    }{%
        \def\@svsechd{%
            #6{%
                \@ifundefined{@runin@to@#1}{\@runin@to}{\csname @runin@to@#1\endcsname}%
                {\hskip#3\relax\H@svsec}{\@svsec}{#8}%
            }%
            \@ifundefined{#1mark}{\@gobble}{\csname #1mark\endcsname}{#7}%
            \addcontentsline{toc}{#1}{%
                \@ifnum{#2>\c@secnumdepth}{%
                    \protect\numberline{}%
                }{%
                    \protect\numberline{\csname the#1\endcsname}%
                }%
                #8}%
        }%
    }%
    \@xsect{#5}}%
\makeatother

\renewcommand{\thefootnote}{\fnsymbol{footnote}}
\newcommand{\e}{\,\mathrm{e}}
\newcommand{\LCDM}{\ensuremath{\Lambda\mathrm{CDM}}}

\newcommand{\citeay}[1]{\citeauthor{#1}~(\citeyear{#1})}%
\DeclareSIUnit\nats{nats}
\DeclareSIUnit\clight{\mathit{c}}
\DeclareSIUnit\parsec{pc}
\DeclareSIUnit\efolds{e\text{-}folds}
\DeclareSIUnit\planckmass{m_\mathrm{p}}
\DeclareSIUnit\plancktime{t_\mathrm{p}}
\DeclareSIUnit\plancklength{\ell_\mathrm{p}}

\begin{document}

\title{Finite inflation in curved space}

\author{L.~T.~Hergt}
    \email{lthergt@phas.ubc.ca}%
    \affiliation{Department of Physics and Astronomy, University of British Columbia, Vancouver, BC~V6T\,1Z1, Canada}%
    \affiliation{Astrophysics Group, Cavendish Laboratory, J.~J.~Thomson Avenue, Cambridge, CB3~0HE, UK}%
    \affiliation{Kavli Institute for Cosmology, Madingley Road, Cambridge, CB3~0HA, UK}%
\author{F.~J.~Agocs}
    \email{fagocs@flatironinstitute.org}%
    \affiliation{Astrophysics Group, Cavendish Laboratory, J.~J.~Thomson Avenue, Cambridge, CB3~0HE, UK}%
    \affiliation{Kavli Institute for Cosmology, Madingley Road, Cambridge, CB3~0HA, UK}%
    \affiliation{Center for Computational Mathematics, Flatiron Institute, 162 Fifth Avenue, New York, New York, USA}
\author{W.~J.~Handley}%
    \email{wh260@mrao.cam.ac.uk}%
    \affiliation{Astrophysics Group, Cavendish Laboratory, J.~J.~Thomson Avenue, Cambridge, CB3~0HE, UK}%
    \affiliation{Kavli Institute for Cosmology, Madingley Road, Cambridge, CB3~0HA, UK}%
\author{M.~P.~Hobson}%
    \email{mph@mrao.cam.ac.uk}%
    \affiliation{Astrophysics Group, Cavendish Laboratory, J.~J.~Thomson Avenue, Cambridge, CB3~0HE, UK}%
\author{A.~N.~Lasenby}%
    \email{a.n.lasenby@mrao.cam.ac.uk}%
    \affiliation{Astrophysics Group, Cavendish Laboratory, J.~J.~Thomson Avenue, Cambridge, CB3~0HE, UK}%
    \affiliation{Kavli Institute for Cosmology, Madingley Road, Cambridge, CB3~0HA, UK}%

\date{\today}

\begin{abstract}
    We investigate the effects of non-zero spatial curvature on cosmic inflation in the light of cosmic microwave background~(CMB) anisotropy measurements from the \textsc{Planck 2018} legacy release and from the 2015 observing season of \textsc{Bicep2} and the \textsc{Keck Array}.
    Even a small percentage of non-zero curvature today would significantly limit the total number of e-folds of the scale factor during inflation, rendering just-enough inflation scenarios with a kinetically dominated or fast-roll stage prior to slow-roll inflation more likely.
    Finite inflation leads to oscillations and a cutoff towards large scales in the primordial power spectrum and curvature pushes them into the CMB observable window.
    Using nested sampling, we carry out Bayesian parameter estimations and model comparisons taking into account constraints from reheating and horizon considerations. 
    We confirm the preference of CMB data for closed universes with Bayesian odds of over $100:1$ and with a posterior on the curvature density parameter of $\Omega_{K,0}=\num{-0.051(17)}$ for a curvature extension of \LCDM\ and $\Omega_{K,0}=\num{-0.031(14)}$ for Starobinsky inflation.
    Model comparisons of various inflation models give similar results as for flat universes with the Starobinsky model outperforming most other models.
\end{abstract}

\maketitle

\begin{spacing}{0.99}
\tableofcontents
\end{spacing}

\section{Introduction}

Measurements of the cosmic microwave background~(CMB) radiation anisotropies with the \textsc{Planck} satellite~\cite{Planck2013I,Planck2015I,Planck2018I} have allowed us to refine our view of the Universe to unprecedented precision. This has led to what can be viewed as the current standard model of cosmology, called \LCDM\ for the two dominant contributions to the overall energy density today: a cosmological constant~$\Lambda$ and cold dark matter (effectively collisionless with no electromagnetic interactions). The \LCDM\ model is characterised through six free parameters, most of which can by now be given to percent-level precision~\cite{Planck2013Parameters, Planck2015Parameters, Planck2018Parameters}. It assumes the Friedmann--Lema\^itre--Robertson--Walker~(FLRW) metric~\cite{Friedmann1922, Lemaitre1933, Robertson1935, Robertson1936a, Robertson1936b, Walker1937} with flat spatial geometry.

Cosmic inflation was originally~\cite{Starobinsky1979,Guth1981,Linde1982,Albrecht1982} developed as a mechanism to explain the observed homogeneity and flatness of our Universe and is the currently preferred description of the generation of primordial density perturbations. In the \LCDM\ model the latter are summarised by two parameters, the scalar power amplitude~$A_\mathrm{s}$ and spectral index~$n_\mathrm{s}$, characterising a power-law primordial power spectrum~(PPS) of density anisotropies.
The small deviation from unity of the scalar spectral index, and thus from scale invariance of the PPS, has been confirmed to $\SI{8}{\sigma}$ precision and is one of the prime successes of cosmic inflation~\cite{Planck2018Parameters}. In its simplest form, the accelerated expansion of the Universe during cosmic inflation is driven by a single scalar field~$\phi$ (called ``inflaton'') with a standard kinetic term that slowly rolls down a potential~$V(\phi)$. This slow-roll scenario in a flat universe has been investigated extensively for various inflationary potentials~\cite{Planck2013Inflation, Planck2015Inflation, Planck2018Inflation, Ijjas2013, EncyclopaediaInflationaris2014, Linde2014, Ijjas2014, Chowdhury2019}.

Despite the success of flat \LCDM\ there has been a persistent preference for positive curvature (closed universes) in CMB temperature and polarisation data~\cite{WMAP1Parameters, Uzan2003, WMAP9Parameters, Planck2013Parameters, Planck2015Parameters, Planck2018Parameters, CamSpec2019}. The 2018 data release from the \textsc{Planck} satellite~\cite{Planck2018Parameters} in particular has sparked discussion around evidence for spatial curvature in the CMB and about a possible discordance or tension with measurements from other sources such as baryon acoustic oscillations~(BAO) or luminosity distance data, e.g.\ type~Ia supernovae or Cepheid variables~\cite{DiValentino2019, Handley2019c, DiValentino2020b, CamSpec2019, Efstathiou2020}.
Such tensions can be caused by incomplete physical models or unknown systematic effects and need to be isolated and corrected in order for the combination of such datasets to improve the overall accuracy of parameter constraints (see e.g.~\cite{Lemos2019,Raveri2019,Raveri2020}). In the case of curved universes the tension with BAO data could be caused by the assumption of a flat fiducial model, e.g.\ for the purpose of data compression (see also~\cite{ODwyer2019}).
Immediately prior to the release of this paper, a new analysis of BAO data without flat fiducial assumptions~\cite{2022arXiv220505892G} demonstrated that in fact BAO data prefer closed universes at $\SI{2}{\sigma}$ significance $\Omega_{K,0} = -0.089^{+0.049}_{-0.046}$. In combination with Planck data, the significance is maintained, although is closer to flat $\Omega_{K,0} = -0.0041^{+0.0026}_{-0.0021}$, with curved models compatible with data with a Bayesian evidence of 2:1 for closure.
In this work we investigate pure CMB constraints, which in the context of studying cosmic inflation has some merit of itself, since inflation provides a mechanism to drive the primordial Universe towards flatness. Thus it is interesting to explore whether the CMB preference for closed universes survives the application to specific inflation models.
Future work will continue to incorporate alternative likelihoods as and when they are updated removing flatness assumptions.

Non-zero spatial curvature affects the CMB anisotropy spectrum on two levels. 
First, curvature terms in the Boltzmann equations will modify the transfer functions, which encode the evolution of linear perturbations through the standard Big Bang epochs of radiation, matter and~$\Lambda$ domination until today. This makes up the curvature effect, which is commonly studied through an extension of the base \LCDM\ model with a variable curvature density parameter~$\Omega_{K,0}$ as mentioned in the abstract. Note that we use the subscript~$0$ to refer to present-day quantities. 
Second, a detection of present-day non-zero curvature would strongly limit the total amount of slow-roll inflation (measured in terms of e-foldings of the scale factor~$a$) and thereby would affect the PPS, particularly on large scales~\cite{Lasenby2005}. This could explain some of the unexplained features in the CMB angular power spectra, such as the lack of power on the largest scales~\cite{Efstathiou2003}. Additionally, non-zero curvature renders scenarios of finite inflation more likely, including those with a phase of fast-roll inflation or kinetic dominance~(KD) (where the inflaton's kinetic energy dominates over its potential energy) preceding slow-roll. See e.g.~\cite{Belinsky1985,Belinsky1988} for early accounts on the generality of slow-roll inflation as an attractor solution and possible preceding stages of KD and fast-roll with and without curvature. 
Other motivations for KD or fast-roll include holographic bounds~\cite{Banks2003, Albrecht2009, Albrecht2011, Phillips2015} or certain potentials that render a preceding phase of KD or fast-roll more likely~\cite{Hergt2019a} or that predict fewer e-folds of inflation~\cite{Remmen2014}.
This form of a finite amount of inflation is often also referred to as just-enough inflation~\cite{Schwarz2009}. The consequences of a preceding KD or fast-roll stage have mostly been studied assuming a flat cosmology, see e.g.~\cite{Linde2001, Contaldi2003, Boyanovsky2006, Boyanovsky2006a, Destri2008, Ramirez2009, Destri2010, Ramirez2012, Ramirez2012a, Handley2014, Lello2014, Cicoli2014, Scacco2015, Hergt2019, Ragavendra2020}. 
There have been a few studies of the large scale curvature effects on the PPS and how they translate to the CMB anisotropy spectra~\cite{White1996, Lasenby2005, Bonga2016c, Bonga2017, Handley2019e}, but these did not go beyond a phenomenological study of some specific parameter combinations.

In this paper we build on the \LCDM\ extension with a non-zero curvature density parameter~$\Omega_{K,0}$, which already accounts for curvature effects on the transfer functions of the Boltzmann equations of cosmology. We further investigate the other implications of non-zero curvature on cosmic inflation, the PPS of scalar and tensor perturbations, and thus the CMB temperature and polarisation angular power spectra. 
To that end we compute the PPS numerically in order to include large scale curvature effects where the slow-roll approximation of inflation no longer holds. We interface our numerical PPS with the Boltzmann code \texttt{CLASS}~\cite{Class1} to compute the CMB anisotropy spectra, which we then use for parameter estimation and model comparison in a fully Bayesian analysis, making use of \texttt{Cobaya}'s~\cite{Cobaya1} interface with the nested sampling code \texttt{PolyChord}~\cite{PolyChord1, PolyChord2}. 
Using \texttt{CLASS} with its fully quantised treatment for closed universes~\cite{Class6}\footnote{Quantised means here that, owing to their finite size of closed universes, the set of possible perturbation modes becomes discrete. Thus, the primordial power spectrum of curvature perturbations is defined on a discrete set of wavenumbers~$k$ rather than being a continuous function.} addresses the concerns raised in~\cite{Efstathiou2020} about the proper treatment of the power spectra on large scales with non-zero spatial curvature in other cosmological codes.
For the post-processing of the nested sampling chains we use \texttt{anesthetic}~\cite{Anesthetic}.

Note that there are two different perspectives one can adopt when applying a prior to the curvature density parameter~$\Omega_{K,0}$. 
On the one hand, one can claim ignorance about the spatial curvature of the universe and apply a uniform prior over some range, typically symmetric about $\Omega_{K,0}=0$. This is the approach taken e.g.\ in~\cite{Planck2013Parameters, Planck2015Parameters, Planck2018Parameters} and also in this paper. A preference for non-zero curvature by the data will then limit the total amount of inflation as described in the previous paragraph. 
On the other hand, one can take the view that \emph{a priori} one expects inflation to produce a large number of e-folds of the scale factor, in which case it has been suggested that one should instead adopt a prior that is strongly peaked at $\Omega_{K,0}=0$, with tails extending to non-zero curvature values~\cite{Efstathiou2020}. 
We leave the exploration of such different prior assumptions to future work and take the more common, curvature agnostic approach in this paper.

The structure of this paper is as follows. We first review our statistical and computational methods in \cref{sec:methods}. In the following \cref{sec:background,sec:finite_inflation,sec:initial_conditions} we review the necessary theoretical background of inflation, including reasons why it might be finite, and the initial conditions we use in such cases of finite inflation. These sections also serve as an introduction of our notation. In \cref{sec:cHH} we present our numerical results for the evolution of the comoving Hubble horizon prior to and during inflation, with a specific focus on the effects of curvature at the very start of inflation. \Cref{sec:conformal_time} focuses on the amount of conformal time passing before versus after the end of inflation. This places an important constraint on primordial parameters, especially primordial curvature, in order to solve the horizon problem. Next, in \cref{sec:reheating}, we investigate another constraint originating from the reheating epoch following inflation, which is particularly relevant for the total duration of inflation. In \cref{sec:pps} we review the computations going into the generation of the PPS. In \cref{sec:potentials} we present some popular choices of inflationary potentials and their predictions for slow-roll parameters such as the scalar spectral index~$n_\mathrm{s}$, its running~$n_\mathrm{run}$, and the tensor-to-scalar ratio~$r$ of primordial perturbations. We compare the slow-roll predictions for these parameters to their corresponding one-parameter extensions of \LCDM, while also allowing for non-zero spatial curvature.
Much of the theory presented up to this point is well-covered in the literature for the flat case. However, since the curved case is considerably more complex when all effects of and constraints associated with curvature are combined, it warrants a lengthier exposition in this paper.
\Cref{sec:parametrisation} gives an overview of our choice of parametrisation used for our nested sampling runs, the results of which we present in \cref{sec:results}. We start with results from parameter extensions of \LCDM\ with the curvature density parameter~$\Omega_{K,0}$ and the tensor-to-scalar ratio~$r$, followed by various single scalar field inflation models using the fully numeric computation of the PPS. We draw conclusions in \cref{sec:conclusion}.

\section{Methods}
\label{sec:methods}

We make use of the same tools as in our preceding analysis~\cite{Hergt2021} of the effects of half-constrained parameters on the Bayesian evidence. Hence, this section follows closely the corresponding ``methods'' section therein.

\subsection{Bayesian inference and nested sampling}

For the estimation of the probability distribution of a set of model parameters~$\theta$ and for the comparison of various models~$M$ with one another, we make use of Bayesian methods rooted in Bayes' theorem. The theorem describes how to update a prior belief~$\pi_M(\theta)$ with the likelihood~$\mathcal{L}_M(\theta)$ of the parameters~$\theta$ in light of new data~$D$: 
\addtolength{\jot}{6pt}
\begin{alignat}{3}
    \Pr(\theta\,|\,D,M)       &\times \Pr(D\,|\,M)         & &=\,\, & \Pr(D\,|\,\theta,M)       &\times \Pr(\theta\,|\,M) , \nonumber \\
    \mathrm{Posterior}    &\times \mathrm{Evidence}& &=\,\, & \mathrm{Likelihood}   &\times \mathrm{Prior} , \nonumber \\
    \mathcal{P}_M(\theta) &\times \mathcal{Z}_M    & &=\,\, & \mathcal{L}_M(\theta) &\times \pi_M(\theta) . 
\end{alignat}
The posterior~$\mathcal{P}_M(\theta)$ is the main quantity of interest in parameter estimation, representing our state of knowledge about the parameters~$\theta$ given a model~$M$ and the data~$D$. The evidence~$\mathcal{Z}_M$ is pivotal for model comparison.

Were we interested only in parameter estimation, it would be sufficient to consider the unnormalised posterior, which is proportional to the product of likelihood and prior, and the Bayesian evidence could be neglected as a mere normalisation factor. However, for the comparison of two models~$A$ and~$B$ the evidence becomes important. Putting the two models on the same footing a priori, i.e.\ $\Pr(A)=\Pr(B)$, the posterior probability ratio is equal to the evidence ratio of the two models:
\begin{align}
    \frac{\Pr(A|D)}{\Pr(B|D)} = \frac{\Pr(D|A)}{\Pr(D|B)} = \frac{\mathcal{Z}_A}{\mathcal{Z}_B}
\end{align}
This ratio can be interpreted as betting odds for the two models. We typically quote this in terms of the log-difference of evidences $\Delta\ln\mathcal{Z} = \ln(\mathcal{Z}_A/\mathcal{Z}_B)$.

The Bayesian evidence is sometimes also referred to as the marginal likelihood and thereby can be interpreted as the prior average of the likelihood:
\begin{align}
    \mathcal{Z}_M = \int \mathcal{L}_M(\theta) \, \pi_M(\theta) \,\dd\theta = \big\langle \mathcal{L}_M \big\rangle_\pi .
\end{align}
The Bayesian evidence is notoriously difficult to calculate because it requires the whole parameter space to be taken into account, unlike the posterior distribution, which typically spans only a small fraction of the sampled parameter space. On the other hand, the evidence takes the complexity of a model into account by automatically applying an Occam penalty to over-parametrised models.
We use \texttt{PolyChord}~\cite{PolyChord1,PolyChord2}, a nested sampler designed to efficiently sample high-dimensional parameter spaces equipped with a speed hierarchy that allows it to oversample the many nuisance parameters that come with experiments such as \textsc{Planck}~\cite{Planck2018CMB} or the \textsc{Bicep2} and \textsc{Keck Array}~\cite{BicepKeck2018BKX}.

Another useful quantity to investigate is the Kullback--Leibler~(KL) divergence defined as
\begin{align}
    \mathcal{D}_{\mathrm{KL},M} = \int \mathcal{P}_M(\theta) \ln\left( \frac{\mathcal{P}_M(\theta)}{\pi_M(\theta)} \right) \dd\theta .
\end{align}
It is also referred to as the relative entropy, describing its role in quantifying the information gain when going from the prior to the posterior distribution. 
Splitting up the log-evidence into KL-divergence and posterior average of the log-likelihood highlights how the KL-divergence can also be viewed as a measure of the Occam penalty that goes into the Bayesian evidence, with the posterior average of the log-likelihood being a measure of the fit of the model (see also~\cite{Hergt2021}):
\begin{alignat}{3}
\label{eq:evidence_fit_occam}
    \ln( \int\!\! \mathcal{L}_M \pi_M \dd\theta\!) \! &= & 
    \int\! \mathcal{P}_M \ln\mathcal{L}_M \dd\theta &-\! 
    \int\! \mathcal{P}_M \ln(\frac{\mathcal{P}_M}{\pi_M}) \dd\theta , \nonumber \\
    \text{(log-)evidence} &= & \text{parameter fit}\,   &- \,\text{Occam penalty} ,\nonumber \\
    \ln\mathcal{Z}_M  &= & \langle \ln\mathcal{L}_M \rangle_\mathcal{P} &- \mathcal{D}_{\mathrm{KL},M} . 
\end{alignat}

Besides the posterior \emph{average} of the log-likelihood, we note that the posterior \emph{variance} of the log-likelihood gives us the Bayesian model dimensionality~$d_M$, a measure of the number of parameters \emph{constrained} by the data, which typically differs from the total number of free sampling parameters~\cite{Handley2019a}:
\begin{align}
    \frac{d_M}{2} 
    &= \int \mathcal{P}_M(\theta) \left( \ln\frac{\mathcal{P}_M(\theta)}{\pi_M(\theta)} - \mathcal{D}_{\mathrm{KL},M} \right)^2 \dd{\theta} \\
    &= \big\langle \left( \ln\mathcal{L}_M \right)^2 \big\rangle_\mathcal{P} - \big\langle \ln\mathcal{L}_M \big\rangle_\mathcal{P}^2 .
\end{align}

\subsection{ODE solvers}

For the integration of the ordinary differential equations~(ODEs) of the primordial inflationary background, to be introduced in \cref{sec:background} in \cref{eq:background1,eq:background2,eq:eom,eq:background3}, we use \texttt{scipy}'s high-order Runge--Kutta~(RK) integrator~\cite{Scipy,Dormand1980}.
The mode \cref{eq:mukhanov-sasaki_scalar,eq:mukhanov-sasaki_tensor} of primordial perturbations are highly oscillatory in time and therefore pose a challenge for standard RK integrators, whose stepsizes are limited to around one wavelength of oscillation. Their runtime therefore scales as $\mathcal{O}(k)$, with $k$ being the characteristic frequency of the mode. Instead, we use \texttt{oscode}~\cite{Oscode,Oscode2} for these equations, whose adaptive algorithm~\cite{Handley2016a} switches automatically between two numerical methods based on whether the solution is slowly varying or oscillatory. When the solution varies slowly, \texttt{oscode} behaves as a fifth order Runge--Kutta solver with an adaptive stepsize. In regions of high-frequency oscillations, \texttt{oscode} switches over to using the asymptotic Wentzel--Kramers--Brillouin~(WKB) approximation valid for oscillatory functions, which allows it to step over many wavelengths of oscillations and have a frequency-independent, $\mathcal{O}(1)$, runtime.

\subsection{Statistical and cosmological software}

We explore the posterior distributions of cosmological and nuisance parameters using \texttt{Cobaya}~\cite{Cobaya1}, which interfaces the sampling with the theory codes and provides both the MCMC sampler developed for \texttt{CosmoMC}~\cite{CosmoMC1,CosmoMC2} with a ``fast dragging'' procedure described in~\cite{CosmoMC3} and the nested sampling code \texttt{PolyChord}~\cite{PolyChord1,PolyChord2}, tailored for high-dimensional parameter spaces, which can simultaneously calculate the Bayesian evidence.
For the cosmological theory code we use the Boltzmann solver \texttt{CLASS}~\cite{Class1, Class2, Class5, Class6}, which computes the theoretical CMB power spectra for temperature and polarisation modes. We extend \texttt{CLASS} with our own code~\texttt{primpy}\footnote{Access to the \texttt{primpy} code can be provided upon request. It can be found on GitHub at \url{https://github.com/lukashergt/primpy} and will potentially be made public at a later stage.} computing the primordial power spectrum for various inflationary potentials, making use of the aforementioned ODE solver \texttt{oscode}~\cite{Oscode}.

The resulting MCMC and nested sampling chains generated for this paper have been published on Zenodo~\cite{ZenodoCurved}.

We use~\texttt{GetDist}~\cite{GetDist} to generate the data tables of marginalised parameter values.
The post-processing of the nested sampling output for the computation of Bayesian evidence, KL-divergence and Bayesian model dimensionality, as well as the plotting functionality for posterior contours is performed using the python module \texttt{anesthetic}~\cite{Anesthetic}.

\subsection{Data}

We use the 2018 temperature and polarisation data from the \textsc{Planck} satellite~\cite{Planck2018CMB}, abbreviated as ``$TT,TE,EE+\mathrm{low}E$'' in the \textsc{Planck} publication.
Note that the explicit writing of ``$\mathrm{low}E$'' but lack of ``$\mathrm{low}T$'' might mislead one to the conclusion that only $E$-mode and no temperature data were used at low multipoles. However, this is \emph{not} the case. The abbreviation implies the inclusion of both high-$\ell$ \emph{and} low-$\ell$ temperature auto-correlation data. 
To save space we will frequently refer to this \textsc{Planck} data release as~``P18''.
For computationally more expensive sampling runs, we sometimes use the ``lite'' version of the likelihood, where most nuisance parameters were marginalised out. We will label this in corresponding figure legends as~``P18lite''.

Additionally, we use $B$-mode data from the 2015 observing season of \textsc{Bicep2} and the \textsc{Keck Array} CMB experiments~\cite{BicepKeck2018BKX} (2018 data release), which we abbreviate as~``BK15''.

Note that there have been new data releases by both the \textsc{Planck}~\cite{Planck2020Joint,Tristram2021} and the \textsc{Bicep/Keck}~\cite{BicepKeck2021} collaborations, but due to the computationally expensive curvature runs, we use their previous data releases in this paper.

As mentioned in the introduction we have chosen not to include data from CMB lensing and from Baryon Acoustic Oscillations~(BAOs), which have been shown to be in some tension for closed universes~\cite{Handley2019c,DiValentino2019,DiValentino2020b}. The reasons for this discordance are unclear, but may be related to the same issue as with the lensing parameter~$A_\mathrm{lens}$ or the fact that the corresponding likelihoods assume a fiducially flat cosmology.

\section{Inflationary background equations}
\label{sec:background}

The Friedmann equations~\cite{Friedmann1922} and the related continuity equation are derived from the Einstein equations of general relativity~\cite{Einstein1916} assuming the homogeneous and isotropic FLRW metric~\cite{Friedmann1922, Lemaitre1933, Robertson1935, Robertson1936a, Robertson1936b, Walker1937}. They govern the dynamics of the Universe in form of the scale factor~$a$ and its energy content given by the energy density~$\rho$ and pressure~$p$:
\begin{align}
    \label{eq:friedmann1}
    H^2 + \frac{K\si{\clight\squared}}{a^2} &= \frac{8\pi G}{3}\rho + \frac{\Lambda \si{\clight\squared}}{3} , \\
    \label{eq:friedmann2}
    \frac{\ddot a}{a} &= - \frac{4\pi G}{3} \left(\rho + \frac{3p}{\si{\clight\squared}} \right) + \frac{\Lambda \si{\clight\squared}}{3} , \\
    \label{eq:continuity}
    0 &= \dot\rho + 3H \left(\rho + \frac{p}{\si{\clight\squared}} \right) , 
\end{align}
where $G$ is Newton's gravitational constant, $H=\frac{\dot{a}}{a}$ is the Hubble parameter, $K\in\{-1,0,+1\}$ is the sign of the spatial curvature\footnote{Note that there are different conventions in the treatment of the spatial curvature parameter~$K$. Here, we absorb any arbitrariness in the magnitude of~$K$ into the radial coordinate and the scale factor~$a$, such that $K$ only takes one of $\{-1,0,+1\}$ (see e.g.~\cite{Hobson2006ch14.6} for more details). Consequently we generally have~$a_0\ne1$, in contrast to the flat case.} corresponding to open, flat and closed universes respectively, and~$\Lambda$ is the cosmological constant. Dots denote derivatives with respect to cosmic time.

\Cref{eq:friedmann1} can be re-expressed in terms of density parameters~$\Omega$ by introducing the critical density 
\begin{align}
\label{eq:Omega_K}
    \rho_\mathrm{crit} &\equiv \frac{3 H^2}{8 \pi G} , &
    \Omega_i &\equiv \frac{\rho_i}{\rho_\mathrm{crit}} , &
    \Omega_K &\equiv -\frac{Kc^2}{(aH)^2} ,
\end{align}
\begin{equation}
    1 = \sum_i \Omega_i , \quad i\in\{\mathrm{r, m, \Lambda, K, \phi}\} ,
\end{equation}
where the index~$i$ runs over all relevant types of fluids such as radiation~r, matter~m, dark energy (cosmological constant)~$\Lambda$, a scalar inflaton field~$\phi$, or curvature~$K$.

We will now switch to \emph{reduced} Planck units with the Planck mass~$\si{\planckmass}\equiv\sqrt{\frac{\si{\hbar\clight}}{8\pi G}}$ and with $\si{\clight}=\hbar=1$. We assume that a single, scalar field~$\phi$, which we call the inflaton, dominates over all other species early in the history of the Universe, with the possible exception of curvature.
The energy density~$\rho$ and pressure~$p$ associated with the inflaton field~$\phi$ are
\begin{align}
    \label{eq:rho_phi}
    \rho_\phi &= \tfrac{1}{2} \dot\phi^2 + V(\phi) , \\
    \label{eq:p_phi}
    p_\phi &= \tfrac{1}{2} \dot\phi^2 - V(\phi) , 
\end{align}
where $V(\phi)$ is the potential of the inflaton field.
Inserting the energy density and pressure into \cref{eq:friedmann1,eq:friedmann2,eq:continuity} and switching to reduced Planck units we get the background equations for the evolution of the inflaton field early on, before contributions from radiation and matter become relevant:
\begin{align}
	\label{eq:background1} 
	H^2 + \frac{K}{a^2} &= \frac{1}{\SI{3}{\planckmass\squared}} \left(\tfrac{1}{2} \dot\phi^2 + V(\phi)\right) , \\
	\label{eq:background2}
    \frac{\ddot a}{a} = \dot H + H^2 &= - \frac{1}{\SI{3}{\planckmass\squared}} \left(\dot\phi^2 - V(\phi)\right) , \\
    \label{eq:eom}
    0 &= \ddot\phi + 3 H \dot\phi + V'(\phi) .
\end{align}
\Cref{eq:background1,eq:background2} can be combined to give a slightly simpler, potential-independent expression:
\begin{align}
    \label{eq:background3}
    \dot H = - \frac{1}{\SI{2}{\planckmass\squared}} \dot\phi^2 + \frac{K}{a^2} . 
\end{align}{}

Inflation is commonly defined as a period of accelerated expansion from which we can derive the following equivalent definitions of inflation:
\begin{align}
    \label{eq:def_acceleration}
    \ddot a &> 0 , \\
    \label{eq:def_cHH}
    \dv{t}(aH)^{-1} &< 0 , \\
    \label{eq:def_inflaton}
    V(\phi) &> \dot\phi^2 . 
\end{align}
\Cref{eq:def_cHH} is a direct consequence of \cref{eq:def_acceleration} and provides a more practical way of defining inflation, since it directly relates to the horizon and flatness problem which motivated inflation in the first place.
\Cref{eq:def_inflaton} can be derived from the previous definitions together with the second Friedmann \cref{eq:background2}. It links the definition of inflation to the inflaton field~$\phi$ and its time derivative~$\dot\phi$, which we will use in the sections to follow for setting initial conditions at the start of inflation.

The equation of state for the inflaton field is
\begin{align}
    w_\phi \equiv \frac{p_\phi}{\rho_\phi} = \frac{\frac{1}{2} \dot\phi^2 - V(\phi)}{\frac{1}{2} \dot\phi^2 + V(\phi)}
\end{align}
and can be used to differentiate between different regimes such as kinetic dominance~(KD) or slow-roll~(SR):
\begin{align}
    \label{eq:eos}
	w_\phi 
    \left\{
    \begin{aligned}
    	&\approx \hphantom{-}1 && \text{kinetic dominance}, &  \dot\phi^2 \gg V(\phi) ,\\
        &> -\tfrac{1}{3}\vphantom{\dot\phi^2} && \text{no inflation} , & \\
        &< -\tfrac{1}{3}\vphantom{\dot\phi^2} && \text{(fast-roll) inflation} , & \\
        &\approx -1 && \text{slow-roll inflation}, &  \dot\phi^2 \ll V(\phi) .
    \end{aligned}
    \right .
\end{align}

\section{Finite inflation}
\label{sec:finite_inflation}

In this paper we seek to investigate further the effect of a finite amount of inflation on various quantities such as the comoving Hubble horizon, conformal time, or the primordial power spectrum of curvature perturbations and gravitational waves. 
Where there is only a finite amount of inflation, there is also a start to inflation, which will influence our choice of initial conditions, as we will discuss further in \cref{sec:initial_conditions}.

\begin{figure}[tb]
    \includegraphics[width=\columnwidth]{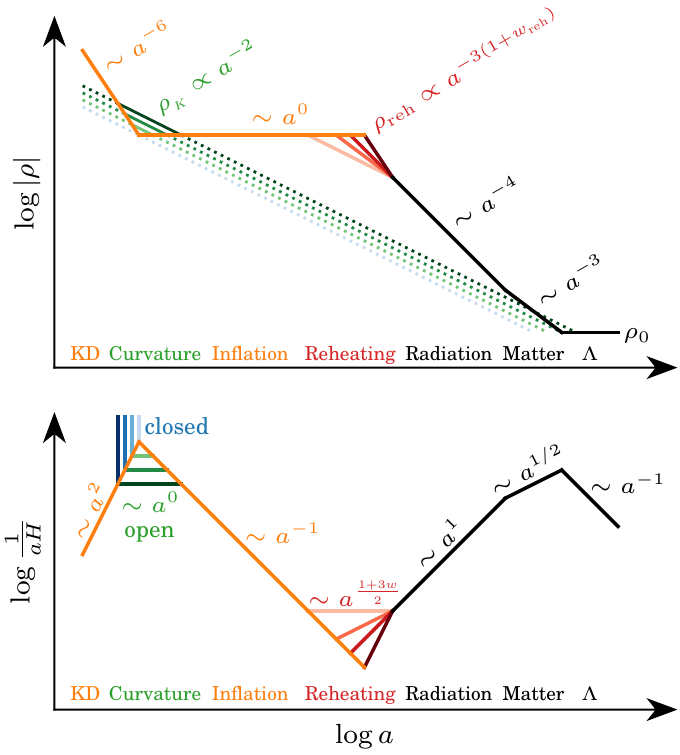}
    \caption{\label{fig:univolution_curvature} Sketch of the evolution of the energy density~$\rho$ and the comoving Hubble horizon~$(aH)^{-1}$. We highlight the possible effect of different levels of curvature in green. Darker shades of green correspond to higher curvature densities. 
    The dotted lines indicate how the curvature becomes increasingly less relevant during inflation.
    If curvature were to dominate over the sum of all other components (inflaton, radiation, matter, dark energy) in a closed universe, it would lead to collapsing universes (diverging comoving Hubble horizon), as sketched in blue. 
    Orange corresponds to evolution in the absence of curvature or where curvature is negligible, starting out in kinetic dominance~(KD) and then transitioning into inflation.
    Black corresponds to the standard Big Bang evolution from radiation, to matter and roughly today to $\Lambda$ (or dark energy) domination.
    The red lines correspond to different reheating scenarios parametrised by the equation of state parameter of reheating~$w_\mathrm{reh}$.}
\end{figure}

There are various mechanisms that can prevent inflation at early times. In this paper we focus on two components: kinetic dominance~(KD) and spatial curvature. Their effects on the energy density~$\rho$ and the comoving Hubble horizon~$a_0/(aH)$ are shown in \cref{fig:univolution_curvature,fig:cHH_curvature} and detailed in the following \cref{sec:KD,sec:CD}. Similarly, inhomogeneities in the inflaton field might prevent inflation at early times, as sketched in \cref{fig:univolution_inhomogeneities}. We briefly comment on inhomogeneities in \cref{sec:inhomogeneities}, but for simplicity we neglect them for the remainder of this paper and focus on KD and curvature as the regimes preceding inflation. 

\subsection{Kinetic dominance (KD)}
\label{sec:KD}

In \begin{NoHyper}\citeay{Hergt2019a}\end{NoHyper}~\cite{Hergt2019a} we worked in a spatially flat universe and compared the effect of setting the initial conditions for inflation during SR with $\dot\phi^2 \nobreak\ll\nobreak V(\phi)$, or KD with $\dot\phi^2 \nobreak\gg\nobreak V(\phi)$ (cf.\ \cref{eq:eos}). In both regimes the initial conditions for the background \cref{eq:background1,eq:background2,eq:eom,eq:background3} can be conveniently expressed in analytic form~\cite{Handley2014,Hergt2019a}.

In \cref{fig:univolution_curvature} (orange line) we schematically illustrate the role of both the KD and SR regime in the overall evolution of the Universe. Note in particular that the energy density scales as
\begin{align}
    \rho_\mathrm{KD} \propto a^{-6} && \text{and} && \rho_\mathrm{SR} \propto a^0 , \nonumber
\end{align}
and the comoving Hubble horizon as
\begin{align}
    \hphantom{M}(aH)^{-1}_\mathrm{KD} \propto a^2 && \text{and} && (aH)^{-1}_\mathrm{SR} \propto a^{-1} . \nonumber
\end{align}

With the inclusion of spatial curvature, we prefer setting the initial conditions at the start of inflation, i.e.\ at the turnover point where the comoving Hubble horizon becomes maximal and changes from growing during KD to shrinking during SR.
This prevents (at the prior level) running into spatially closed universes that collapse (the comoving Hubble horizon diverges) even before inflation has actually started. 
Nevertheless, we can integrate the inflationary background \cref{eq:background1,eq:background2,eq:eom,eq:background3} both forwards and backwards in time to recover the SR and the KD regime respectively. Forwards in time, SR inflation is an attractor solution, meaning that regardless of where in $(\phi, \dot\phi)$~phase-space the inflaton starts from, it will end up in the SR regime, where the inflaton ``slowly rolls down the potential'', motivating its name (see also \cite{Belinsky1985, Belinsky1988, Mishra2018, Goldwirth1992, Hergt2019, Chowdhury2019}). Integrating backwards in time, the attractor solution is kinetic dominance~\cite{Handley2014, Hergt2019}.

\subsection{Curvature domination}
\label{sec:CD}

\begin{figure}[tb]
    \includegraphics[scale=1.10]{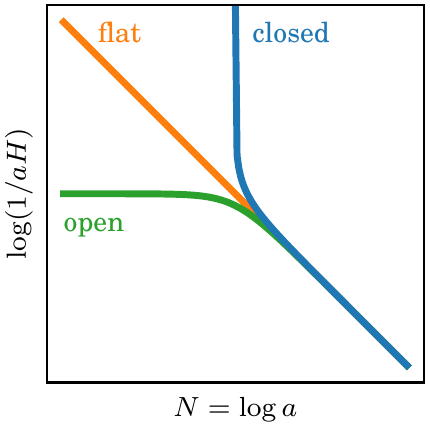}
    \caption{\label{fig:cHH_curvature} Schematic view of effects from significant amounts of spatial curvature on the evolution of the comoving Hubble horizon preceding and during inflation, where we assume slow-roll conditions, such that any term involving~$\dot\phi$ is neglected and the potential~$V$ is constant. The comoving Hubble horizon then becomes $(aH)^{-1}=(V\e^{2N}-\SI{3}{\planckmass\squared}K)^{-1/2}$. Open universes are constrained by \cref{eq:openconstraint}. For significantly closed universes the comoving Hubble horizon diverges.}
\end{figure}

If the energy density of spatial curvature~$\rho_K$ is of the same order as that of the inflaton field, then this leads to visible effects in the comoving Hubble horizon in the transition region between KD and SR. This is shown schematically in \cref{fig:univolution_curvature,fig:cHH_curvature}.

\Cref{fig:cHH_curvature} highlights the symmetry between open (green) and closed (blue) universes around the flat case (orange). During inflation and going backwards in time, the comoving Hubble horizon grows in proportionality to the scale factor~$a$. Due to the direct connection between the comoving Hubble horizon~$(aH)^{-1}$ and the curvature density parameter~$\Omega_K$ (by definition of $\Omega_K$ in \cref{eq:Omega_K}), this implies that the curvature density grows as well. Once the curvature density becomes dominant, the comoving Hubble horizon diverges for a closed universe and levels off for an open universe. Thus, we can only really refer to a phase of curvature \emph{domination} in the open case. Nonetheless, effects of curvature on the comoving Hubble horizon can already be seen for lower (i.e.\ large but not dominating) levels of curvature, as we show in more detail in \cref{fig:cHH}.
An actual divergence in the closed case could have its origin in a coasting or bouncing universe. Note that bouncing universe models typically involve some modification of general relativity, a formulation of quantum gravity or additional assumptions about the inflaton field (see~\cite{Battefeld2015} for a review). 
Viewed from the opposite end by starting in kinetic dominance, a diverging comoving Hubble horizon would in almost all cases correspond to a collapsing universe. 
The plateau in the comoving Hubble horizon in the open case could in principle reach back indefinitely, if the universe started exactly with $\Omega_K=1$. Curvature below that value would mean a preceding phase of KD.

\Cref{fig:univolution_curvature} illustrates schematically the role of curvature in the early evolution of the Universe. 
With $\rho_K \propto a^{-2}$ the curvature density drops slower than the kinetic energy density from the inflaton field and therefore becomes more relevant in the vicinity of the local maximum of the comoving Hubble horizon, causing the horizon to flatten or sharpen 
in the open or closed case, respectively. Once the inflaton potential comes to dominate over the kinetic term, its energy density, which scaled as~$a^{-6}$ during KD, becomes constant during inflation, thereby quickly exceeding the curvature density.

\subsection{Inhomogeneities}
\label{sec:inhomogeneities}

\begin{figure}[tb]
    \includegraphics[width=\columnwidth]{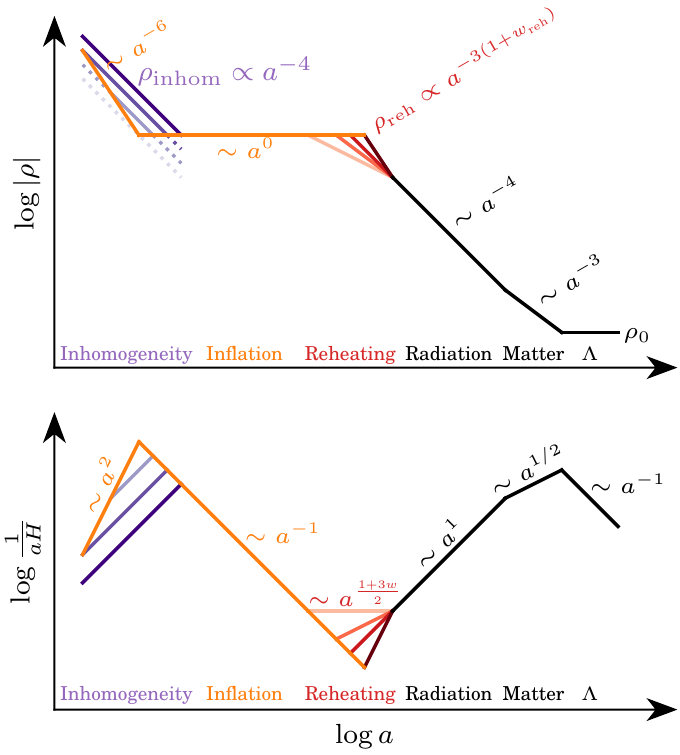}
    \caption{\label{fig:univolution_inhomogeneities} Sketch of the evolution of the energy density~$\rho$ and the comoving Hubble horizon~$(aH)^{-1}$ as in \cref{fig:univolution_curvature}. This time we highlight in purple the possible effect of different levels of inhomogeneities prior to the start of inflation.}
\end{figure}

Various analyses~\cite{Goldwirth1990,Goldwirth1991,Goldwirth1992,East2016,Clough2017,Chowdhury2019} have found that inhomogeneities cause a pre-inflationary phase where the energy density scales as 
\begin{align}
    \rho_\mathrm{inhom} \propto a^{-4} .
\end{align}
This is sketched in \cref{fig:univolution_inhomogeneities}, which also demonstrates how high levels of primordial inhomogeneities might hide a period of kinetic dominance. However, like kinetic dominance, inhomogeneities will lead to a definite start to inflation.

\section{Initial Conditions for inflation}
\label{sec:initial_conditions}


In \cref{sec:KD} we have already mentioned that KD leads to analytic expressions for the evolution of background variables, which may serve as initial conditions for the inflaton field~$\phi$. However, it is equally possible to set the initial conditions at the start of inflation and then numerically integrate both backwards in time towards kinetic dominance and forwards in time into inflation. This gives us a better handle on new parameters that are introduced through the addition of curvature such as the primordial curvature density parameter at the start of inflation~$\Omega_{k,\mathrm{i}}$. We will label parameters referring to the start of inflation with a subscript~``$\mathrm{i}$'', mnemonic for ``initial'' or ``inflation''.

At the start of inflation we can set the inequalities in the definitions for inflation from \cref{eq:def_acceleration,eq:def_cHH,eq:def_inflaton,eq:eos} to equality, leading to the following expressions:
\begin{align}
    \label{eq:inflationstart_def_dda}
    \ddot a_\mathrm{i} &= 0 , \\
    \label{eq:inflationstart_def_cHH}
    \dv{t}(a_\mathrm{i}H_\mathrm{i})^{-1} &= 0 , \\
    \label{eq:inflationstart_def_V_dphi2}
    V(\phi_\mathrm{i}) &= \dot\phi_\mathrm{i}^2 , \\
    \label{eq:inflationstart_def_w}
    w_{\phi,\mathrm{i}} &= - \tfrac{1}{3}. 
\end{align}
This simplifies the background \cref{eq:background1} and, together with \cref{eq:Omega_K} for the curvature density parameter, allows us to relate the scale factor~$a_\mathrm{i}$, the inflaton field~$\phi_\mathrm{i}$, and the curvature density parameter~$\Omega_{K,\mathrm{i}}$ at the start of inflation:
\begin{align}
	\label{eq:backgroundi} 
	\begin{split}
	H_\mathrm{i}^2 
	&\overset{(\ref{eq:background1})}{=} \frac{V(\phi_\mathrm{i})}{\SI{2}{\planckmass\squared}} - \frac{K}{a_\mathrm{i}^2} , \\
	&\overset{(\ref{eq:Omega_K})}{=} \frac{-K}{a_\mathrm{i}^2 \Omega_{K,\mathrm{i}}} . 
	\end{split}
\end{align}

Numerically integrating the background \cref{eq:background1,eq:background2} and the equation of motion for the inflaton field \cref{eq:eom} requires initial values for the variables $\{N, \phi, \dot\phi\}$, where $N = \ln a$ is the number of e-folds of the scale factor and $a$ is measured in reduced Planck units. \Cref{eq:inflationstart_def_V_dphi2} links the initial value for the time derivative of the inflaton field~$\dot\phi_\mathrm{i}$ to the potential of the inflaton field~$V(\phi_\mathrm{i})$ at the start of inflation. This leaves the e-folds~$N_\mathrm{i}$ and the inflaton field~$\phi_\mathrm{i}$ as free parameters. The initial value of the curvature density parameter~$\Omega_{K,\mathrm{i}}$ can be derived using \cref{eq:backgroundi} and hence could be varied in place of either~$N_\mathrm{i}$ or~$\phi_\mathrm{i}$:
\begin{align}
\label{eq:inflationstart_Omega_Ki}
    \Omega_{K,\mathrm{i}} &= \left[ 1 - \frac{V(\phi_\mathrm{i})}{\SI{2}{\planckmass}K} \e^{2N_\mathrm{i}} \right]^{-1}, \\
\label{eq:inflationstart_N_i}
	N_\mathrm{i} &= \tfrac{1}{2} \ln\left[ \frac{2\si{\planckmass\squared}K}{V(\phi_\mathrm{i})} \left( 1 - \Omega_{K,\mathrm{i}}^{-1} \right) \right] .
\end{align}
From this we can derive the condition that the primordial curvature density needs to be
\begin{align}
    \label{eq:openconstraint}
    \Omega_{K,\mathrm{i}} < 1
\end{align}
in order for inflation to start after the Big Bang. Equality would correspond to $N_\mathrm{i}\nobreak\rightarrow\nobreak-\infty$ or $a_\mathrm{i}=0$, i.e.\ inflation starting at the Big Bang.

The initial value for the inflaton field~$\phi_\mathrm{i}$ determines the amount of e-folds of inflation. Hence, it can be useful to infer~$\phi_\mathrm{i}$ from a desired number of e-folds of inflation. Going forward, we will consider the total e-folds of inflation~$N_\mathrm{tot}$ and the e-folds of inflation~$N_\dagger$ \emph{before} and~$N_\ast$ \emph{after} the pivot scale $k_\ast=\SI{0.05}{\per\mega\parsec}$ crosses the comoving Hubble horizon.

\begin{figure}[tb]
    \includegraphics[width=\columnwidth]{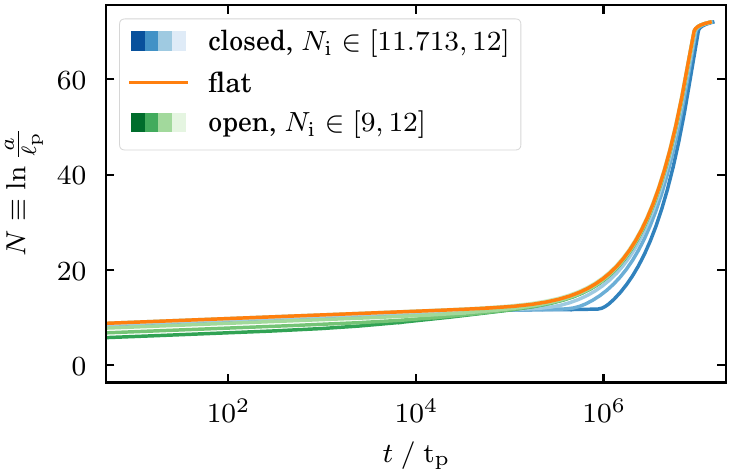}
    \vspace{-1em}
    \caption{\label{fig:ivar} Evolution of the logarithm of the scale factor~$N\equiv\ln(a/\si{\plancklength})$ with respect to physical time~$t$ for closed ($\Omega_K<0$), flat ($\Omega_K=0$) and open ($\Omega_K>0$) universes. Both of these variables are used as independent variables in \cref{fig:background} where we also detail the generation of these curves. The scale factor and physical time are given in the reduced Planck units for length~$\si{\plancklength}$ and time~$\si{\plancktime}$ respectively.}
\end{figure}

\begin{figure*}[p]
    \includegraphics[width=\textwidth]{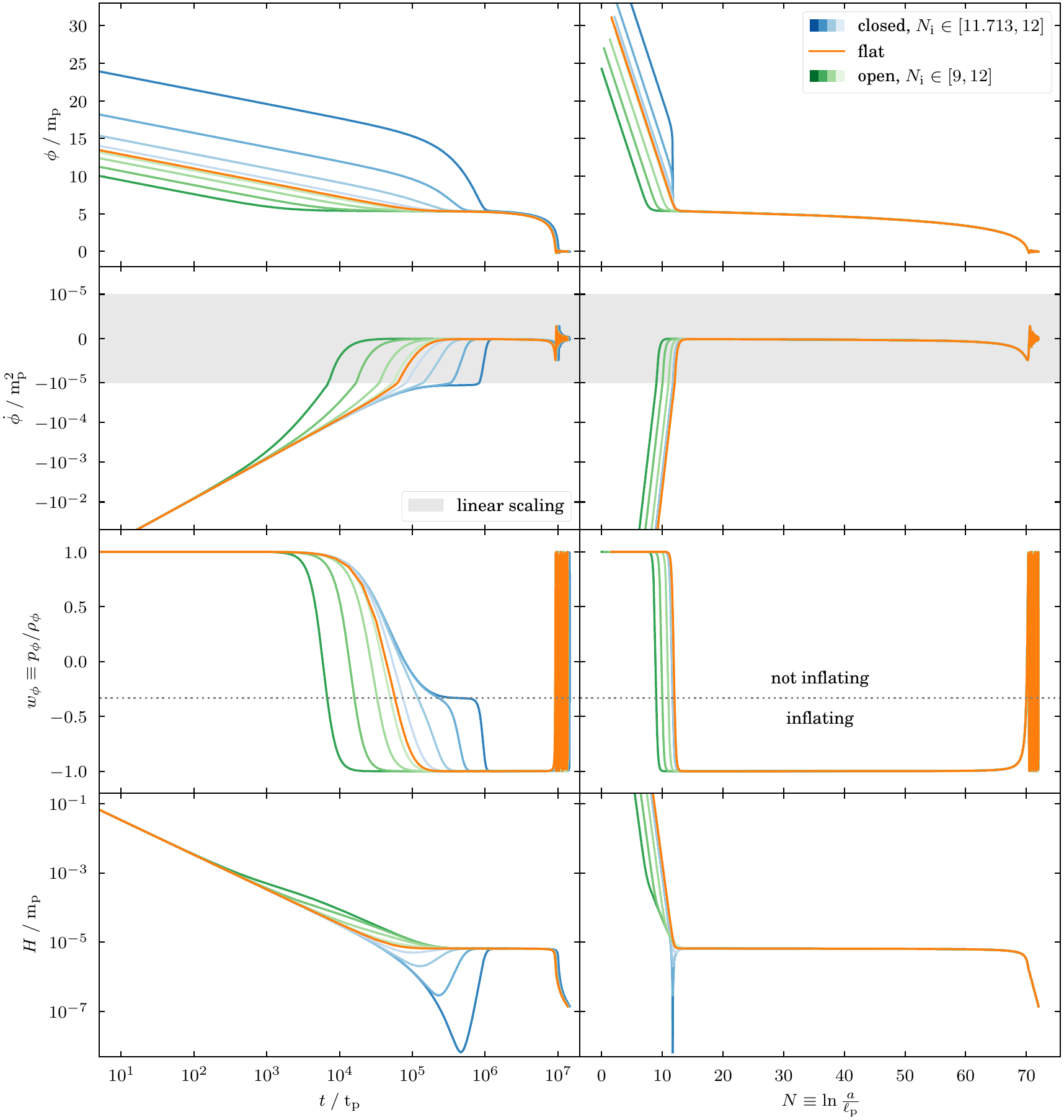}
    \caption{\label{fig:background}Evolution of background variables during inflation for closed ($\Omega_K<0$), flat ($\Omega_K=0$) and open ($\Omega_K>0$) universes. The rows show the inflaton field~$\phi$, its first time derivative~$\dot\phi$ with a semi-logarithmic scaling with threshold at $|\dot\phi|=\num{e-5}$, the equation-of-state parameter~$w_\phi$ with the horizontal dotted line indicating the threshold $w_\phi=-\frac{1}{3}$ between inflating and non-inflating, and the Hubble parameter~$H$, respectively. The left column is with respect to physical time~$t$ and the right column with respect to the natural logarithm of the scale factor~$N\equiv\ln(a/\si{\plancklength})$ (see also \cref{fig:ivar} for their interdependence). The curves were initialised at the start of inflation where $V(\phi_\mathrm{i})=\dot\phi_\mathrm{i}^2$ and from there integrated backwards and forwards in time. For visualisation purposes inflation was specified to end at $N_\mathrm{end}=70$ and the start of inflation was varied uniformly for the closed case within $N_\mathrm{i}\in[9.4, 10]$ and for the open case within $N_\mathrm{i}\in[7, 10]$. These ranges can be converted to the primordial curvature density parameter corresponding roughly to $\Omega_{K,\mathrm{i}}\in[-300, -3]$ and $\Omega_{K,\mathrm{i}}\in[0.997, 0.451]$, respectively. Note that these plots were generated using the Starobinsky potential, but the general picture remains qualitatively mostly the same independent of the choice of potential.}
\end{figure*}

Using these initial conditions we can integrate \cref{eq:background1,eq:background2,eq:eom,eq:background3} forwards and backwards with respect to cosmic time~$t$ or with respect to the number of e-folds~$N$ of the scale factor. The connection between the independent variables~$t$ and~$N$ is shown in \cref{fig:ivar}, illustrating the exponential growth of the universe during inflation. We can also compute various other quantities such as the inflaton field~$\phi$, its time derivative~$\dot\phi$, the equation of state parameter~$w_\phi$ or the Hubble parameter~$H$, all of which are shown in \cref{fig:background} with respect to cosmic time in the left column and with respect to e-folds in the right column. We show the solutions for a flat universe in orange. In green and blue we show the slightly different evolution of open and closed universes respectively. We show these curved cases for different amounts of primordial curvature, which we achieve by varying the starting point~$N_\mathrm{i}$. For a clean visualisation we chose the initial conditions such that inflation ends at~$N_\mathrm{end}=70$.

\section{Linking primordial to present-day scales}
\label{sec:calibrations}

In order to link primordial to present-day scales we need to first calibrate the scale factor~$a$ and the wavenumber~$k$ associated with curvature perturbations, which we briefly review in this section. 

Curved universes have an advantage over flat universes when discussing scales in that \cref{eq:Omega_K} provides a direct link between the curvature density parameter and the scale factor. Given today's curvature density parameter~$\Omega_{K,0}$, this allows a calibration of the scale factor without any knowledge of the evolution of the universe. 
Otherwise, as is the case for flat universes, we would have to make additional assumptions, e.g.\ by introducing a free parameter on the observable e-folds~$N_\ast$ from horizon crossing to the end of inflation or by making specific assumptions about the evolution of the Universe between the end of inflation and before the standard Big Bang evolution, i.e.\ about the epoch of reheating.

\subsection{Calibration of the present-day scale factor}
\label{sec:calibrate_a0}

In order to calculate the comoving Hubble horizon or the primordial power spectrum, we need to first calibrate the scale factor~$a$, which in this paper we do by deriving the present-day scale factor~$a_0$ from the present-day curvature density parameter~$\Omega_{K,0}$ and Hubble parameter~$H_0$. This follows directly from \cref{eq:Omega_K}:
\begin{align}
    \label{eq:a0}
    a_0 = \frac{1}{H_0} \sqrt{\frac{-K}{\Omega_{K,0}}} .
\end{align}

\subsection{Calibration of the wavenumber of primordial perturbations}

As is standard practice, we formulate the condition for horizon crossing in terms of the comoving Hubble horizon (as opposed to the particle horizon). 
We will evolve the gauge-invariant curvature perturbations~$\mathcal{R}_k$ for a given wavenumber~$k$.
Its reciprocal~$1/k$ (ignoring possible factors of $2\pi$ that, one could argue, should be introduced) can be thought of as the comoving wavelength scale of the perturbation itself~\cite{Hobson2006ch16.15}.
Whilst the length-scale~$1/k$ of perturbations is smaller than the comoving Hubble horizon, the curvature perturbations oscillate. From the definition for inflation in \cref{eq:def_cHH} we know that the comoving Hubble horizon shrinks during inflation. Once it drops below $1/k$, the oscillations stop and the curvature perturbations ``freeze'', as the corresponding modes have become larger than the characteristic length-scale over which physical processes operate coherently.
We use the transition point, which we refer to as horizon crossing, to link any given curvature perturbation observable today to the comoving Hubble horizon:
\begin{align}
    k = \frac{aH}{a_0} .
\end{align}
This allows us to draw the dotted line in \cref{fig:cHH} representing the pivot scale~$k_\ast=\SI{0.05}{\per\mega\parsec}$.

\section{The comoving Hubble horizon}
\label{sec:cHH}

\begin{figure*}[tb]
    \subfloat{\includegraphics[scale=1.05]{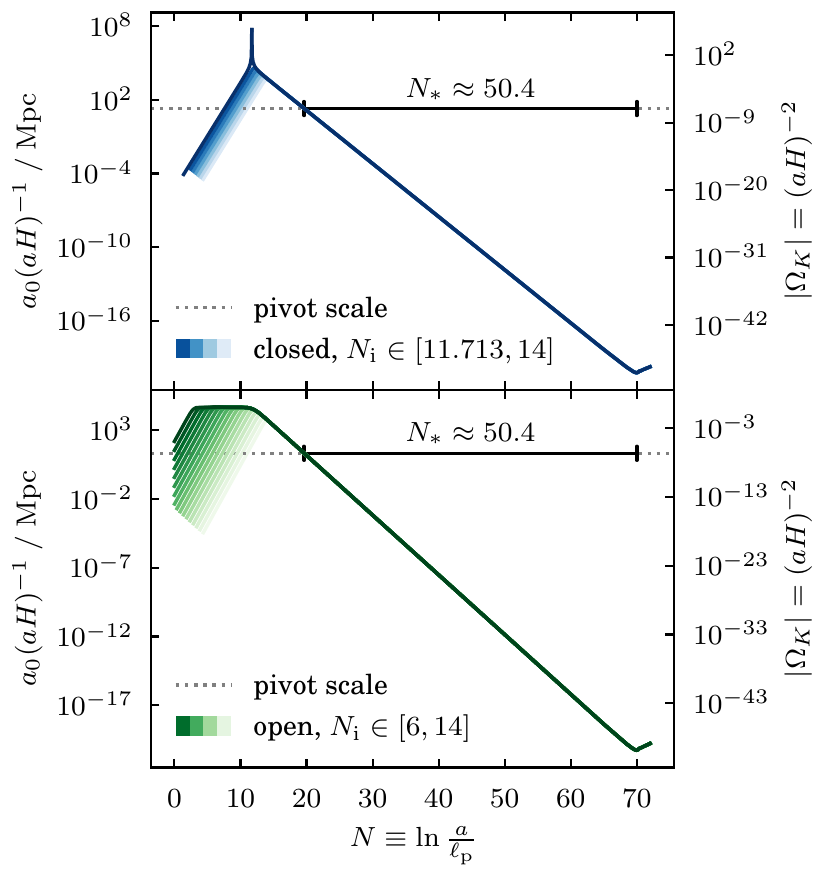}} 
    \hfill
    \subfloat{\includegraphics[scale=1.05]{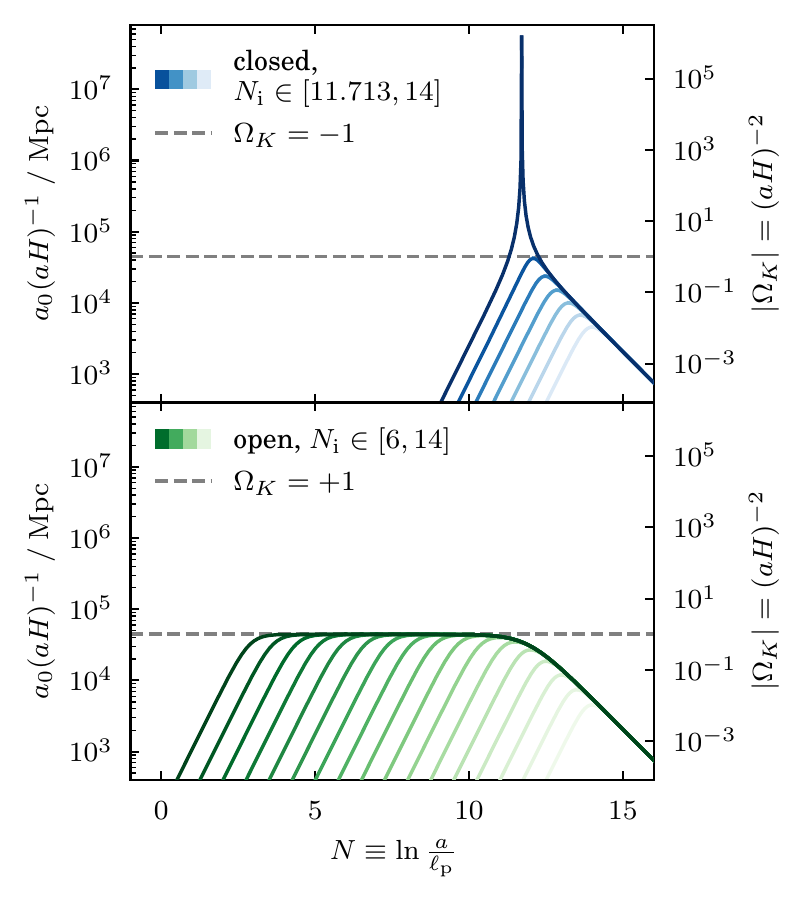}} 
    \vspace{-1em}
    \caption{\label{fig:cHH}Evolution of the comoving Hubble horizon~$a_0 (aH)^{-1}$ with respect to number of e-folds of the scale factor~$N\equiv\ln a$. The secondary y-axis relates the comoving Hubble horizon to the absolute value of the curvature density parameter~$|\Omega_K|=(aH)^{-2}$ for curved universes. As in \cref{fig:background}, the initial conditions were set such that inflation ends at $N_\mathrm{end}=70$ and with a varying start of inflation~$N_\mathrm{i}$. For the comoving Hubble parameter we also need to specify today's scale factor~$a_0$, which can be derived from today's curvature density parameter~$\Omega_{K,0}$ and \cref{eq:a0}. Here, we have set $|\Omega_{K,0}|=0.01$. This effectively fixes the number of e-folds~$N_\ast$ from horizon crossing of the pivot scale~$k_\ast=\SI{0.05}{\mega\parsec}$ to the end of inflation. The right hand plot is a zoom-in into the region of the start of inflation where the shape for closed universes~(upper panels) differs to the one of open universes~(lower panels). The dashed line marks the limit of the constraint for open universes~$\Omega_K<1$ from \cref{eq:openconstraint}. Note that these plots were generated using the Starobinsky potential, but the general picture remains qualitatively the same independent of the choice of potential.}
\end{figure*}

\Cref{fig:cHH} contrasts the evolution of the comoving Hubble horizon in closed and open universes for varying amounts of primordial curvature. For visualisation purposes, we calibrate today's scale factor~$a_0$ as described in \cref{sec:calibrate_a0} by fixing today's curvature density parameter~$\abs{\Omega_{K,0}}=0.01$. Fixing~$\Omega_{K,0}$ yields a linear relation between the evolution of the comoving Hubble horizon and the evolution of the curvature density parameter, and we therefore plot both, on opposite $y$-axes. This makes it apparent how inflation solves the flatness problem, as the shrinking comoving Hubble horizon during inflation (by definition in \cref{eq:def_cHH}) corresponds to the shrinking of the curvature density parameter, such that the standard Big Bang evolution thereafter starts out with a sufficiently small curvature density parameter. For any given total number of e-folds of inflation~$N_\mathrm{tot}$ the choice of $\abs{\Omega_{K,0}}$ decides how many e-folds pass \emph{before} versus \emph{after} horizon crossing of the pivot scale~$k_\ast$. We refer to these numbers of e-folds with~$N_\dagger$ and~$N_\ast$ respectively. The e-folds~$N_\ast$ (i.e. after horizon crossing of the pivot scale) we also call the \emph{observable} number of e-folds of inflation because of their direct connection to primordial cosmological parameters (e.g.\ the scalar spectral index~$n_\mathrm{s}$) in flat slow-roll inflation models, where the total number of e-folds is typically assumed to be much larger but ultimately unknown.

In \cref{fig:cHH}, we vary the start of inflation~$N_\mathrm{i}$ while keeping the end of inflation fixed to $N_\mathrm{end} = 70$ the same way as in \cref{fig:ivar,fig:background}.
This effectively also fixes the number of e-folds after horizon crossing of the pivot scale to $N_\ast \approx 50.4$. The total number of e-folds on the other hand shrinks with larger $N_\mathrm{i}$ as $N_\mathrm{tot} = N_\mathrm{end} - N_\mathrm{i}$ and thus the initial value for the inflaton field~$\phi_\mathrm{i}$ decreases as well.

\Cref{eq:inflationstart_Omega_Ki} links the primordial curvature density parameter~$\Omega_{K,\mathrm{i}}$ to the e-folds~$N_\mathrm{i}$ at inflation start. Thus, a smaller~$N_\mathrm{i}$ means a larger~$|\Omega_{K,\mathrm{i}}|$ and in turn a larger comoving Hubble horizon at inflation start. For open universes this gets capped by the constraint from \cref{eq:openconstraint}, meaning for very early starts of inflation the primordial curvature density parameter tends to unity: 
\begin{align}
    \label{eq:inflationstart_limit_open}
    \Omega_{K,\mathrm{i}} &\rightarrow +1 &&\Longleftrightarrow& N_\mathrm{i} &\rightarrow -\infty . 
\intertext{
For closed universes on the other hand, the primordial curvature density parameter diverges as $N_\mathrm{i}$ is pushed to earlier times:
}
    \label{eq:inflationstart_limit_closed}
    \Omega_{K,\mathrm{i}} &\rightarrow -\infty &&\Longleftrightarrow& N_\mathrm{i} &\rightarrow \tfrac{1}{2} \ln(\frac{\SI{2}{\planckmass\squared}K}{V(\phi)}) .
\end{align}
Note how a small amplitude of the inflationary potential (as expected from data) in \cref{eq:inflationstart_limit_closed} pushes inflation start until \emph{after} the Planck epoch for closed universes: $N_\mathrm{i} > N_\mathrm{p} = 0$.
For small levels of primordial curvature, the shape of the curve in \cref{fig:cHH} is the same for open and closed universes and matches that of a flat universe. With increasing primordial curvature the curve becomes flatter for open universes and pointier for closed universes, moulding to the limits expressed in \cref{eq:inflationstart_limit_open,eq:inflationstart_limit_closed} and sketched out in \cref{fig:cHH_curvature}.

It has frequently been proposed that it would be more natural to count the number of e-folds during inflation in terms of the comoving Hubble horizon~$(aH)^{-1}$ instead of the scale factor~$a$, because of its direct relation with the flatness and the horizon problem~\cite{Liddle1994,Liddle2003,Civiletti2020}.
Where curvature effects are negligible, i.e.\ where $\Omega_{K,\mathrm{i}}<1$, these measures are actually closely related due to the comoving Hubble horizon scaling as~$a^{-1}$ during slow-roll inflation, independently of the geometry of the universe. With the primordial curvature~$\Omega_{K,\mathrm{i}}$ approaching unity at the start of inflation, this common scaling breaks down.
The behaviour of the comoving Hubble horizon in a closed universe is inverse to that in an open universe, as shown in \cref{fig:cHH_curvature}.

For closed universes, the number of e-folds of the scale factor~$a$ are in fact more informative than e-folds of the comoving Hubble horizon~$(aH)^{-1}$ when it comes to effects of finite inflation on features in the primordial power spectrum~(PPS) of curvature perturbations. Finite inflation leads to a cutoff and oscillations towards large scales in the PPS (more on this later in \cref{sec:pps}). The position of the cutoff is governed by the number of e-folds~$N_\dagger$ of the scale factor before horizon crossing of the pivot scale. So, although for a late inflation start this number is closely related to the maximum of the comoving Hubble horizon at inflation start or the ratio~$f_\mathrm{i}=\Omega_{K,\mathrm{i}}/\Omega_{K,0}$ of primordial to present-day curvature, in the limit from \cref{eq:inflationstart_limit_closed} even a very large change in~$(a_\mathrm{i}H_\mathrm{i})^{-1}$ and~$f_\mathrm{i}$ will hardly affect~$N_\dagger$ and the cutoff position will cease shifting.

\section{Conformal time}
\label{sec:conformal_time}

\begin{figure}[tb]
    \includegraphics[width=\columnwidth]{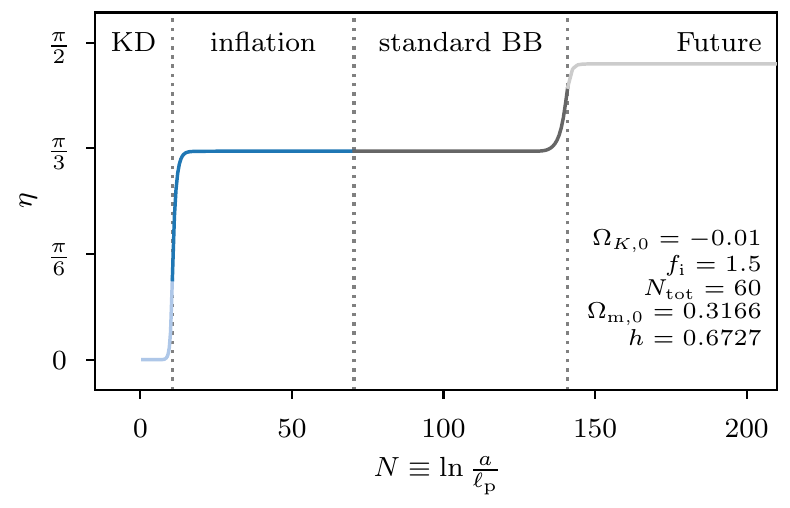}
    \vspace{-1em}
    \caption{\label{fig:conformal_time_evolution} 
    Evolution of conformal time~$\eta$ through different stages of the universe: kinetic dominance (KD) in light blue, inflation in blue, standard Big Bang evolution (radiation, matter and $\Lambda$ domination) in grey, and from today onwards in light grey.
    Note that this plots was generated using a Quadratic potential, but the general picture remains qualitatively the same independent of the choice of potential.
    }
\end{figure}

In order to solve the horizon problem, the amount of conformal time passed during inflation has to match or exceed the conformal time passing thereafter until today (see also~\cite{Liu2020}). Conformal time can be expressed in terms of the comoving Hubble horizon $(aH)^{-1}$ and the e-folds of the scale factor $N \equiv \ln a$:
\begin{align}
    \label{eq:conformal_time}
    \eta = \int \frac{\dd{t}}{a(t)} = \int \frac{\dd{\ln a}}{aH} . 
\end{align}
Comparing this expression to \cref{fig:cHH} and bearing in mind that the comoving Hubble horizon in the figure is shown on a logarithmic scale, it is clear that the largest contribution to the amount of conformal time passing prior to the end of inflation comes from the peak around the start of inflation. Analogously, the majority of conformal time passing after the end of inflation comes from the peak around the present-day comoving Hubble horizon. This is clear from the jumps in \cref{fig:conformal_time_evolution}, which shows the accumulation of conformal time from before inflation start until some future time. The regions where conformal time plateaus correspond to the regions where either the inflaton~$\phi$ or the cosmological constant~$\Lambda$ have made the comoving Hubble horizon shrink so much that there is almost no contribution to the integral in \cref{eq:conformal_time}.
This also holds for the post-inflationary epoch of reheating, which consequently can be neglected with regards to conformal time, which we will do throughout this section.

In this section we will be focusing on two quantities in particular: the total amount of conformal time~$\eta_\mathrm{total}$ passing from the Big Bang prior to inflation up to the future conformal boundary, and the ratio $\eta_\mathrm{before}/\eta_\mathrm{after}$ of conformal time passing before to after the end of inflation. 

\begin{figure*}[tb]
    \subfloat[\label{subfig:conformal_time_total_Ok}Total conformal time~$\eta_\mathrm{total}$.]
        {\includegraphics[width=0.95\columnwidth]{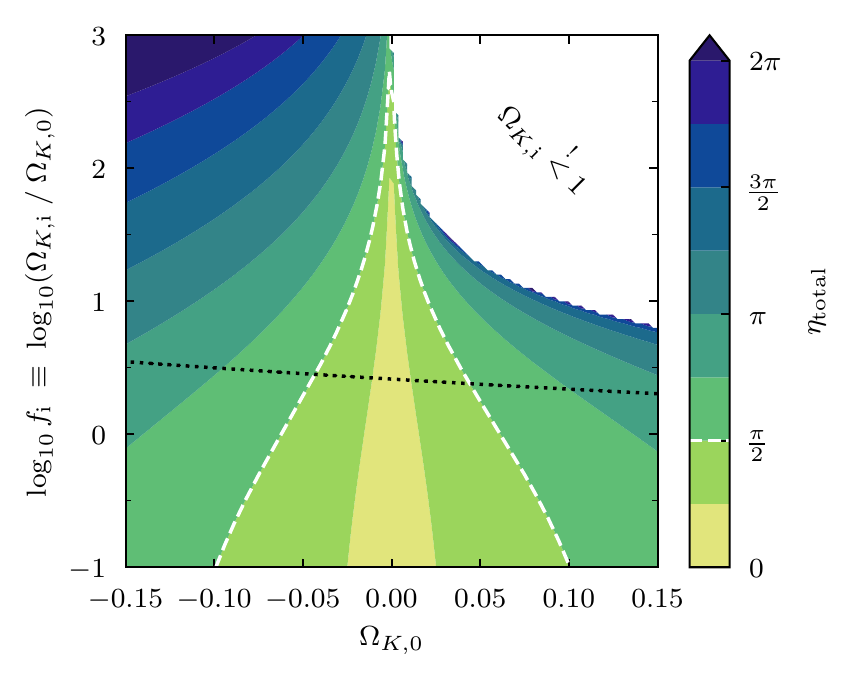}} 
    \hfill
    \subfloat[\label{subfig:conformal_time_total_Oi}Total conformal time~$\eta_\mathrm{total}$.]
        {\includegraphics[width=0.95\columnwidth]{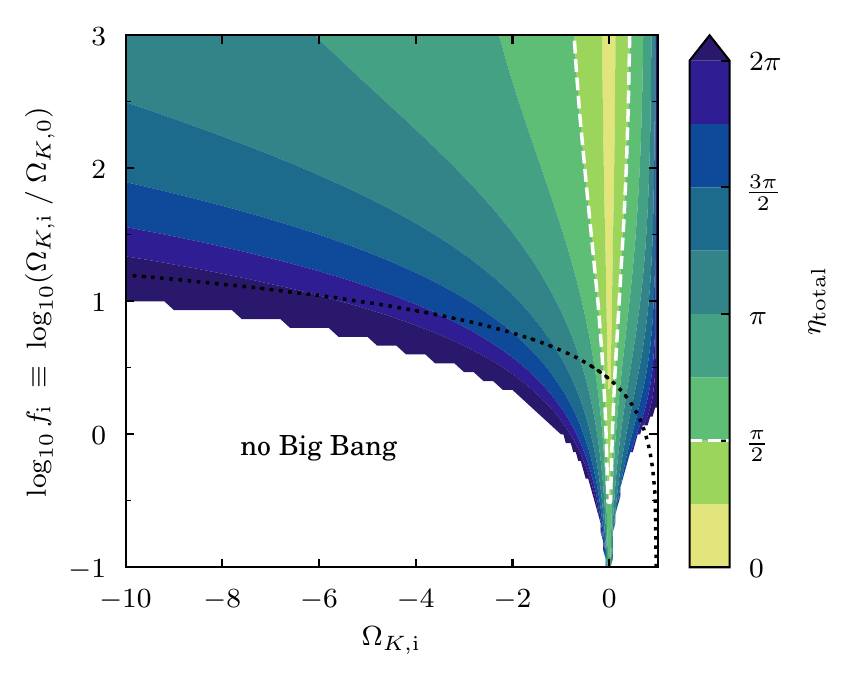}} 
    \\
    \subfloat[\label{subfig:conformal_time_ratio}Conformal time ratio~$\eta_\mathrm{before}/\eta_\mathrm{after}$.]
        {\includegraphics[width=0.95\columnwidth]{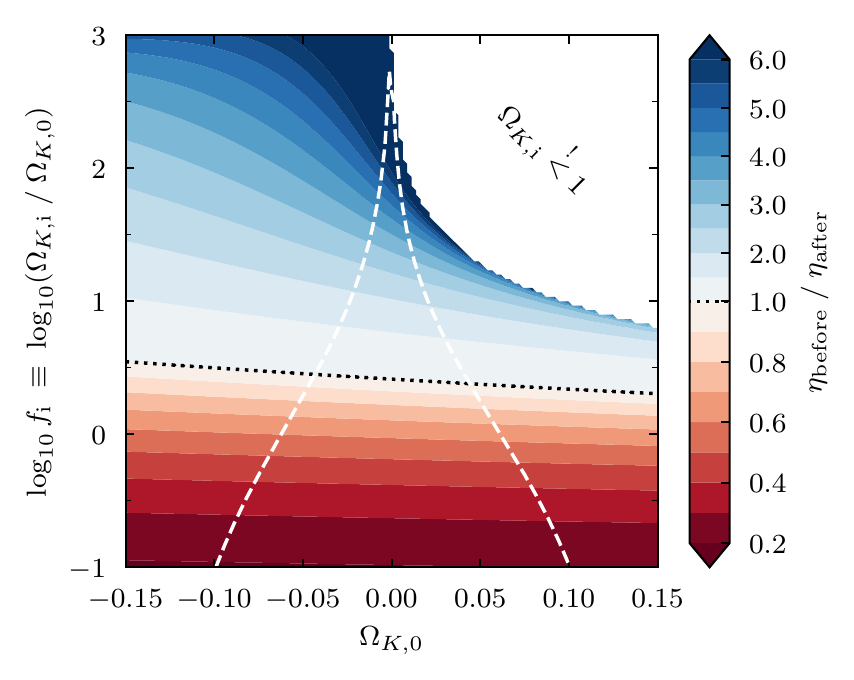}}
    \hfill
    \subfloat[Conformal time ratio~$\eta_\mathrm{before}/\eta_\mathrm{after}$.]
        {\includegraphics[width=0.95\columnwidth]{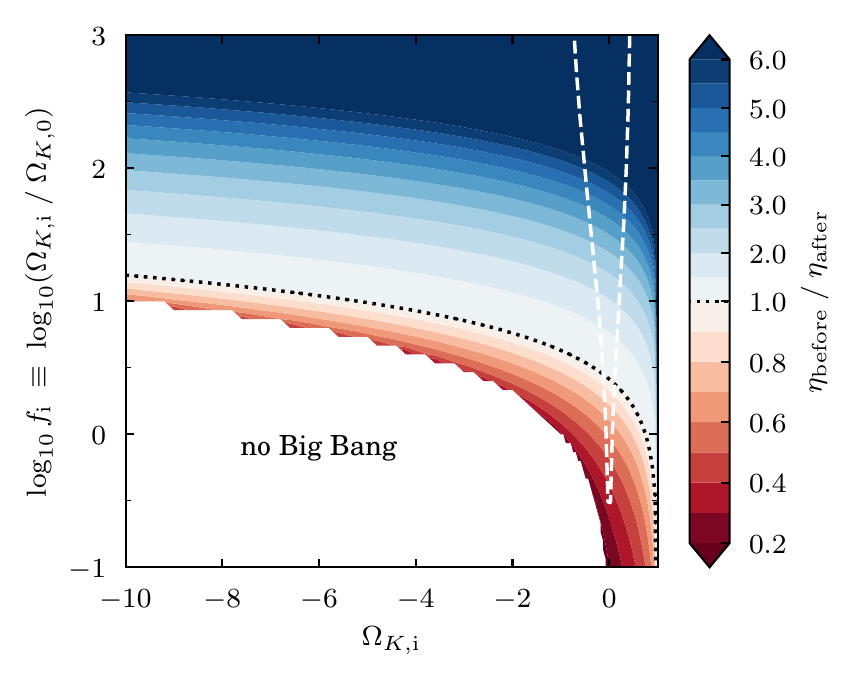}} 
    \caption{\label{fig:conformal_time} 
    The total conformal time~$\eta_\mathrm{total}$ from before inflation until the future conformal boundary and the ratio of conformal time \emph{before} to \emph{after} the end of inflation are shown dependent on the primordial and present-day density parameters, $\Omega_{K,\mathrm{i}}$ and $\Omega_{K,0}$ respectively. 
    The \textbf{white dashed line} highlights the value of $\eta_\mathrm{total}=\frac{\pi}{2}$, which in some closed universe theories~\cite{Lasenby2005} is predicted to be a constraint.
    The \textbf{black dotted line} indicates where~$\eta_\mathrm{after}=\eta_\mathrm{before}$. Thus, the \textbf{blue area} highlights where the horizon problem is solved and the \textbf{red area} where inflation was insufficient in order to solve the horizon problem.
    In all these cases the following parameters were fixed: the total number of e-folds of inflation~$N_\mathrm{tot}=60$, today's matter density parameter~$\Omega_{\mathrm{m},0}=0.3166$ and the dimensionless Hubble parameter~$h=0.6727$.
    Note that these plots were generated using a Quadratic potential, but the general picture remains qualitatively the same independent of the choice of potential.
    }
\end{figure*}

\begin{figure*}[tbp]
    \subfloat[\label{subfig:conformal_time_ratio_Nt} (in-)dependence on total number of e-folds~$N_\mathrm{tot}$]
        {\includegraphics[width=\columnwidth]{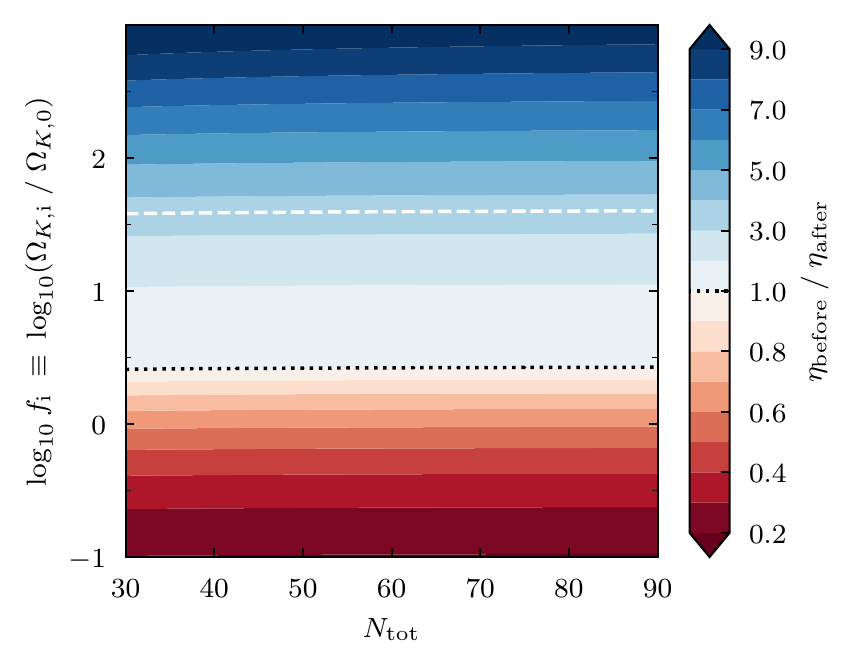}} 
    \hfill
    \subfloat[\label{subfig:conformal_time_ratio_Om} (weak) dependence on matter density~$\Omega_{\mathrm{m},0}$]
        {\includegraphics[width=\columnwidth]{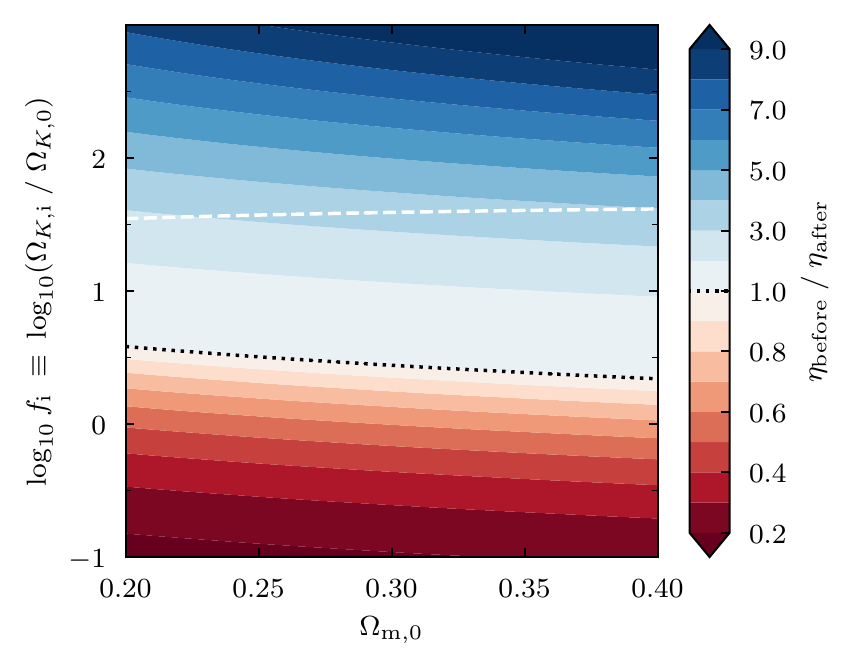}} 
    \caption{\label{fig:conformal_time_Ntot} 
    Conformal time ratio~$\eta_\mathrm{before}/\eta_\mathrm{after}$ as in \cref{fig:conformal_time}, but here exploring the (in-)dependence on the total number of e-folds of inflation~$N_\mathrm{tot}$ (left) and the present-day matter density parameter~$\Omega_{\mathrm{m},0}$ (right).
    The following parameters were fixed: today's curvature density parameter~$\Omega_{K,0}=-0.01$, the dimensionless Hubble parameter~$h=0.6727$, today's matter density parameter~$\Omega_{\mathrm{m},0}=0.3166$ (only in left panel), and the total number of e-folds of inflation~$N_\mathrm{tot}=60$ (only in right panel).
    Note that these plots were generated using a Quadratic potential, but the general picture remains qualitatively the same independent of the choice of potential.
    }
\end{figure*}

The total amount of conformal time~$\eta_\mathrm{total}$ is important for the closed universe theory described by \begin{NoHyper}\citeauthor{Lasenby2005}\end{NoHyper} in~\cite{Lasenby2005}. Here a natural boundary condition on the transition to the final asymptotic de Sitter state is found that requires the total conformal time available to the universe to equal~$\pi/2$. 

Also, as stated earlier, the ratio $\eta_\mathrm{before}/\eta_\mathrm{after}$ is important for addressing the horizon problem. In order to solve the horizon problem we require that more conformal time has passed before than after the end of inflation, which we will refer to as the horizon constraint:
\begin{align}
\label{eq:horizon_constraint}
    \frac{\eta_\mathrm{before}}{\eta_\mathrm{after}} > 1 .
\end{align}

\Cref{fig:conformal_time} illustrates how both the total amount of conformal time~$\eta_\mathrm{total}$ and the ratio $\eta_\mathrm{before}/\eta_\mathrm{after}$ depend on the primordial and present-day curvature density parameters, $\Omega_{K,\mathrm{i}}$ and $\Omega_{K,0}$ respectively. We plot both these parameters against the ratio of primordial to present-day curvature
\begin{align}
    f_\mathrm{i} \equiv \frac{\Omega_{K,\mathrm{i}}}{\Omega_{K,0}} .
\end{align}
The parameter~$f_\mathrm{i}$ will prove useful also later on for decoupling the effects of primordial and present-day curvature on the primordial power spectrum. 
Here, it is useful when looking at the black dotted line, which separates the plots into red regions where inflation was insufficient to solve the horizon problem, and blue regions where it was, i.e.\ the dotted line corresponds to $\eta_\mathrm{before}/\eta_\mathrm{after}=1$. As is particularly clear in \cref{subfig:conformal_time_ratio}, this separation depends primarily on $f_\mathrm{i}$ when considering a prior range of $\Omega_{K,0}\in[-0.15,0.15]$. From this we can infer that in order to solve the horizon problem we require
\begin{align}
    \label{eq:conformal_time_ratio_requirement}
    \log_{10} f_\mathrm{i} \gtrsim 0.5 .
\end{align}

The white regions in the left panels with $\log_{10}f_\mathrm{i}$ versus $\Omega_{K,0}$ correspond to the constraint for open universes from \cref{eq:openconstraint}, also seen in \cref{fig:cHH}. The white regions in the right panels correspond to universes that would have collapsed in the past (labelled ``no Big Bang'') or that would collapse in the future before reaching the future conformal boundary.

The possible constraint of a total conformal time of $\eta_\mathrm{total}=\pi/2$ (white dashed lines) can be satisfied while also resolving the horizon problem, as part of the white dashed line lies in the blue region. This would push the present day universe close to flat (see \Cref{subfig:conformal_time_total_Ok}) and the primordial curvature density parameter close to unity (see \cref{subfig:conformal_time_total_Oi}).

Besides the primordial and present-day curvature density parameters, there are some other parameters (the number of e-folds~$N_\mathrm{tot}$, matter density~$\Omega_\mathrm{m,0}$ and Hubble parameter~$H_0$) that enter into the calculation of both the total conformal time as well as the conformal time ratio. However, their contribution to conformal time is negligible compared to that of the curvature parameters as seen in \cref{fig:conformal_time_Ntot}. 

For the inflationary part of the calculation we additionally need to consider the mass of the inflaton and the duration of inflation. The mass of the inflaton (or the amplitude of the inflationary potential) can be mapped to the amplitude of the primordial power spectrum, which, as we will see in \cref{fig:AsfoH_cHH_PPS}, has no effect on the comoving Hubble horizon and is thus irrelevant for the calculations of conformal time. The total inflationary e-folds~$N_\mathrm{tot}$ only influence the comoving Hubble horizon towards the end of inflation (see also \cref{fig:AsfoH_cHH_PPS}). At that point the comoving Hubble horizon is many orders of magnitude smaller than at its start and consequently this contribution to the integral for conformal time in \cref{eq:conformal_time} is negligible. \Cref{subfig:conformal_time_ratio_Nt} illustrates how the conformal time ratio is almost independent of the total amount of inflation~$N_\mathrm{tot}$.
Note that this goes against the rule of thumb of requiring order \SI{60}{\efolds} of inflation to solve the horizon problem, which is valid when the start of inflation is fixed. \Cref{subfig:conformal_time_ratio_Nt} tells us that we can solve the horizon problem equally well for only \SI{30}{\efolds}. The essential thing is that the comoving Hubble horizon (or the curvature density parameter) needs to have been sufficiently large at the start of inflation compared to today, reinforcing the requirement from \cref{eq:conformal_time_ratio_requirement} that $\log_{10} f_\mathrm{i} \gtrsim 0.5$. Note, however, that we are investigating conformal time completely isolated from other possible constraints from reheating, here. We will investigate constraints from reheating in the following \cref{sec:reheating} and later in \cref{sec:result_reheating}.

For the calculation of conformal time after the end of inflation and throughout radiation, matter and $\Lambda$ domination we need to further consider today's matter density parameter~$\Omega_{\mathrm{m},0}$ and Hubble parameter~$H_0$. Through \cref{eq:a0} the Hubble parameter mostly serves as a normalisation factor to the scale factor~$a_0$ and therefore primarily only shifts the comoving Hubble horizon along~$\ln a$, which does not affect the  integral for conformal time in \cref{eq:conformal_time}. Increasing the present-day matter density parameter~$\Omega_{\mathrm{m},0}$ increases the matter contribution to the comoving Hubble horizon, which therefore becomes larger during matter domination in general and at the end of matter domination in the late-time Universe in particular, when the comoving Hubble horizon peaks (see also \cref{fig:univolution_curvature}). Thus, there is a dependence of the conformal time ratio~$\eta_\mathrm{before}/\eta_\mathrm{after}$ on the matter density parameter. However, for the range of $\Omega_{\mathrm{m},0} \in [0.2, 0.4]$ this dependence is weak compared to the dependence on~$f_\mathrm{i}$, as seen in \cref{subfig:conformal_time_ratio_Om}.

\section{Reheating}
\label{sec:reheating}

\begin{figure*}[tb]
    \includegraphics[scale=1.05]{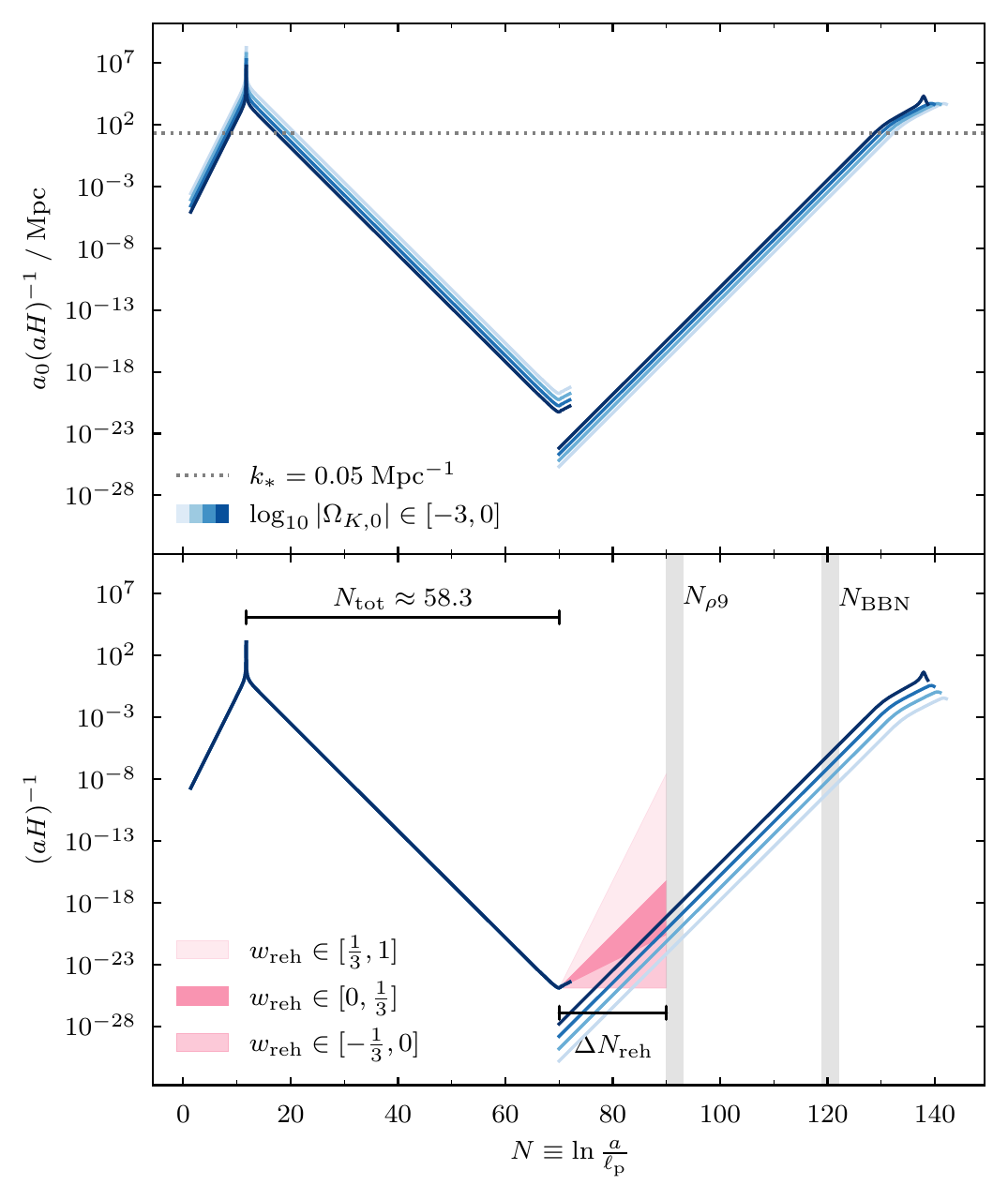}
    \vspace{-1em}
    \caption{\label{fig:reheating}The top panel shows the comoving Hubble horizon before, during and after inflation. Different from \cref{fig:cHH}, here, we vary today's curvature density parameter~$\Omega_{K,0}$ which governs the calibration between primordial and today's scales. Therefore we cannot simultaneously draw a $y$-axis for the evolution of the curvature density parameter anymore. The axis would be different for every $\Omega_{K,0}$. If we forego the linking to today's scales, we can actually map the curves during inflation onto one another as done in the lower panel, which allows for a better visualisation of the reheating period after inflation and before the standard Big Bang evolution (radiation, matter and $\Lambda$ domination). The shaded regions span the range of allowed values for the effective equation-of-state parameter~$w_\mathrm{reh}$ during reheating. The duration of reheating~$\Delta N_\mathrm{reh}$ is bounded at least by the requirement that reheating should have finished before Big Bang nucleosynthesis~(BBN) happens. Taking a stricter view, one might require that reheating has ended by the time~$N_{\rho9}$, when the energy density has dropped to $\rho_\mathrm{reh}^{1/4}=\SI{e9}{\giga\eV}$ (see also \cref{fig:univolution_curvature} for a sketch on the evolution of the energy density in parallel to the comoving Hubble horizon).}
\end{figure*}

While the start of inflation plays a crucial role for considerations of the conformal time and thereby also the horizon problem, the end of inflation is important for the period of reheating. Reheating links the primordial evolution of the Universe to the standard Big Bang evolution, comprised of radiation, matter, and~$\Lambda$ domination. Going back to \cref{fig:univolution_curvature}, we can see this schematically for the energy density~$\rho$ and the comoving Hubble horizon~$(aH)^{-1}$. For the reheating period we plot four characteristic equation-of-state parameters~$w$: the lower limit~$w_\mathrm{reh}=-1/3$, matter domination $w_\mathrm{reh}=0$, radiation domination $w_\mathrm{reh}=1/3$ and the upper limit $w_\mathrm{reh}=1$.

The link between primordial and standard Big Bang evolution becomes particularly important in the case of the Universe having non-zero curvature, since the latter informs us about the overall scale of the Universe, as established in \cref{sec:calibrate_a0,sec:cHH}.
This dependence on the curvature density parameter~$\Omega_{K,0}$ is illustrated in \cref{fig:reheating} showing the comoving Hubble horizon (upper panel). 
Different from \cref{fig:cHH_curvature}, in \cref{fig:reheating} we fix both start~$N_\mathrm{i}$ and end~$N_\mathrm{end}$ of inflation while allowing today's curvature density parameter~$\Omega_{K,0}$ to vary. Consequently we can no longer show the evolution of the comoving Hubble horizon~$a_0/(aH)$ and the curvature density parameter~$\Omega_K$ in one plot, since today's curvature density parameter~$\Omega_{K,0}$ serves as calibrator for today's scale factor~$a_0$ which in turn calibrates the comoving Hubble horizon. We therefore also plot an uncalibrated version~$(aH)^{-1}$, where the primordial evolution collapses onto a single line, whereas the standard Big Bang evolution shifts vertically with~$\Omega_{K,0}$.

We parametrise the epoch of reheating through an effective equation-of-state parameter~$w_\mathrm{reh}$, a duration~$\Delta N_\mathrm{reh}$ and an energy scale~$\rho_\mathrm{reh}$ where thermalisation is guaranteed to have occurred (see also~\cite{Adshead2011} for more details on this reheating parametrisation). Note first that~$w_\mathrm{reh}$ is an effective parameter. During the inflaton's oscillations around a potential minimum at the end of inflation, the equation-of-state parameter also oscillates rapidly between~$\pm1$. For the \emph{effective} equation-of-state parameter we consider the time-averaged value. For a monomial potential with exponent~$p$ this gives $w_\mathrm{reh}=(p-2)/(p+2)$, e.g.\ we have $w_\mathrm{reh}=0$ for a quadratic and $w_\mathrm{reh}=1/3$ for a quartic potential. Since reheating is by definition a post-inflationary epoch, we at the very least expect that on average $w_\mathrm{reh}>-1/3$ (otherwise we would have more inflation, cf.\ \cref{eq:eos}). Additionally, the equation of state is typically capped at $w_\mathrm{reh}<1$ to avoid a super-luminal sound speed~\cite{Easther2012}. Together this leads to our first reheating constraint:
\begin{align}
\label{eq:reheating_constraint_w}
    -\tfrac{1}{3} < w_\mathrm{reh} < 1 .
\end{align}
Second, we note that there is little information on the energy scale of thermalisation~$\rho_\mathrm{th}$. Hence, we also view the energy scale~$\rho_\mathrm{reh}$ as an effective parameter by which thermalisation must have happened, but not necessarily equal to~$\rho_\mathrm{th}$. Thus, the case $\rho_\mathrm{th}>\rho_\mathrm{reh}$ will effectively be reflected in the equation-of-state parameter~$w_\mathrm{reh}$ incorporating part of the radiation dominated epoch and thereby tending towards $w=1/3$. In order for reheating not to affect any confirmed observations of the standard Big Bang cosmology, we require at the very least that the epoch of reheating must have happened before Big Bang nucleosynthesis~(BBN):
\begin{align}
\label{eq:reheating_constraint_N}
    N_\mathrm{reh} < N_\mathrm{BBN} = \ln(\frac{a_0}{1 + z_\mathrm{BBN}}) ,
\end{align}
where $a_0$ is inferred from the present-day curvature density parameter (see \cref{sec:calibrate_a0}) and where we use $z_\mathrm{BBN}=\num{e9}$ as a rough estimate of the epoch of BBN. 

The pink shaded regions in \cref{fig:reheating} subdivide the range for~$w_\mathrm{reh}$ from \cref{eq:reheating_constraint_w}, with the dividing lines given by $w_\mathrm{reh}=0$ corresponding to a matter dominated epoch of reheating and $w_\mathrm{reh}=1/3$ corresponding to a radiation dominated epoch of reheating. 
Requiring matter domination exactly, i.e.\ fixing $w_\mathrm{reh}=0$, is an often used model for reheating, because most single-field inflationary potentials can be approximated by the quadratic potential close to their minimum, and thus predict $w_\mathrm{reh}=0$. The duration of reheating~$\Delta N_\mathrm{reh}$ (or equivalently the energy scale of reheating~$\rho_\mathrm{reh}$) is still a free parameter in this case.
Radiation domination, on the other hand, would seamlessly continue into the standard Big Bang evolution and is therefore also referred to as instant reheating and often used as the most restrictive but simplest case of reheating, since it leads to $\Delta N_\mathrm{reh}=0$ and $\rho_\mathrm{reh}=V_\mathrm{end}=V(\phi_\mathrm{end})$. 

From the marginal variation of the standard Big Bang evolution owing to curvature, we can already deduce that the latter will barely affect the equation of state of reheating. Much more important is the role of the total amount of inflation~$N_\mathrm{tot}$, which determines whether inflation ends before or after the primordial curve crosses the radiation domination line in \cref{fig:reheating}. The crucial role curvature plays in this scenario is through the linking of scales between primordial and standard Big Bang evolution.

In the very permissive scenario outlined by \cref{eq:reheating_constraint_w,eq:reheating_constraint_N} linking primordial and late-time evolution will practically always be possible if inflation ends early, before crossing the radiation domination line. Otherwise an equation-of-state parameter $w_\mathrm{reh}>1/3$ will be required to catch up in time with the standard Big Bang evolution. In more restrictive settings such an equation-of-state parameter is typically excluded at the prior level~\cite{Planck2013Inflation,Planck2015Inflation,Planck2018Inflation}, which we will explore further in \cref{sec:result_reheating}.

\section{The primordial power spectrum~(PPS)}
\label{sec:pps} 

\subsection{Power-law PPS}
\label{sec:pps_power_law} 

In the base \LCDM\ cosmological model the primordial power spectrum of scalar curvature perturbations~$\mathcal{R}$ is phenomenologically described via two of its six free parameters in form of a simple power law:
\begin{align}
\label{eq:powerlaw_pps}
    \mathcal{P}_\mathcal{R}(k) = A_\mathrm{s} \left( \frac{k}{k_\ast} \right)^{n_\mathrm{s}-1} ,
\end{align}
where the power amplitude~$A_\mathrm{s}$ and the spectral index~$n_\mathrm{s}\equiv1+\dd\ln\mathcal{P}_\mathcal{R}(k_\ast)/\dd\ln k$ are the free parameters with the subscript ``s'' referring to scalar perturbations and where $k_\ast$ is a pivot scale in the window of observable wavenumbers~$k$. We choose to work with the commonly used pivot scale of~$k_\ast=\SI{0.05}{\per\mega\parsec}$. Note that we are using the primordial power spectrum in its dimensionless form.

One of the prime successes of large-field inflation so far is the prediction of a spectral index~$n_\mathrm{s}$ slightly smaller than unity, where unity would correspond to a scale-invariant power spectrum. This deviation from scale-invariance has been confirmed by the measurements of the \textsc{Planck} satellite to high precision~\cite{Planck2013Parameters,Planck2015Parameters,Planck2018Parameters}.

Another major prediction of inflation is the presence of primordial gravitational waves, typically parametrised by a (non-zero) tensor-to-scalar ratio~$r=A_\mathrm{t}/A_\mathrm{s}$. The PPS for gravitational waves is defined analogously to \cref{eq:powerlaw_pps} but the tensor spectral index~$n_\mathrm{t}$ is typically given without the `$-1$' in the exponent:
\begin{align}
\label{eq:powerlaw_pps_tensor}
    \mathcal{P}_\mathrm{t}(k) = r A_\mathrm{s} \left( \frac{k}{k_\ast} \right)^{n_\mathrm{t}}
\end{align}
For the tensor-to-scalar ratio~$r$ and the tensor spectral index~$n_\mathrm{t}$ we assume the inflation consistency relation for a single scalar field with a standard kinetic term~\cite{Lidsey1997,Planck2015Inflation}:
\begin{align}
    \label{eq:inflation_consistency}
    n_\mathrm{t} = -\frac{r}{8} \left( 2 - \frac{r}{8} - n_\mathrm{s} \right) . 
\end{align}

Besides the tensor-to-scalar ratio, another common extension to \cref{eq:powerlaw_pps} is an expansion to higher orders in $\dd\ln k$, introducing the running of the spectral index
\begin{align}
    n_\mathrm{run} = \dv{n_\mathrm{s}}{\ln k} = \dv[2]{\ln\mathcal{P}_\mathcal{R}(k_\ast)}{(\ln k)} .
\end{align}

In \cref{sec:potentials} we introduce various large-field inflation models and their predictions for the spectral index~$n_\mathrm{s}$, its running~$n_\mathrm{run}$ and the tensor-to-scalar ratio~$r$.

\subsection{The slow-roll approximate PPS}
\label{sec:pps_sr} 

Using just the background quantities~$a$, $H$, and~$\dot\phi$ from the solution to the background \cref{eq:background1,eq:background2,eq:eom} we can compute a slow-roll~(SR) approximation to the primordial power spectrum~(see e.g.~\cite{Lyth2009ch25} for a derivation):
\begin{align}
	\mathcal{P_R}(k) \approx \left( \frac{H^2}{2\pi \dot\phi} \right)^2_{k=aH} , 
    \label{eq:PPSapprox}
\end{align}
where the subscript expresses that the quantities need to be evaluated where each mode crosses the comoving Hubble horizon, i.e.\ where $k=aH$.  
This approximation is accurate on sufficiently small scales (large~$k$), where the PPS takes the form of an almost scale-invariant power-law, which motivates the phenomenological power-law spectrum from \cref{eq:powerlaw_pps}.

We can use the slow-roll approximation to make an estimate~$A_\mathrm{SR}$ of the amplitude parameter~$A_\mathrm{s}$ of the primordial scalar power spectrum in \cref{eq:powerlaw_pps}. We can further relate this to the amplitude parameter of the inflaton potential, which we will refer to as~$\Lambda$ (see \cref{eq:monomial,eq:starobinsky,eq:doublewell} for some specific potentials):
\begin{align}
    \label{eq:A_SR}
    A_\mathrm{SR} &= \frac{1}{\SI{12}{\pi\squared\planckmass\tothe6}} \left.\frac{V^3}{V'\,^2}\right|_{\phi=\phi_\ast}, \\
    \Lambda^4 &= \SI{12}{\pi\squared\planckmass\tothe6} A_\mathrm{SR} \left.\frac{v'\,^2}{v^3}\right|_{\phi=\phi_\ast}, \\ 
    &\quad \text{with $v(\phi)\equiv\frac{V(\phi)}{\Lambda^4}$}. \nonumber
\end{align}
The subscript asterisk indicates evaluation at the pivot scale.

In models of finite inflation the slow-roll approximation breaks down on large scales (small~$k$), where the modes have not started out from sufficiently well within the comoving Hubble horizon, and where the primordial power spectrum then exhibits a cutoff towards large scales. This cutoff behaviour can already be observed qualitatively in the approximate PPS, however to properly quantify this cutoff, we need to perform a full numerical integration of the primordial perturbations.

\Cref{fig:AsfoH_cHH_PPS} shows the approximate PPS and its dependence on various input parameters, which will be discussed in \cref{sec:parametrisation} in more detail.

\subsection{The full numerical PPS}

In order to solve the PPS numerically, we need to integrate the Mukhanov--Sasaki equation for the curvature perturbation~$\mathcal{R}_k$, which can be written as a damped harmonic oscillator with respect to cosmic time (adapted from \cite{Handley2019e}):
\begin{align}
    \label{eq:mukhanov-sasaki_scalar}
    & 0 = \mathcal{\ddot R}_k + 2 \gamma \mathcal{\dot R}_k + \omega^2 \mathcal{R}_k , \\
    \nonumber\\
    & \text{with damping} \quad 2 \gamma = \left( 3 + \xi \right) H , \\
    & \text{and frequency}\quad \omega^2 = \frac{\kappa^2}{a^2} - \frac{K}{a^2} \left(1+\xi\right) , 
\end{align}
where damping and frequency share the term 
\begin{equation*}
    \xi = \frac{2 \kappa^2}{\kappa^2+K\mathcal{E}} \left( \mathcal{E} + \frac{\ddot\phi}{H \dot\phi} + \Omega_K \right) ,
\end{equation*}
with 
\begin{equation*}
    \mathcal{E} = \frac{\dot\phi^2}{2 H^2}, \quad \frac{\ddot\phi}{H \dot\phi} = -3 - \frac{V'(\phi)}{H \dot\phi}, \quad \Omega_K = -\frac{K}{(a H)^2} . 
\end{equation*}
For the wavenumber~$k$ we use the following effective expression in curved spaces: 
\begin{align}
    &\kappa^2 = k^2 + k K (K + 1) - 3 K , \\
    &\quad\text{with}\quad
\begin{cases}
    k \in \mathbb{R}, k > 0 \quad\text{ if }\quad K=0, -1 , \\
    k \in \mathbb{Z}, k > 2 \quad\text{ if }\quad K=+1 .
\end{cases} \nonumber
\end{align}
We get the expression for the effective wavenumber~$\kappa$ from Fourier transforming the $\nabla_i\nabla^i$~operator in curved space~\cite{Handley2019e}. Note how the wavenumber becomes discrete with $k>2$ for positively curved (closed) universes.

In the small-scale limit $k \rightarrow \infty$ or in the absence of curvature $K = 0$ we recover the better known terms from the flat universe case (compare e.g.\ with equation 16.45 in~\cite{Hobson2006ch16.15}):
\begin{align}
    \label{eq:reduce_flat_1}
    \kappa^2 &\quad\longrightarrow\quad k^2 , \\
    \label{eq:reduce_flat_2}
    \omega^2 &\quad\longrightarrow\quad \frac{k^2}{a^2} , \\
    2 \gamma &\quad\longrightarrow\quad \frac{\dot\phi^2}{H} + 3 H + \frac{2 \ddot\phi}{\dot\phi} .
\end{align}

For tensor modes, the modification of the Mukhanov--Sasaki equation from the flat case to curvature is much simpler. The equivalent form of \cref{eq:mukhanov-sasaki_scalar} for tensor modes is 
\begin{align}
    \label{eq:mukhanov-sasaki_tensor}
    \ddot h + 3 H \dot h + \left( \frac{\kappa^2}{a^2} + \frac{5 K}{a^2} \right) h = 0 ,
\end{align}
which again reduces as expected to \cref{eq:reduce_flat_1,eq:reduce_flat_2} in the limit of small scales.

Using \texttt{oscode}'s~\cite{Oscode} efficient algorithm for oscillatory ordinary differential equations, we can integrate \cref{eq:mukhanov-sasaki_scalar,eq:mukhanov-sasaki_tensor} for each mode~$k$ from the start of inflation until well past horizon crossing for that given mode, where the frozen values of the primordial perturbations can be read off. 
We can then compute the primordial power spectra for scalar and tensor perturbations according to:
\begin{align}
    \label{eq:scalarpower}
	\mathcal{P_R}(k) &= \frac{k^3}{2\pi^2} \left| \mathcal{R}_k \right|^2 , \\
    \label{eq:tensorpower}
	\mathcal{P}_\mathrm{t}(k) &= 2 \cdot \mathcal{P}_h(k)  = 2 \cdot \frac{k^3}{2\pi^2} \left| h_k \right|^2 , 
\end{align}
where the factor 2 in the tensor spectrum comes from the two possible polarisation states of gravitational waves.

We initialise~$\mathcal{R}_k$ and~$h_k$ in their vacuum state defined as the state which minimises energy density via the renormalised stress-energy tensor~\cite{Handley2016,Handley2019e}:
\begin{align}
    \mathcal{R}_{k,\mathrm{i}} &= \frac{1}{z\sqrt{2 k}} , &
    \dot{\mathcal{R}}_{k,\mathrm{i}} &= - \frac{i k}{a} \mathcal{R}_{k,\mathrm{i}} , \\
    h_{k,\mathrm{i}} &= \frac{1}{a}\sqrt{\frac{2}{k}} , &
    \dot{h}_{k,\mathrm{i}} &= - \frac{i k}{a} h_{k,\mathrm{i}} . 
\end{align}
We prefer these initial conditions over similar formulations such as the commonly used Bunch Davies vacuum~\cite{Bunch1978}, because their predictions are stable across different choices of dependent or independent variables, i.e.\ they are invariant under canonical transformations~\cite{Agocs2020}.

\section{Inflationary potentials and slow-roll predictions}
\label{sec:potentials}

\begin{figure}[tb]
    \includegraphics[width=\columnwidth]{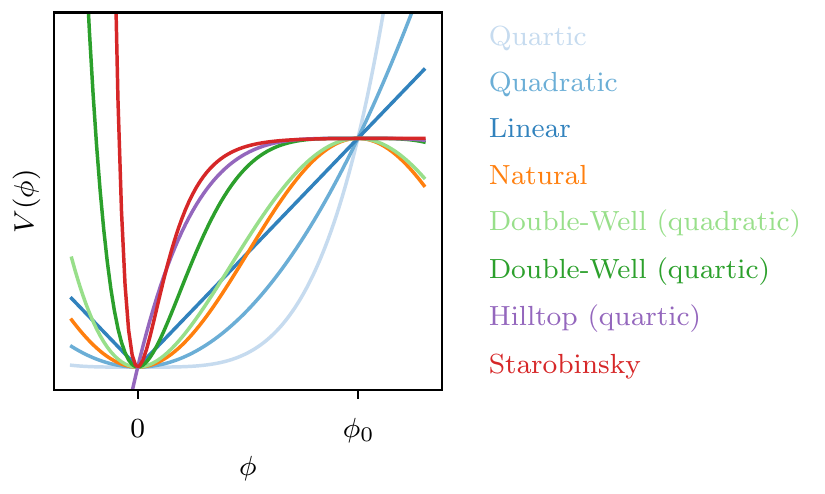}
    \vspace{-1em}
    \caption{\label{fig:potentials} Schematic view of various inflationary potentials that we investigate. If necessary, potentials are shifted such that the potential minimum and hence the end stage after inflation lies at the origin at $\phi=0$. Potentials with a local maximum are defined such that that maximum lies at the potential parameter~$\phi_0$ to the right of the origin. The potentials were plotted with the same potential amplitude~$\Lambda$, except for the quartic, quadratic and linear potentials which were rescaled for visualisation purposes such that they meet the other potentials in the point $V(\phi_0)$. In this form, the linear potential illustrates the categorisation into convex and concave potentials, which is frequently used in $n_\mathrm{s}$-$r$ plots, e.g.\ in \cref{fig:nsr_slow_roll}.}
\end{figure}

In this section we briefly review a few scalar single-field inflation models. \Cref{fig:potentials} shows a schematic view of the various inflationary potentials used in this paper. 
To ease the computation of the inflation models and their comparison with one another, we try to unify the notation by rewriting traditional formulations as follows.
They will share a potential amplitude parameter~$\Lambda$ (not to be confused with the cosmological constant~$\Lambda$) in units of the reduced Planck mass, $[\Lambda]=\si{\planckmass}$. The potential minimum $V=0$ will be located at the origin $\phi=0$ and any potential local maximum (for the natural, hilltop and double-well potentials) will be located at a parameter~$\phi_0$. The amplitude parameter, common to all inflationary potential, is linked directly to the power amplitude~$A_\mathrm{s}$ of scalar primordial perturbations.

Note that we only consider large-field inflation in this paper, i.e.\ models where the field excursion of the inflaton takes values greater than the Planck scale. Small-field inflation predicts a tensor-to-scalar ratio so small that it will remain unobservable for the near future~\cite{Lyth1997}.

\Cref{fig:nsr_slow_roll} illustrates the slow-roll~(SR) predictions for the tensor-to-scalar ratio~$r$, the spectral index~$n_\mathrm{s}$ and its running $n_\mathrm{run}=\dd n_\mathrm{s}/\dd\log k$.
Due to the wide dynamic range predicted for the tensor-to-scalar ratio, we show~$r$ scaled both linearly in the upper left and logarithmically in the lower left plot. In the \textsc{Planck} inflation papers~\cite{Planck2013Inflation, Planck2015Inflation, Planck2018Inflation} a linear scaling in~$r$ was preferred, however with upcoming CMB experiments such as the Simons Observatory~\cite{SimonsObservatoryScience}, the LiteBIRD satellite~\cite{LiteBird2020WhitePaper} or CMB\=/S4~\cite{CMBS4Science} pushing to a tensor-to-scalar ratio of about $r\sim\num{e-3}$, a logarithmic scaling of~$r$ allows for better visualisation and sampling of the smaller scales. For a recent discussion on uniform versus logarithmic priors on~$r$ and their effects on Bayesian model comparison see~\cite{Hergt2021}.
The slow-roll predictions for the running of the spectral index from all inflation models considered here, on the other hand, only span a small fraction of the posterior distribution, which can be seen in the upper right plot of \cref{fig:nsr_slow_roll}. For a better comparison of the predictions of the individual models we zoom in on the $n_\mathrm{run}$-range in the lower right plot. This highlights how the uncertainty of the running of the spectral index is far too large for the purpose of distinguishing between these inflation models.

Note that the P18 and P18+BK15 contours in \cref{fig:nsr_slow_roll} come from an extension of the base \LCDM\ cosmology not only with the tensor-to-scalar ratio and the running of the spectral index, but also with the spatial curvature parameter~$\Omega_{K,0}$. As such, the contours differ from those in the \textsc{Planck} inflation papers \cite{Planck2013Inflation,Planck2015Inflation,Planck2018Inflation} and in the BK15 paper~\cite{BicepKeck2018BKX}, which we elaborate on in our results in \cref{sec:result_extensions}. Note further that we only look at the SR predictions in this section, comparing them to the $n_\mathrm{s}$-$r$ and $n_\mathrm{s}$-$n_\mathrm{run}$ contours. The results from our nested sampling analysis follow later in \cref{sec:result_inflation}.

\subsection{Monomial potential}

The monomial potentials\footnote{Inflation models with a monomial potential are also referred to as large-field inflation (e.g.\ in the encyclop\ae{}dia inflationaris~\cite{EncyclopaediaInflationaris2014}), however, since large-field displacements are not unique to the monomial potential, we prefer naming the inflation models after their potential shapes, here. Yet another name often associated with the monomial potential is chaotic inflation~\cite{Linde1983}, but similarly chaotic inflation at its core actually pertains to the idea that the inflaton started from a chaotic initial state varying wildly from one place to another, rather than describing a specific potential model. See~\cite{Vilenkin2004} for a helpful discussion of terminology.}  are one of the simplest classes of inflationary potentials, given by: 
\begin{equation}
\label{eq:monomial}
    V(\phi) = \Lambda^4 \left( \frac{\phi}{\si{\planckmass}} \right)^p
\end{equation}

Inflation ends when the value of the inflaton field drops to 
\begin{equation}
    \phi_\mathrm{end} = \frac{\si{\planckmass}\,p}{\sqrt{2}} .
\end{equation}
With that we can approximate the number of e-folds of inflation from some~$\phi$ until the end of inflation to be
\begin{equation}
    N(\phi) \simeq \frac{\phi^2}{\SI{2}{\planckmass\squared}\,p}-\frac{p}{4} ,
\end{equation}
showing that the number of e-folds of inflation grow quadratically with the inflaton field value for monomial potentials. This scaling carries through to the spectral index, to the tensor-to-scalar ratio and to the running of the spectral index, where we get to leading order in~$1/N_\ast$:
\begin{align}
    n_\mathrm{s} &\simeq 1-\frac{2+p}{2N_\ast}, & 
    r &\simeq \frac{4p}{N_\ast}, & 
    n_\mathrm{run} &\simeq -\frac{2+p}{2N_\ast^2} .
\end{align}

\begin{figure*}[tb]
    \flushright
    \subfloat{\includegraphics[]{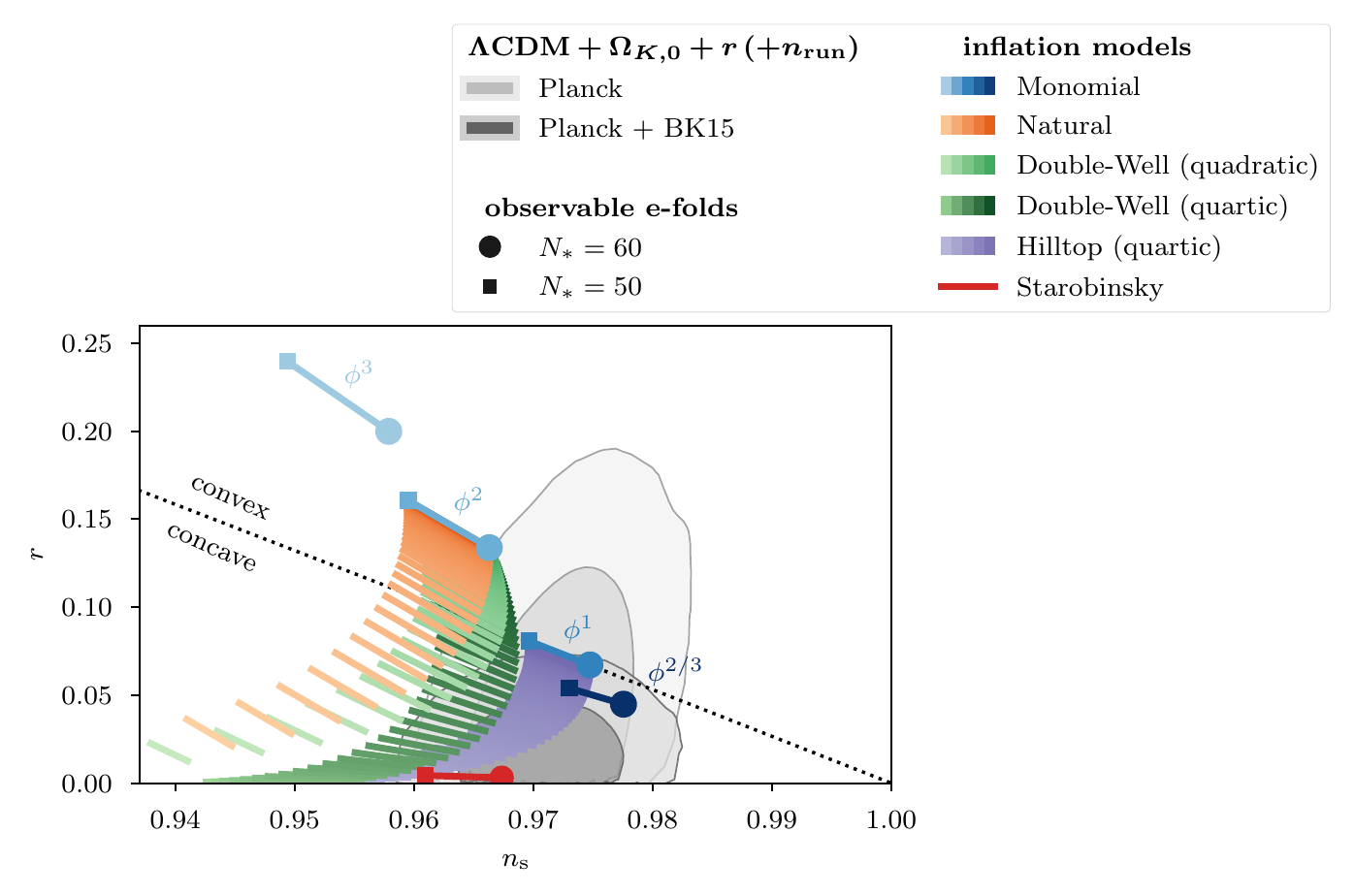}} \hspace*{-4.1cm}
    \subfloat{\includegraphics[]{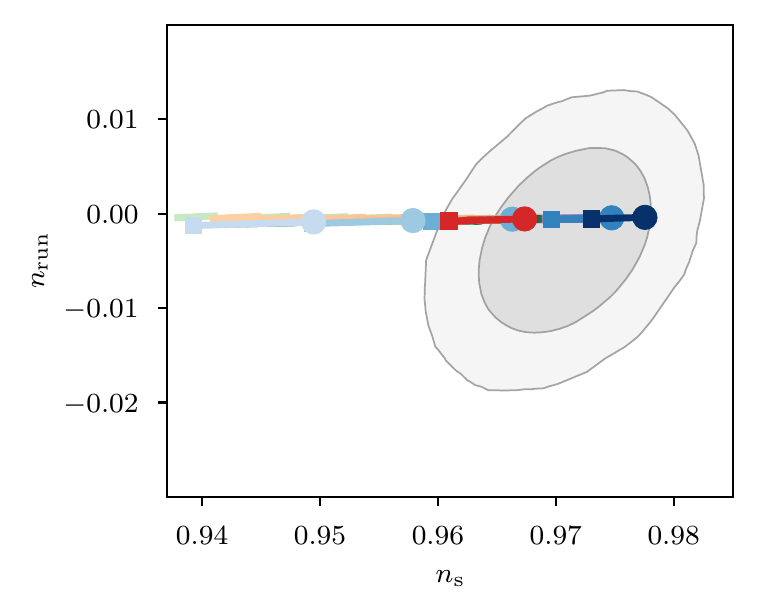}} 
    \\ \vspace{-0.85cm}
    \subfloat{\includegraphics[]{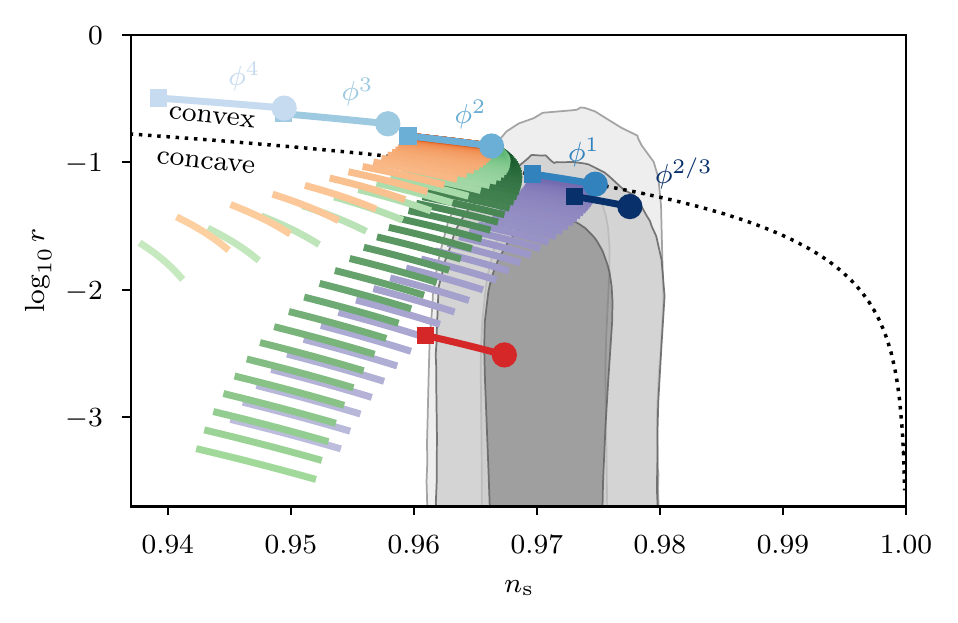}} 
    \subfloat{\includegraphics[]{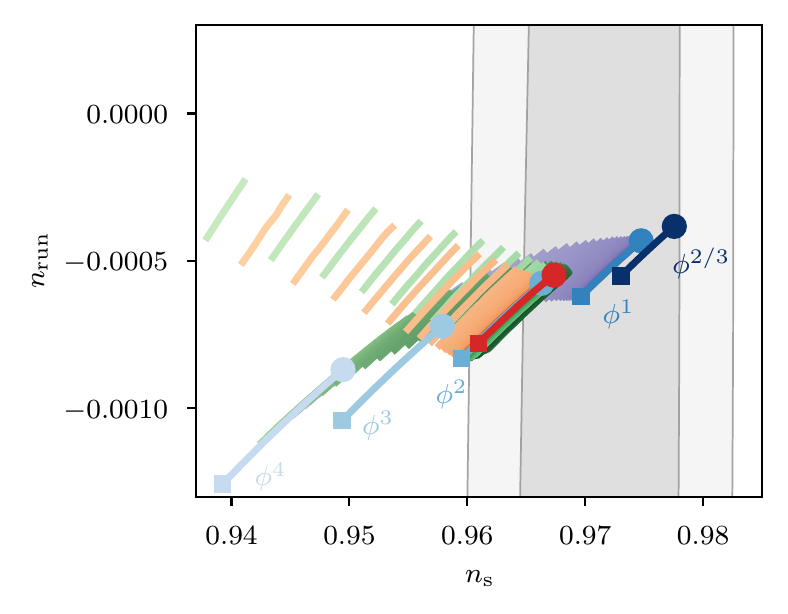}} 
    \vspace{-0.5cm}
    \caption{\label{fig:nsr_slow_roll} Comparison of slow-roll inflation model predictions for the spectral index~$n_\mathrm{s}$ against the tensor-to-scalar ratio~$r$ on the left and the running of the spectral index~$n_\mathrm{run}$ on the right. The top left plot is linear and the bottom left plot logarithmic in~$r$. The top right plot shows the full joint contour for~$n_\mathrm{s}$ and~$n_\mathrm{run}$ and the bottom right plot a zoom-in into the region relevant for the inflationary potentials.
    Note that for all these plots the curvature density parameter~$\Omega_{K,0}$ is one of the sampling parameters.
    For each inflation model we show the line(s) delimited by the requirement of producing~\SIrange[range-units=single]{50}{60}{\efolds} of observable inflation~$N_\ast$, which is the rough range needed for viable reheating scenarios. 
    In blue we show the slow-roll predictions for the monomial potentials from \cref{eq:monomial} and in red that of Starobinsky (or~$R^2$) inflation from \cref{eq:starobinsky}. 
    For the other inflation models we show a range of predictions for a range of values of the potential hill parameter~$\phi_0$, spaced logarithmically. 
    We show natural inflation from \cref{eq:natural} in orange, quartic hilltop inflation from \cref{eq:hilltop} in purple and both quadratic and quartic double-well potentials from \cref{eq:doublewell} in light and dark green respectively.
    }
\end{figure*}

In \cref{fig:nsr_slow_roll} we show in blue the slow-roll predictions for the quartic, cubic, quadratic and linear monomial as well as for the monomial with $p=2/3$.

\paragraph*{Quartic and cubic potentials:} Similarly to the flat case in the \textsc{Planck} inflation papers, the predictions for quartic and cubic inflation in light blue lie far outside the \SI{95}{\percent} contours from the \LCDM\ extensions in grey.

\paragraph*{Quadratic potential:} The quadratic potential with $p=2$ has long been used as the simplest realisation of single-field inflation. It is often given in the following (slightly different) form with an additional pre-factor of one half:
\begin{equation}
\label{eq:quadratic}
    V(\phi) = \tfrac{1}{2} m^2 \phi^2 ,
\end{equation}
where~$m$ is referred to as the inflaton mass, which can be related to the potential amplitude~$\Lambda$ in \cref{eq:monomial} directly. Although allowing for spatial curvature to vary in \cref{fig:nsr_slow_roll}, which significantly stretches the P18 contours to larger~$r$, this stretching coincides with a shift to larger~$n_\mathrm{s}$ such that the prediction for quadratic inflation ends up just outside the \SI{95}{\percent} P18 contours, just like in the flat case~\cite{Planck2018Inflation}. With the addition of BK15 data, the SR prediction lies far outside the contour irrespective of any curvature effects.

\paragraph*{Linear and $p=2/3$ potential:} These two potentials are motivated by axion monodromy~\cite{Silverstein2008,McAllister2010} and agree better with the P18 and P18+BK15 contours. Both in fact profit from the shift to a larger spectral index that comes with varying curvature.

\subsection{Natural potential} 

Natural inflation is motivated by particle physics considerations~\cite{Freese1990} to naturally accommodate the very flat potentials required for inflation. It is given by the periodic potential
\begin{equation}
    V(\phi) = \Lambda^4 \left[ 1 +  \cos(\frac{\phi}{f}) \right] , 
\end{equation}
where~$f$ corresponds to the global symmetry-breaking scale and governs the slope of the potential.

We can rewrite the potential such that the local maximum lies at~$\phi_0=\pi f$ and is given by the potential amplitude $V(\phi_0)=\Lambda^4$. From this unstable maximum the inflaton rolls down to the minimum at the origin $\phi=0$ (see also \cref{fig:potentials}):
\begin{equation}
    \label{eq:natural}
    V(\phi) = \frac{\Lambda^4}{2} \left[ 1 -  \cos(\pi\,\frac{\phi}{\phi_0}) \right] .
\end{equation}

In order to produce sufficient e-folds of (large-field) inflation, we require a potential hill parameter~$\phi_0\agt\SI{10}{\planckmass}$ (or correspondingly for~$f$).
In the limit of very large~$\phi_0\agt\SI{100}{\planckmass}$, the spectral index and tensor-to-scalar ratio of the natural potential tend to those of the quadratic potential. 

While natural inflation still overlaps with the P18 contours in a flat universe~\cite{Planck2018Inflation}, because of its ability to accommodate a smaller tensor-to-scalar ratio, it only touches the \SI{95}{\percent} contours in the curved case due to the shift in the spectral index.

\subsection{Double-Well potential}

Similarly to the natural potential, we define the double-well potential such that the local maximum lies at~$\phi_0$ with the maximum potential value given by the potential amplitude $V(\phi_0)=\Lambda^4$ (see also \cref{fig:potentials}):
\begin{equation}
\label{eq:doublewell}
    V(\phi) = \Lambda^4 \left[ 1 -  \left( \frac{\phi-\phi_0}{\phi_0} \right)^p \right]^2 .
\end{equation}
where~$p$ can in principle take any positive value. We will consider the quadratic ($p=2$) and quartic ($p=4$) double-well in particular.

Double-well potentials are typically associated with small-field inflation. However, inflation with large field displacements~$\phi_0>\si{\planckmass}$ is also possible. In that case the spectral index and tensor-to-scalar ratio tend to that of the quadratic potential for very large $\phi_0\agt\SI{100}{\planckmass}$, similarly to natural inflation and irrespective of the parameter~$p$. For smaller values of the potential hill parameter~$\phi_0\agt\SI{10}{\planckmass}$, both the spectral index and tensor-to-scalar ratio decrease. The SR predictions for the quadratic double-well are very close to those of the natural potential, which is to be expected considering their similar shapes (cf.\ \cref{fig:potentials}). The quartic double-well with its flatter hill leads to a faster drop in~$r$, and therefore a greater overlap with the P18 and P18+BK15 contours.

\subsection{Hilltop potential}

Closely related to double-well potentials, hilltop potentials are given by:
\begin{equation}
\label{eq:hilltop}
    V(\phi) = \Lambda^4 \left[ 1 -  \left( \frac{\phi-\phi_0}{\phi_0} \right)^p + \dots \right] ,
\end{equation}
in which only the first order in~$\phi^p$ is retained and higher order terms (indicated by the ellipsis) are neglected, since the latter only become relevant towards the end of inflation.
For small values of the potential hill parameter~$\phi_0$ the SR predictions are close to those of the double-well potential, but for larger values the spectral index, running, and tensor-to-scalar ratio will tend towards those of the linear potential (monomial with $p=1$) instead of the quadratic potential, since that is what the \cref{eq:hilltop} approximates to close to $V(\phi)=0$. However, this asymptotic behaviour would have to be different if the higher order terms were present, which are required to ensure the positiveness of the potential. We therefore prefer to work with the double-well potential for the scope of this paper.
The asymptotic behaviour does, however, mean that the SR predictions agree better with the P18 and P18+BK15 contours than those of the double-well potentials.

\subsection{Starobinsky potential}

The Starobinsky potential, given in the Einstein frame by
\begin{equation}
\label{eq:starobinsky}
    V(\phi) = \Lambda^4 \left[ 1 -  \e^{-\sqrt{\frac{2}{3}} \frac{\phi}{\si{\planckmass}}} \right]^2 ,
\end{equation}
was the first proposed inflationary potential and motivated by an extension of the Einstein--Hilbert action with a term quadratic in the Ricci tensor~\cite{Starobinsky1980,Planck2013Inflation}. Therefore this model of inflation is frequently also referred to as $R^2$~inflation.

It can be shown that inflation generated by the Higgs field of the (particle physics) standard model can be reduced to the potential from \cref{eq:starobinsky} in the Einstein frame, where all parameters connected to the Higgs boson are included in the amplitude parameter~$\Lambda$~\cite{EncyclopaediaInflationaris2014}. This motivates yet another name for this type of potential: Higgs inflation.

Due to the shift in spectral index from varying curvature, the SR predictions for the Starobinsky potential no longer lie as spot-on in the centre of the \SI{68}{\percent} contour lines as in the flat case. Nevertheless, they remain in excellent agreement with the P18 and P18+BK15 contours.



\newcolumntype{+}{D{,}{\,\pm\,}{-1}}
\newcolumntype{,}{D{,}{,}{1}}
\begin{table*}[tb!]
\renewcommand{\arraystretch}{1.5}
    \caption{\label{tab:params} Overview of the cosmological parameters with their fiducial values used for visualisation purposes (e.g.\ in \cref{fig:AsfoH_cHH_PPS,fig:AsfoH_PPS_Cls}) and their prior ranges used in our Bayesian analysis. We list the base parameters of the \LCDM\ cosmology in the first block, the \LCDM\ extension parameters in the second block, and primordial parameters pertaining to a full inflationary analysis in the third block. 
    The primordial parameters $A_\mathrm{s}$, $A_\mathrm{SR}$, $n_\mathrm{s}$, $r$, and $N_\ast$ all refer to the pivot scale $k_\ast=\SI{0.05}{\per\mega\parsec}$.
    Note that the prior ranges for $N_\ast$ and $f_\mathrm{i}$ are effectively further restricted by the horizon constraint in \cref{eq:horizon_constraint} and the reheating constraint in \cref{eq:reheating_constraint_w} respectively. 
    }
    \begin{ruledtabular}
        

\begin{tabular}{ l c c l }
    Parameter                                                      & fiducial value   & \multicolumn{1}{c}{Prior range}         & Definition                                         \\ \hline
    $\omega_\mathrm{b}\equiv h^2\Omega_\mathrm{b}$                 & 0.022632         & $0.019 < \omega_\mathrm{b} < 0.025$     & Baryon density today                               \\
    $\omega_\mathrm{c}\equiv h^2\Omega_\mathrm{c}$                 & 0.11792          & $0.025 < \omega_\mathrm{c} < 0.471$     & Cold dark matter density today                     \\
    $H_0$                                                          & $h=0.7$          & $0.2 < h < 1.0$                         & Hubble parameter with $H_0=100\,h\,\si{\km\per\s\per\mega\parsec}$                          \\
    $\theta_\mathrm{s}$                                            & 0.01041338       & $1.03 < 100\,\theta_\mathrm{s} < 1.05$  & Angular size of sound horizon at last scattering   \\
    $\tau_\mathrm{reio}$                                           & 0.0495           & $0.01 < \tau_\mathrm{reio} < 0.40$      & Optical depth to reionization                      \\
    $A_\mathrm{s}$                                                 & $2\times10^{-9}$ & $2.5 < \ln(10^{10}A_\mathrm{s})  < 3.7$ & Amplitude of the scalar power spectrum             \\
    $n_\mathrm{s}$                                                 & 0.97235          & $0.885 < n_\mathrm{s} < 1.040$          & Primordial scalar spectral index                              \\ \hline
     \multirow{2}{*}{$r$}                                          &                  & $0 < r < 1$                             &  \multirow{2}{*}{Primordial tensor-to-scalar power ratio}     \\
                                                                   &                  & $-5 < \log_{10}r < 0$                   &                                                    \\
    $\Omega_{K,0}$                                                 & $-0.01$          &                                         & Curvature density today                            \\ 
    $\omega_{K,0}\equiv h^2\Omega_{K,0}$                           & $-0.005$         & $-0.04 < \omega_{K,0} < 0.04$           &                                                    \\ \hline
    $A_\mathrm{SR}$                                                & $2\times10^{-9}$ & $2.5 < \ln(10^{10}A_\mathrm{SR}) < 3.7$ & Inflationary slow-roll estimate of $A_\mathrm{s}$  \\
    $N_\ast$                                                       & $55$             & $20 < N_\ast < 90$                      & Inflationary e-folds \emph{after} horizon crossing \\
    $f_\mathrm{i}\equiv\frac{\Omega_{K,\mathrm{i}}}{\Omega_{K,0}}$ & 5                & $-1 < \log_{10}f_\mathrm{i} < 5$        & Fraction of primordial to present-day curvature    \\
\end{tabular}

    \end{ruledtabular}
\end{table*}

\section{Choice of parametrisation}
\label{sec:parametrisation}

\begin{figure*}[p]
    \flushleft
    \subfloat{\includegraphics{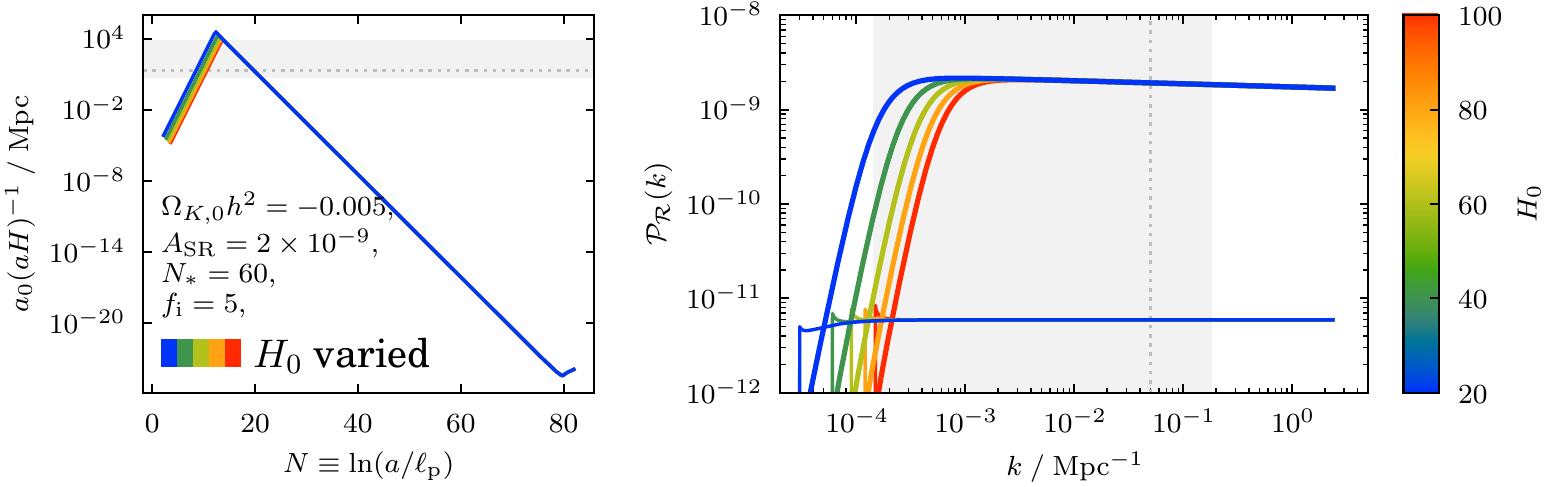}} 
    \vspace{-0.4cm} \\
    \subfloat{\includegraphics{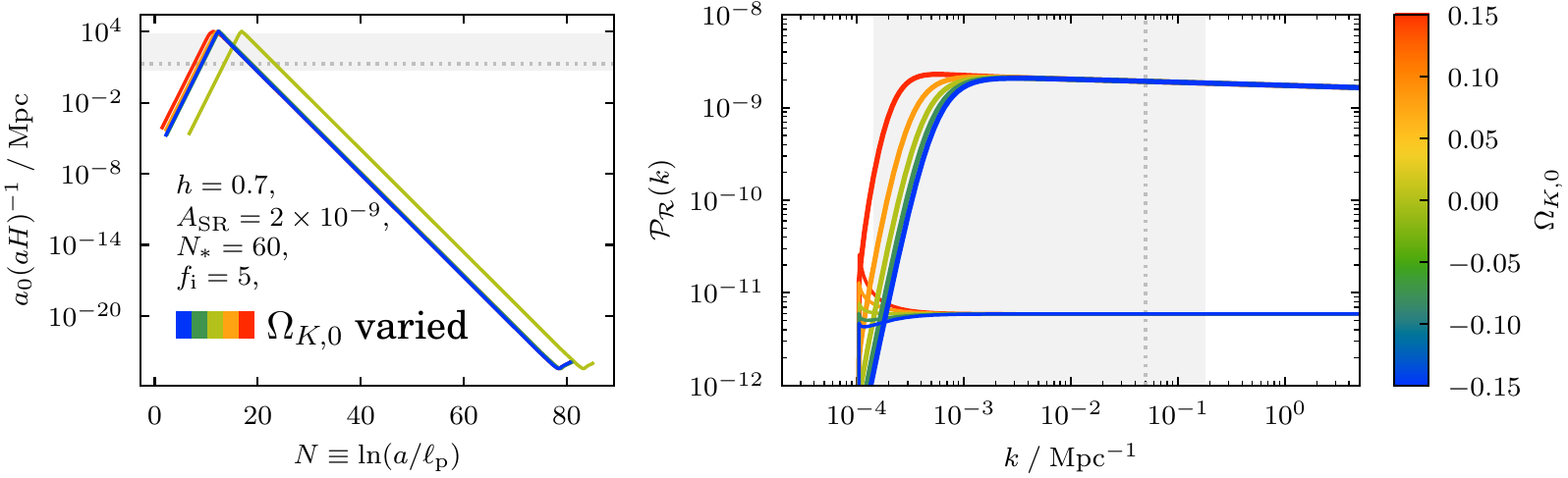}} 
    \vspace{-0.4cm} \\
    \subfloat{\includegraphics{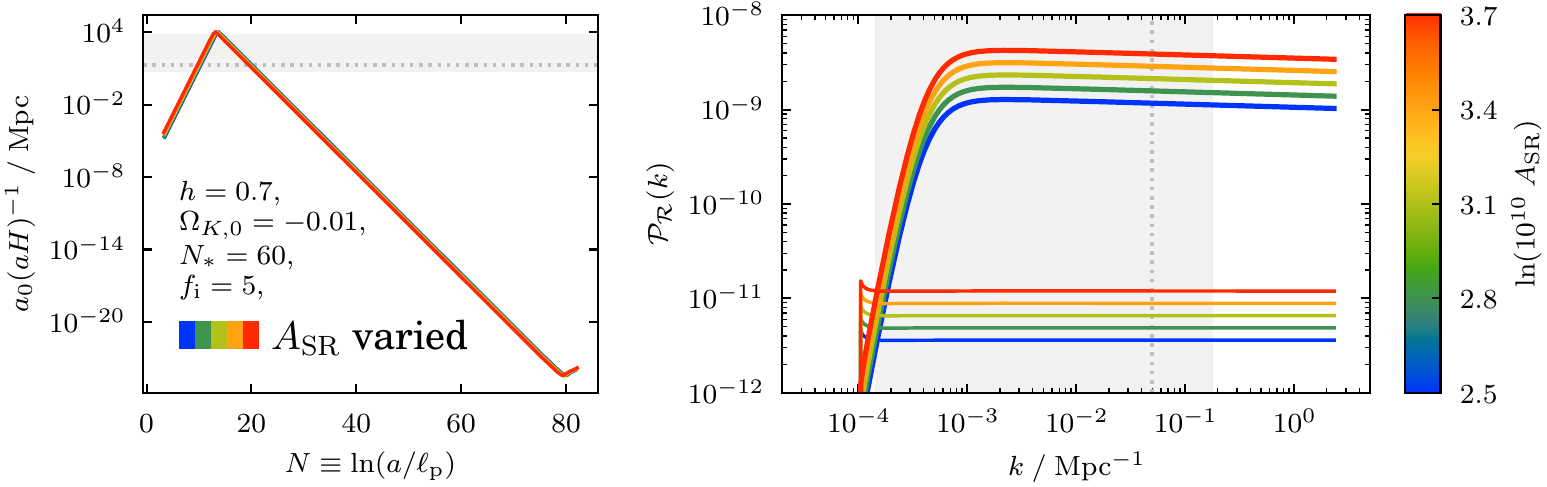}} 
    \vspace{-0.4cm} \\
    \subfloat{\includegraphics{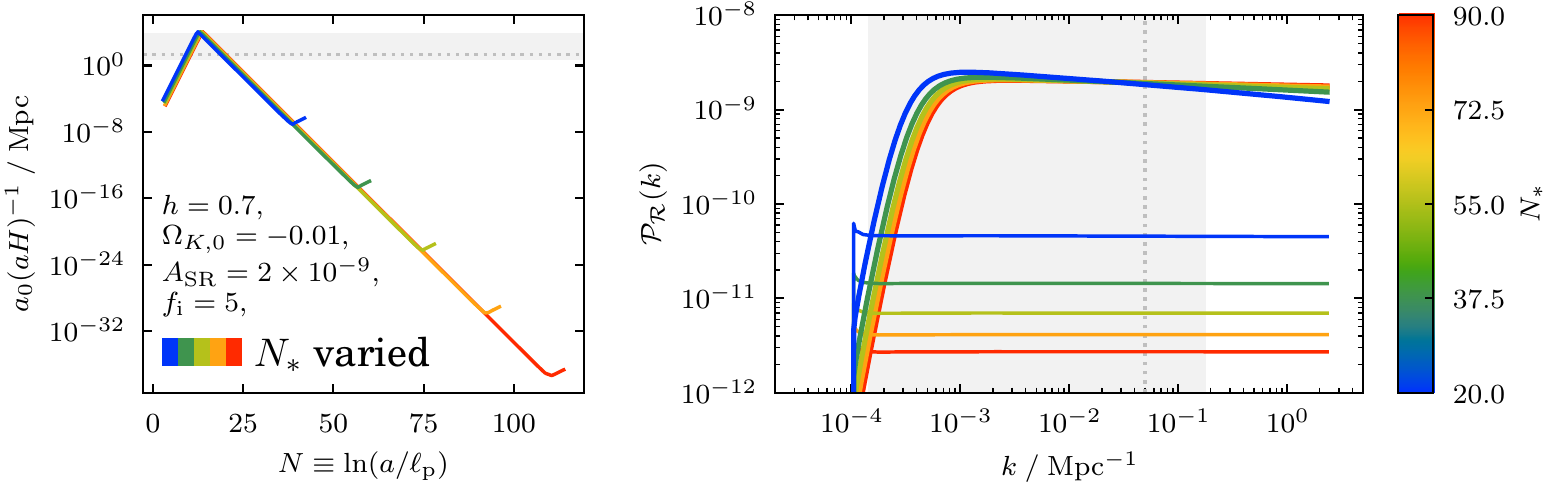}} 
    \vspace{-0.2cm} 
    \caption{\label{fig:AsfoH_cHH_PPS} Parameter dependence of the comoving Hubble horizon in the left column and of the slow-roll approximation of the primordial power spectrum~(PPS) in the right column on the sampling parameters for our Bayesian analysis: Hubble parameter~$H_0$, present-day curvature density~$\Omega_{K,0}$, approximate power amplitude~$A_\mathrm{SR}$, and number of e-folds of inflation after horizon crossing~$N_\ast$, where one parameter is varied in each row, while the others stay fixed.
    The upper and heavier lines in the PPS plots correspond to scalar, the lower and thinner lines to tensor perturbations.
    We used the Starobinsky potential to generate these plots, explaining the fairly big gap between scalar and tensor modes.
    The corresponding plots for the fully numerically integrated PPS and for the CMB power spectrum are shown in \cref{fig:AsfoH_PPS_Cls}.}
\end{figure*}

\begin{figure*}[p]
    \flushleft
    \subfloat{\includegraphics[]{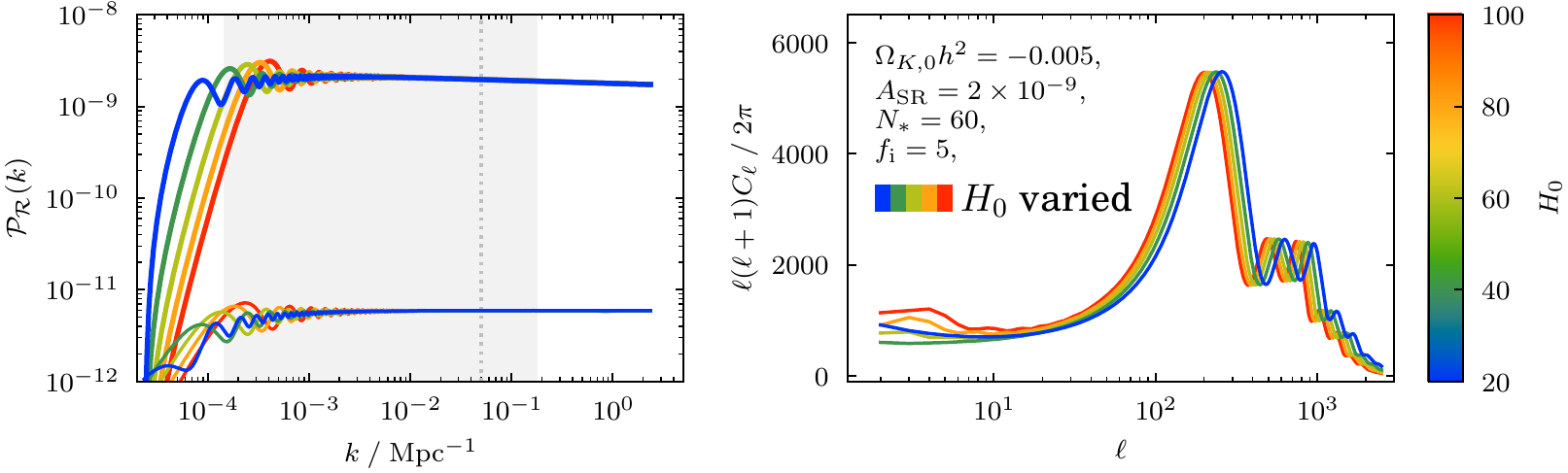}}
    \vspace{-0.4cm} \\
    \subfloat{\includegraphics[]{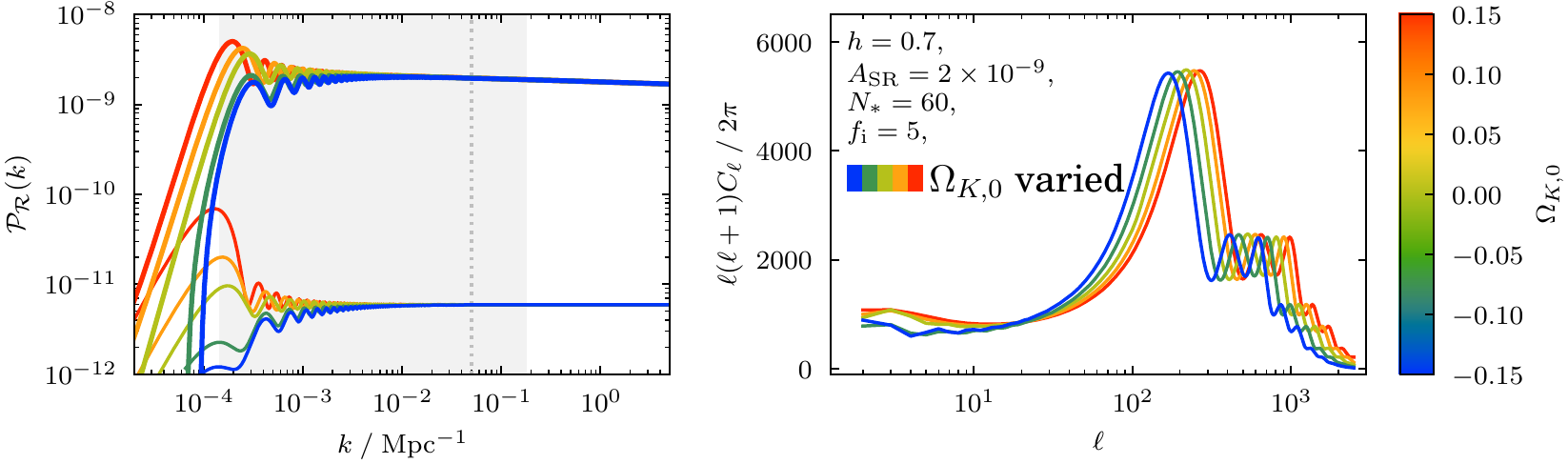}}
    \vspace{-0.4cm} \\
    \subfloat{\includegraphics[]{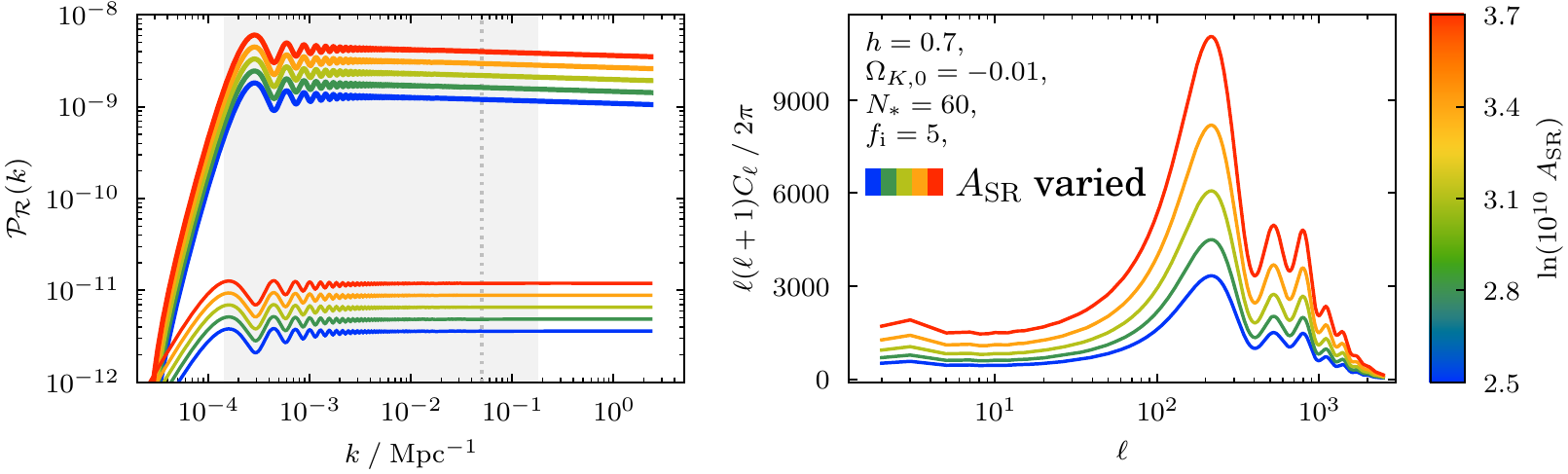}}
    \vspace{-0.4cm} \\
    \subfloat{\includegraphics[]{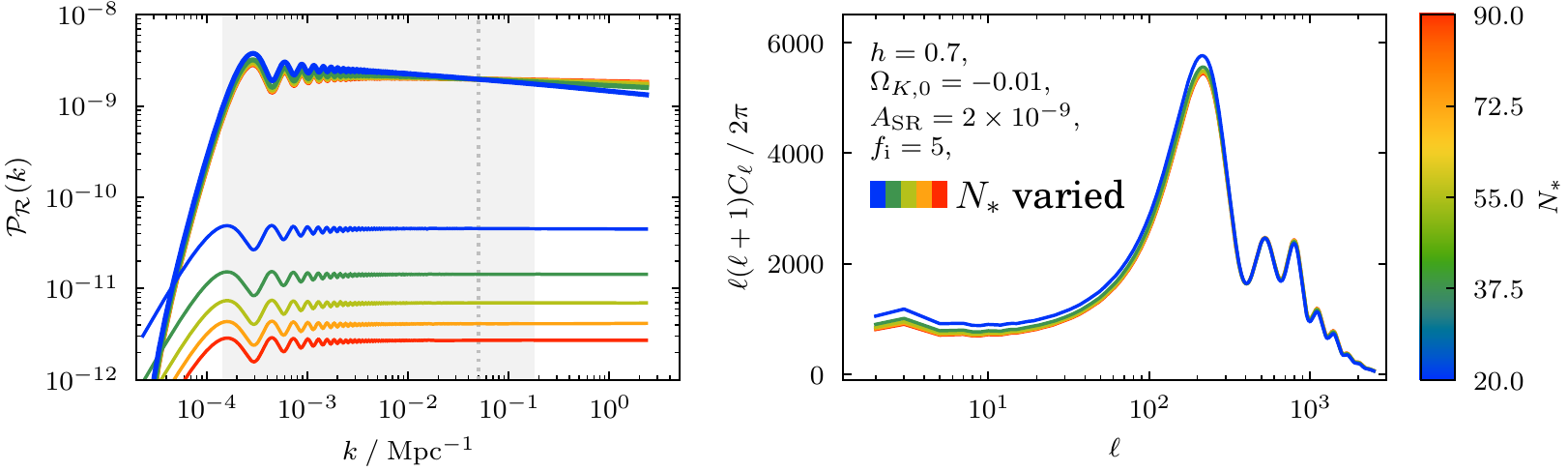}}
    \vspace{-0.2cm}
    \caption{\label{fig:AsfoH_PPS_Cls} Parameter dependence of the numerically integrated primordial power spectrum~(PPS) in the left column and of the CMB power spectrum in the right column on the sampling parameters for our Bayesian analysis: Hubble parameter~$H_0$, present-day curvature density~$\Omega_{K,0}$, approximate power amplitude~$A_\mathrm{SR}$, and number of e-folds of inflation after horizon crossing~$N_\ast$, where one parameter is varied in each row, while the others stay fixed.
    The upper and heavier lines in the PPS plots correspond to scalar, the lower and thinner lines to tensor perturbations.
    We used the Starobinsky potential to generate these plots, explaining the fairly big gap between scalar and tensor modes. 
    The corresponding plots for the comoving Hubble horizon and for the slow-roll approximation of the PPS are shown in \cref{fig:AsfoH_cHH_PPS}.
    For the CMB spectra we fix the other cosmological parameters to their \textsc{Planck}~2018 best-fit values~\cite{Planck2018Parameters}: $\omega_\mathrm{b}=0.022632$, $\omega_\mathrm{cdm}=0.11792$, $\tau_\mathrm{reio}=0.0495$.
    }
\end{figure*}

\Cref{tab:params} lists the sampling parameters used in our Bayesian analysis  
together with their prior ranges and fiducial values which they are fixed to for visualisation purposes in some figures, such as the \cref{fig:AsfoH_cHH_PPS,fig:AsfoH_PPS_Cls}.

For the base \LCDM\ model, we use the following six sampling parameters: 
\begin{itemize}
    \item $\omega_\mathrm{b}=h^2\Omega_\mathrm{b}$: Baryon density today
    \item $\omega_\mathrm{c}=h^2\Omega_\mathrm{c}$: Cold dark matter density today
    \item $100\,\theta_\mathrm{s}$: Angular size of sound horizon at last scattering
    \item $\tau_\mathrm{reio}$: Optical depth to reionization
    \item $\ln(10^{10} A_\mathrm{s})$: Scalar power spectrum amplitude
    \item $n_\mathrm{s}$: Scalar spectral index
\end{itemize}
Additionally we consider the following parameter extensions to the base \LCDM\ model:
\begin{itemize}
    \item $r$: Tensor-to-scalar power ratio
    \item $\Omega_{K,0}$: Spatial curvature parameter today
\end{itemize}
All primordial parameters $\{A_\mathrm{s}, n_\mathrm{s}, r\}$ are taken at the pivot scale of $k_\ast=\SI{0.05}{\per\mega\parsec}$.

The primordial parameters~$A_\mathrm{s}$ and~$n_\mathrm{s}$ refer to the simplified power-law spectrum from \cref{eq:powerlaw_pps}. When we derive the primordial power spectrum from an inflationary potential as outlined in the previous sections \cref{sec:pps_power_law,sec:pps_sr}, these parameters and also~$r$ turn into derived parameters.
In our analysis of individual inflationary potentials, we keep $\omega_\mathrm{b}$, $\omega_\mathrm{c}$, and $\tau_\mathrm{reio}$ as sampling parameters, but change the following sampling parameters: 
\begin{itemize}
    \item Instead of~$\theta_\mathrm{s}$ we sample over the Hubble parameter~$H_0$. This simplifies the computational complexity, as~$H_0$ can be directly used by both our primordial inflation code as well as the Boltzmann theory code (\texttt{CLASS}), without the need to first infer it from the angular size of the sound horizon~$\theta_\mathrm{s}$.
    \item Instead of $\Omega_{K,0}$ we sample over $\omega_{K,0}\equiv h^2\Omega_{K,0}$, which turns the banana shaped dependence between the Hubble parameter and curvature density parameter into a more linear dependence and thereby improves the sampling efficiency. The (small) effect of this parameter change on model comparisons is documented in \cref{sec:appendix,fig:stats_theta_vs_H0} for curvature extensions of the \LCDM\ base model. 
    \item Instead of $\ln(10^{10} A_\mathrm{s})$ we sample over its inflationary slow-roll approximation $\ln(10^{10} A_\mathrm{SR})$ from \cref{eq:A_SR} (for a check of the goodness of the approximation see \cref{fig:As_vs_ASR} in the appendix).
\end{itemize}
In addition to these \LCDM\ related parameters, we sample over the following parameters in our analysis of individual inflationary potentials:
\begin{itemize}
    \item $N_\ast\equiv\ln\left(\frac{a_\mathrm{end}}{a_\ast}\right)$:\\
    Inflationary e-folds \emph{after} horizon crossing of the pivot scale~$k_\ast$.
    \item $\log_{10}f_\mathrm{i}\equiv\log_{10}\frac{\Omega_{K,\mathrm{i}}}{\Omega_{K,0}}$:\\
    Fraction of primordial to present-day curvature.
\end{itemize}

Instead of the power amplitude~$A_\mathrm{s}$, we could have used the amplitude parameter for the inflationary potential~$\Lambda$. The two are directly related to one another, since the background \cref{eq:background1,eq:background2,eq:eom} are invariant under a simultaneous rescaling of the time coordinate and the inflaton potential:
\begin{align} 
\begin{split}
\label{eq:rescaling}
    t &\mapsto \sigma^{-1} t  \\
    V(\phi) &\mapsto \sigma^2 V(\phi)  \\
    \Rightarrow \mathcal{P_R}(k) &\mapsto \sigma^2 \mathcal{P_R}(k) .
\end{split}
\end{align}
However, we prefer sampling over~$A_\mathrm{s}$, as it allows for a more direct comparison with the base \LCDM\ model and its extensions. Also, as opposed to~$\Lambda$, the power amplitude~$A_\mathrm{s}$ is not correlated with the other primordial parameters that affect the comoving Hubble horizon or the e-folds of inflation. This becomes very clear in the third row of \cref{fig:AsfoH_cHH_PPS} showing the variation of the comoving Hubble horizon with respect to the logarithm of~$A_\mathrm{SR}$ on the left, and that of the primordial power spectrum~(PPS) on the right. While~$A_\mathrm{SR}$ governs the amplitude of the PPS by definition, it leaves the comoving Hubble horizon invariant. In inflation models such as natural or double-well inflation, with a local maximum separated from the global minimum by~$\phi_0$, the potential amplitude~$\Lambda$ is also strongly correlated with~$\phi_0$, and sampling~$A_\mathrm{SR}$ instead of~$\Lambda$ avoids having to navigate that degeneracy.

Similarly, we choose to use the present-day spatial curvature parameter $\Omega_{K,0}$ for a better comparison with the $\Omega_{K,0}$-extension of \LCDM. Alternatively, one could use reheating parameters to track the evolution of energy densities in the universe and infer the present-day scale factor~$a_0$ and curvature density. We defer exploring this option to future work.

There is considerable freedom in the choice between the primordial parameters at the start of inflation, i.e. parameters related to the inflaton field~$\phi_\mathrm{i}$, the e-folds~$N_\mathrm{i}$, or the primordial curvature~$\Omega_{K,\mathrm{i}}$, which are all linked via \cref{eq:inflationstart_Omega_Ki,eq:inflationstart_N_i}. And these parameters are connected to e.g.\ the total number of e-folds of inflation~$N_\mathrm{tot}$ or the e-folds of inflation before~($N_\dagger$) and after~($N_\ast$) horizon crossing of the pivot scale. 
We choose to work with~$N_\ast$, because it allows a better comparison across different inflationary potentials and because of its direct link to both the scalar spectral index~$n_\mathrm{s}$ and the tensor-to-scalar ratio~$r$. 

The fraction~$f_\mathrm{i}$ of primordial curvature is a useful sampling parameter for two reasons: firstly, due to it governing the ratio of conformal time passing before and after the end of inflation which we explored in the previous \cref{sec:conformal_time} (see especially \cref{subfig:conformal_time_ratio}), and secondly because of its major role in governing the cutoff position in the primordial power spectrum, which we explore in more detail in the following.

\begin{figure*}[p]
    \flushright
    \subfloat{\includegraphics[scale=0.98]{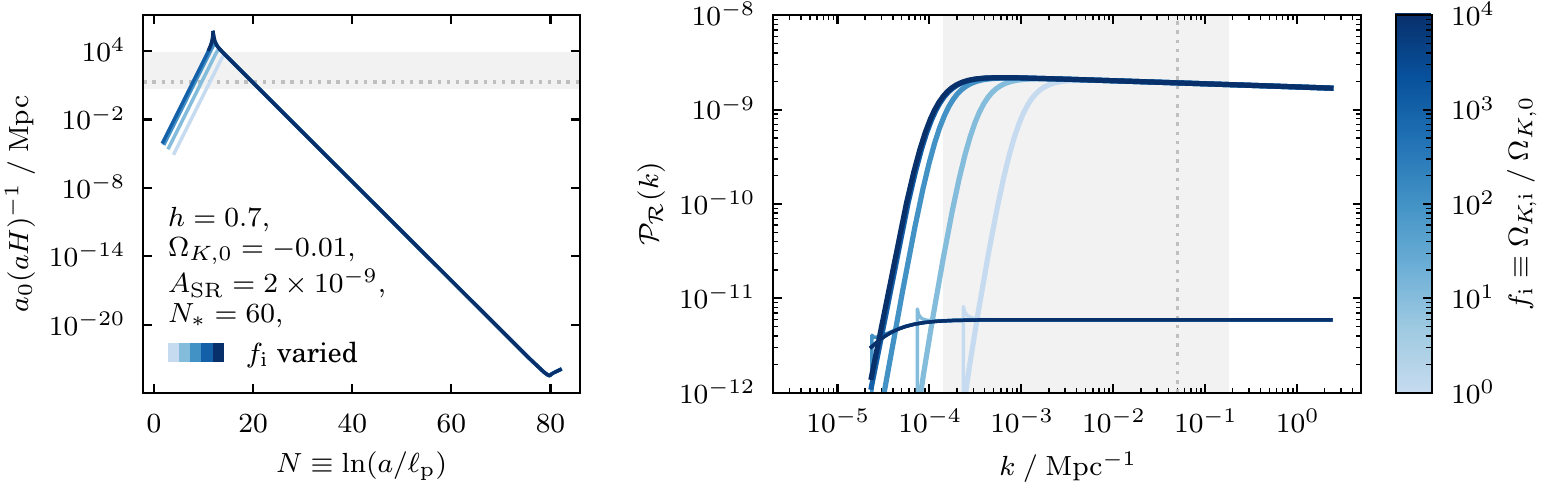}}
    \vspace{-0.4cm} \\
    \subfloat{\includegraphics[scale=0.98]{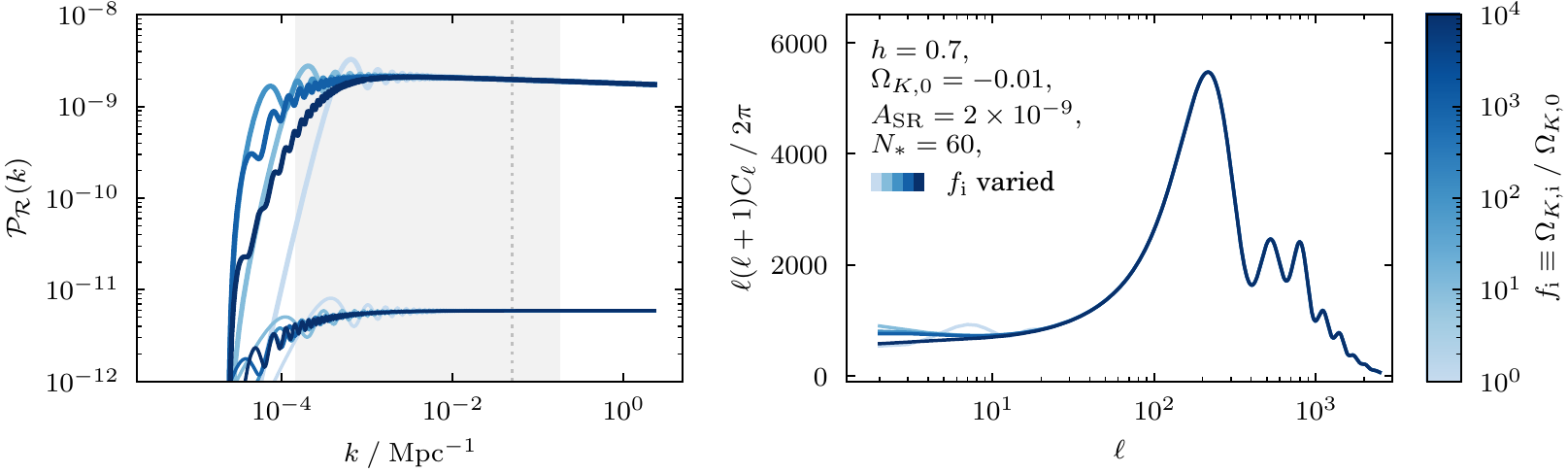}}
    \vspace{-0.4cm} \\
    \subfloat{\includegraphics[scale=0.98]{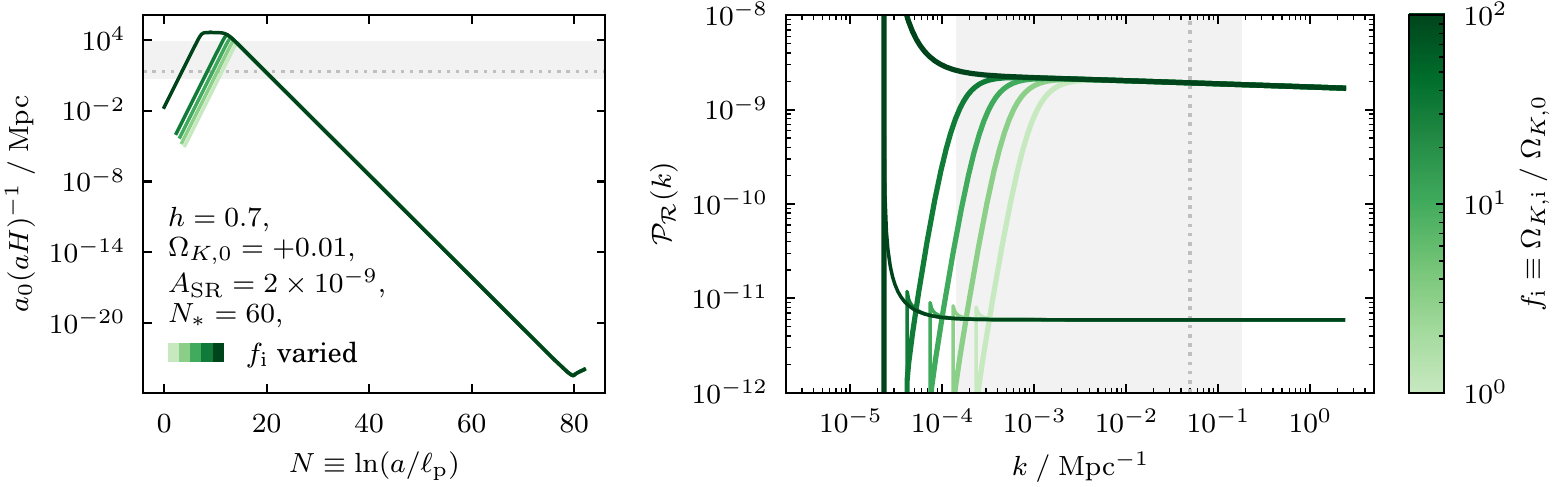}}
    \vspace{-0.4cm} \\
    \subfloat{\includegraphics[scale=0.98]{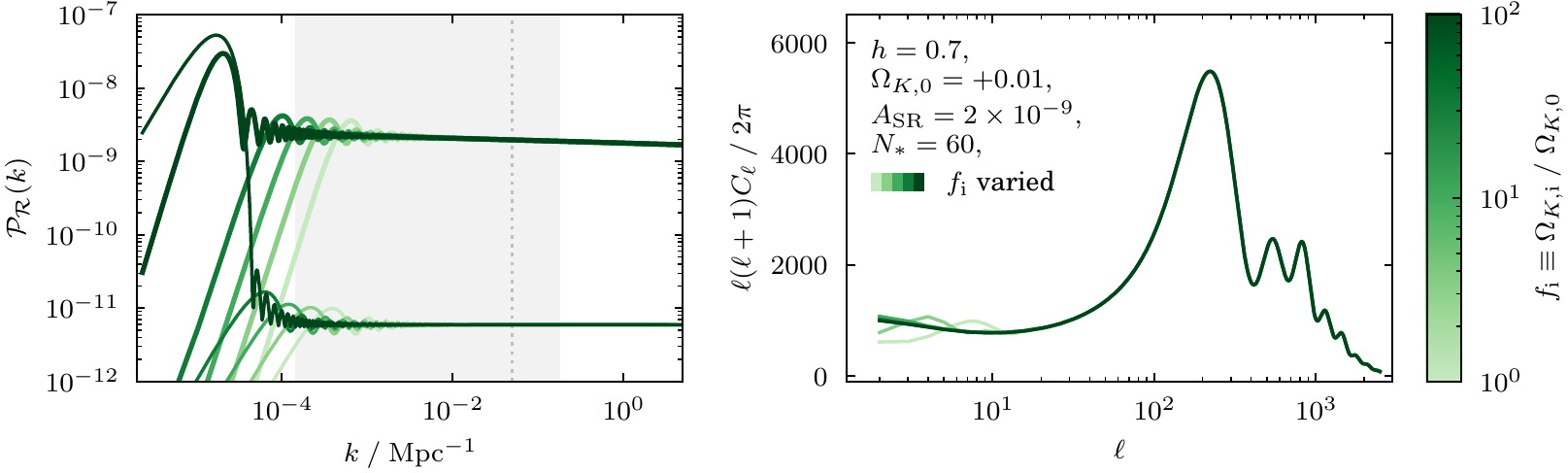}}
    \vspace{-0.2cm}
    \caption{\label{fig:AsfoH_fi} Similar to \cref{fig:AsfoH_cHH_PPS,fig:AsfoH_PPS_Cls} we show the parameter dependence of the comoving Hubble horizon, the slow-roll approximation of the primordial power spectrum~(PPS), the fully numerically integrated PPS and the CMB power spectrum on the fraction of primordial to present-day curvature $f_\mathrm{i}\equiv\Omega_{K,\mathrm{i}}/\Omega_{K,0}$. 
    In the upper two rows in blue the present-day curvature was fixed to $\Omega_{K,0}=-0.01$, in the lower two rows in green to $\Omega_{K,0}=+0.01$, thus showing the effects of a closed and open universe respectively.
    We used the Starobinsky potential to generate these plots, explaining the fairly big gap between scalar and tensor modes. 
    The upper and heavier lines in the PPS plots correspond to scalar, the lower and thinner lines to tensor perturbations.
    For the CMB spectra we fix the other cosmological parameters to their \textsc{Planck}~2018 best-fit values~\cite{Planck2018Parameters}: $\omega_\mathrm{b}=0.022632$, $\omega_\mathrm{cdm}=0.11792$, $\tau_\mathrm{reio}=0.0495$.
    }
\end{figure*}

\Cref{fig:AsfoH_cHH_PPS,fig:AsfoH_PPS_Cls,fig:AsfoH_fi} show the effect of our sampling parameters on the comoving Hubble horizon (left column of \cref{fig:AsfoH_cHH_PPS}), on the slow-roll approximation of the primordial power spectrum (right column of \cref{fig:AsfoH_cHH_PPS}), on the fully numerically integrated primordial power spectrum (left column in \cref{fig:AsfoH_PPS_Cls}) and on the CMB power spectrum (right column in \cref{fig:AsfoH_PPS_Cls}).

The cutoff and oscillations in the PPS towards large scales (small~$k$) are features of a kinetically dominated or fast-roll stage prior to inflation, already known and studied for flat universes~\cite{Scacco2015,Hergt2019}. However, in flat universes these features can easily be pushed outside the observable window by large amounts of inflation, which is no longer the case with non-zero spatial curvature.

The first row in \cref{fig:AsfoH_cHH_PPS,fig:AsfoH_PPS_Cls} shows the effects of varying the Hubble parameter~$H_0$ while keeping $h^2\Omega_{K,0}$ fixed. This will affect the starting value of the comoving Hubble horizon and thereby influence the large-scale cutoff position in the PPS. This effect is translated through to the CMB power spectrum, but additionally the Hubble parameter shifts the CMB power spectrum horizontally. This horizontal shift is not attributed to the PPS but an effect already present in the \LCDM\ model (an effect on the transfer function, not the PPS).

The different shapes of the PPS for closed and open universes can be seen more directly in the second row of \cref{fig:AsfoH_cHH_PPS,fig:AsfoH_PPS_Cls}, where the present-day curvature density~$\Omega_{K,0}$ is varied, showing the transition from large-scale power \emph{suppression} for closed universes to \emph{amplification} for open universes. 
The ability of positive curvature to suppress large-scale power is particularly interesting in light of the lack of power on large scales found in full-sky CMB data (see~\cite{Efstathiou2003} for an early discussion on this).
Similarly to the Hubble parameter~$H0$, the CMB power spectrum shifts horizontally with~$\Omega_{K,0}$ resulting in a degeneracy between these two parameters.

As already mentioned the power amplitude~$A_\mathrm{s}$ and its slow-roll approximation~$A_\mathrm{SR}$ have the straightforward effect of vertically shifting the PPS and the CMB power spectrum, while leaving the comoving Hubble horizon and the e-folds of inflation unaffected.

The last row of \cref{fig:AsfoH_cHH_PPS,fig:AsfoH_PPS_Cls} shows the variation of the comoving Hubble horizon, PPS, and CMB power spectrum with respect to the number of e-folds of inflation~$N_\ast$ after horizon crossing of the pivot scale~$k_\ast=\SI{0.05}{\per\mega\parsec}$.
Looking at the variation of the comoving Hubble horizon one can observe how increasing~$N_\ast$ stretches the duration of inflation to a later end, while leaving the evolution prior to horizon crossing of the pivot scale invariant. 
An advantage of~$N_\ast$ over alternative parameters is its direct link to the scalar spectral index~$n_\mathrm{s}$ and the tensor-to-scalar ratio~$r$. Fixing~$N_\ast$ while varying the other parameters will leave the slope and the ratio of tensor to scalar power invariant. 

\begin{figure*}[tbp]
    \includegraphics[scale=1.05]{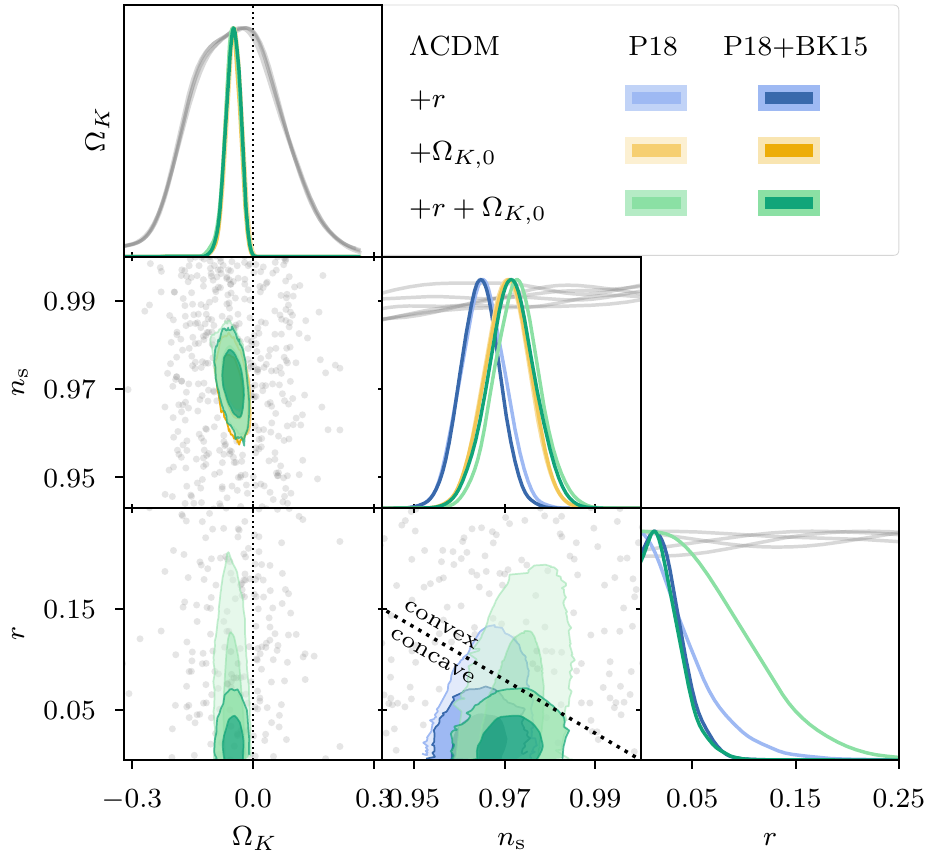}
    \hfill
    \includegraphics[]{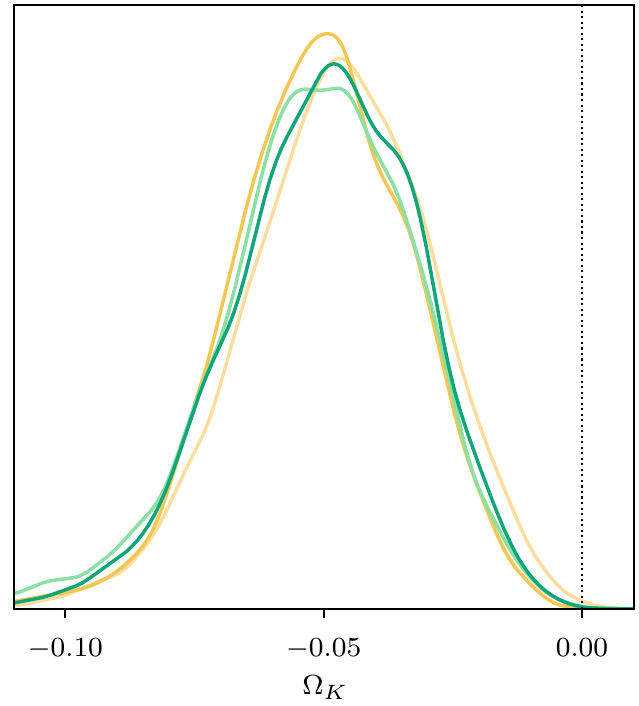}
    \caption{\label{fig:posterior_Onr} Posterior constraints on the \LCDM\ extension parameters, the tensor-to-scalar ratio~$r$ and the curvature density parameter~$\Omega_{K,0}$. We also include the spectral index~$n_\mathrm{s}$, since the $n_\mathrm{s}$-$r$ plot is typically the main plot of interest when investigating inflation models. The dotted line splits the $n_\mathrm{s}$-$r$ plot into the regions of convex or concave inflationary potentials. The lighter hue corresponds to using \textsc{Planck}~2018 $TT,TE,EE+\mathrm{low}E$ data only~\cite{Planck2018CMB}, whereas the darker hue corresponds to additionally using data from the \textsc{Bicep2} and \textsc{Keck Array}~\cite{BicepKeck2018BKX}. For the single-parameter extension, the posterior on the curvature density parameter amounts to~$\Omega_{K,0}=\num{-0.051(17)}$. The grey lines and dots illustrate the prior distributions.}
\end{figure*}

We show the variation of comoving Hubble horizon, PPS, and CMB power spectrum with respect to the fraction~$f_\mathrm{i}$ in \cref{fig:AsfoH_fi}.
For a fixed \emph{present-day} curvature density, varying~$f_\mathrm{i}$ is equivalent to a variation of the \emph{primordial} curvature density~$\Omega_{K,\mathrm{i}}$ and thus also to a variation of the initial size of the comoving Hubble horizon at the start of inflation. 
Since fixing~$N_\ast$ decorrelates~$f_\mathrm{i}$ from the spectral index and the tensor to scalar ratio, this isolates the effect of~$f_\mathrm{i}$ on the large-scale (small~$k$) cutoff position and the shape of the PPS, independent of slopes or amplitudes. \Cref{fig:AsfoH_fi} contrasts this behaviour for closed universes in blue (top two rows) and for open universes in green (bottom two rows), in which~$f_\mathrm{i}$ affects the shape differently. In both cases increasing~$f_\mathrm{i}$ initially (i.e.\ for small~$f_\mathrm{i}$) shifts the PPS cutoff to larger scales, out of the CMB observable window. However, once~$\Omega_{K,\mathrm{i}}$ gets close to or exceeds unity, this shift is replaced by a suppression of perturbation modes in the closed case and an amplification in the open case, for large scales just about smaller than the PPS cutoff.
Note how looking at the slow-roll approximation of the PPS only may be misleading when trying to gauge the effect of~$f_\mathrm{i}$ on the PPS. The shift of the cutoff to larger scales is similar in both the approximate and full numerical PPS as long as the curvature density is comparably small. However, once primordial curvature plays a significant role, the approximate PPS stops changing. The secondary, geometry-dependent effects on large scales are only visible in the fully numerically integrated PPS. This is not surprising considering that the slow-roll approximation is only valid for modes that were well within the comoving Hubble horizon at the start of inflation.

\section{Nested sampling results}
\label{sec:results}

In this section we present the results from our Bayesian analysis using nested sampling. We start by investigating one- and two-parameter extensions to the base \LCDM\ model in \cref{sec:result_extensions}. In \cref{sec:result_inflation} we then change the phenomenological description of the primordial power spectrum from the \LCDM\ model to that of specific inflationary models with a full numerical integration of the mode \cref{eq:mukhanov-sasaki_scalar,eq:mukhanov-sasaki_tensor}.

\begin{figure*}[tbp]
    \includegraphics[width=\textwidth]{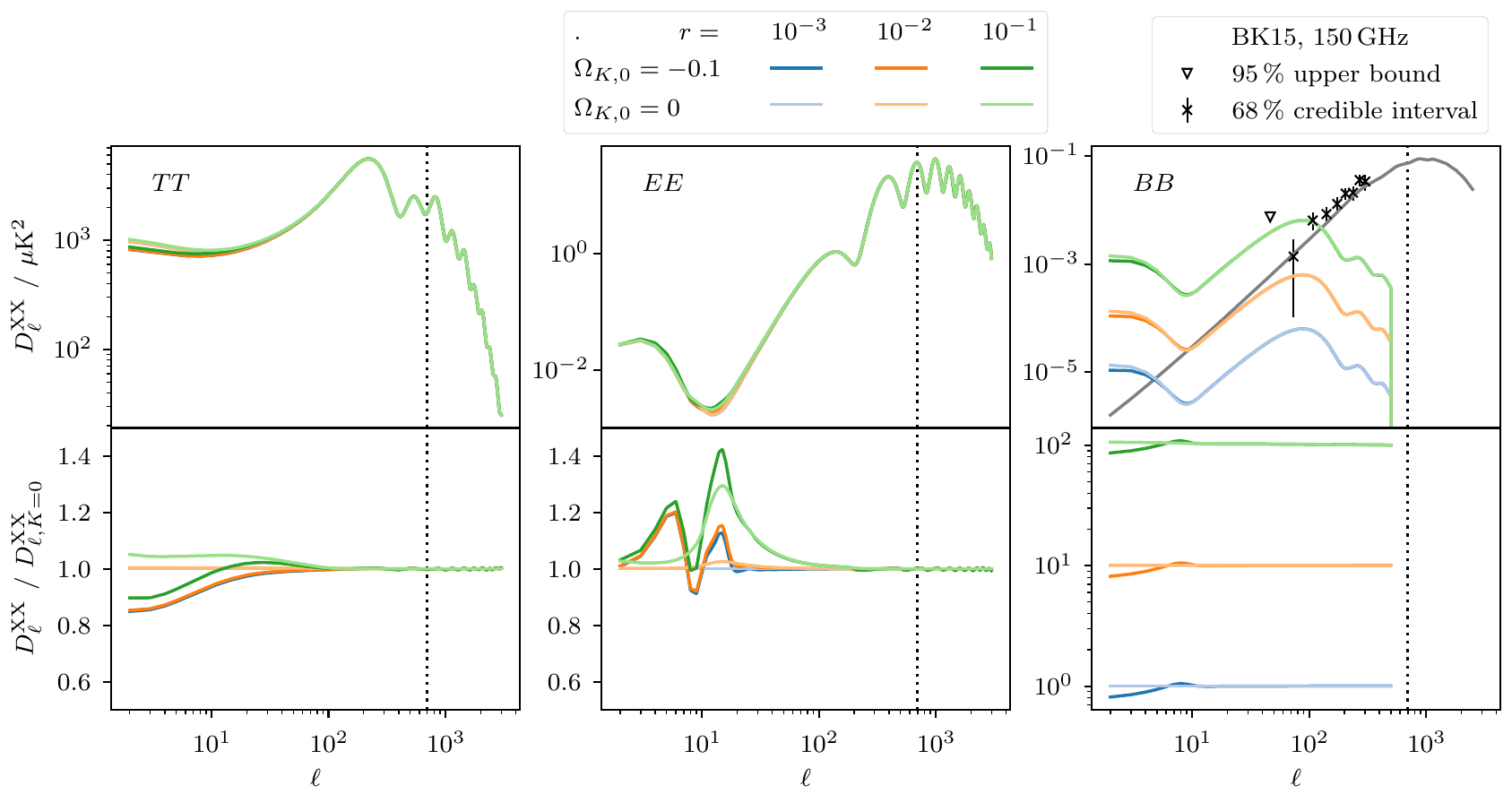}
    \caption{\label{fig:cl_Or} CMB power spectra of the temperature ($TT$) and polarisation ($EE$ in the middle and $BB$ on the right) anisotropies for different parameter values of the curvature density parameter~$\Omega_{K,0}$ and the tensor-to-scalar ratio~$r$. 
    The lighter hue corresponds to a flat universe, whereas the darker hue assumes a closed universe with $\Omega_{K,0}=-0.1$.
    The bottom plots show the relative difference, where we use the spectrum with $\Omega_{K,0}=0$ and $r=\num{e-3}$ as reference in the denominator.
    Note that we are using the \emph{unlensed} spectra for visualisation of the effects of the tensor-to-scalar ratio, here.
    The black line in the $B$-mode power spectrum on the right is the contribution of \emph{lensed} $E$-modes to the $B$-mode spectrum for $\Omega_{K,0}=0$, $r=0$.}
\end{figure*}

\subsection[\texorpdfstring{\LCDM}{LCDM} extensions]{Nested sampling results: \texorpdfstring{\LCDM}{LCDM} extensions}
\label{sec:result_extensions}

Since this paper focuses on cosmic inflation in curved universes, one obvious extension of the base \LCDM\ model to investigate is an extension with the present-day curvature density parameter~$\Omega_{K,0}$. 
Cosmic inflation governs the primordial Universe. Therefore we additionally look at extensions with primordial parameters.
Possible parameter extensions to the base \LCDM\ model include the running of the spectral index~$n_\mathrm{run}$ and the tensor-to-scalar ratio~$r$. In the following sections we focus on the tensor-to-scalar ratio~$r$, which is  more strongly constrained by current datasets than the running~$n_\mathrm{run}$. Hence, in what follows we present the results of a Bayesian analysis of the \LCDM\ model extended by~$r$ and~$\Omega_{K,0}$, both independently and jointly.

\subsubsection{Posteriors of \texorpdfstring{\LCDM}{LCDM} extensions}
\label{sec:posterior_extensions}

In \cref{fig:posterior_Onr} we show the one-dimensional and the pairwise joint two-dimensional posterior distributions for the present-day curvature density parameter~$\Omega_{K,0}$, the spectral index~$n_\mathrm{s}$ and the tensor-to-scalar ratio~$r$. We present the results using both \textsc{Planck}~2018 $TT,TE,EE+\mathrm{low}E$ data only, and using the data from the \textsc{Bicep2} and \textsc{Keck Array} in addition.

Because of the importance of the degree of compression from prior to posterior distribution for model comparison, we also include the prior distributions in grey in \cref{fig:posterior_Onr}, which are the same for all models. For visualisation purposes we illustrate the two-dimensional prior distribution in form of scatter points, as contours are better suited for constrained distributions. Note that while in principle all three parameters are sampled uniformly across their prior range, some parameter combinations need to be excluded at the prior level in order to compute a viable cosmological model, e.g.\ parameter combinations with large dark energy density~$\Omega_\Lambda$ and small matter density~$\Omega_\mathrm{m}$ leading to universes that had no Big Bang in the first place.
This leads to effectively non-uniform priors, the non-uniformity being somewhat visible for the spectral index~$n_\mathrm{s}$ and very clear for the curvature density parameter~$\Omega_{K,0}$ with a clear prior preference of close to flat universes.

CMB results for the \emph{one}-parameter extensions have been investigated thoroughly in previous analyses~\cite{Planck2018Parameters,BicepKeck2018BKX}, giving a mostly closed universe for the $\Omega_{K,0}$~extension and an upper bound on the tensor-to-scalar ratio for the $r$-extension. Joint analyses of~$r$ and~$\Omega_{K,0}$ have been briefly discussed in~\cite{Planck2013Inflation, Planck2015Inflation, Planck2018Inflation}. 
While the curvature parameter is little affected by the tensor-to-scalar ratio, the inverse is not true. The uncertainty on the tensor-to-scalar ratio increases considerably when allowing non-zero curvature with P18 data. This difference vanishes when BK15 data is taken into consideration, though, giving essentially the same distribution on~$r$ as without curvature. However, the shift in the spectral index~$n_\mathrm{s}$ from curvature is retained when allowing for a non-zero tensor-to-scalar ratio. This is important for inflation models and reheating bounds, as will be explored in the later \cref{sec:result_inflation,sec:result_reheating}.

\begin{figure*}[tb]
    \includegraphics[width=\textwidth]{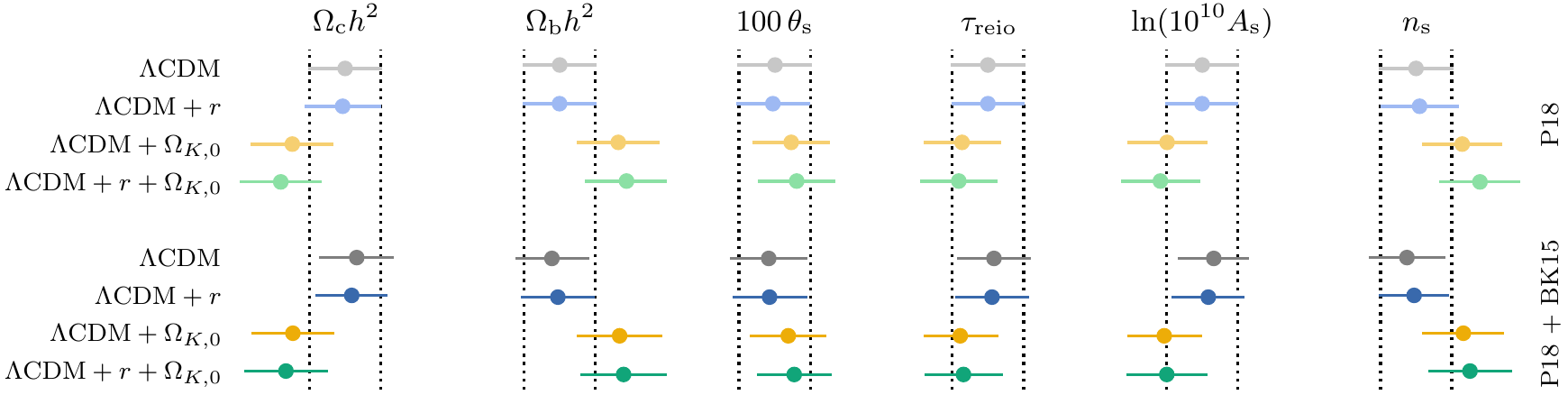}
    \caption{\label{fig:paramstability_lcdm} Parameter (in-)stability for extensions to the \LCDM\ base model using \textsc{Planck}~2018 $TT,TE,EE+\mathrm{low}E$~\cite{Planck2018CMB} data only (top four, light hue) and data from both \textsc{Planck} and from the \textsc{Bicep2} and \textsc{Keck Array}~\cite{BicepKeck2018BKX} (bottom four, dark hue). The dot represents the parameter mean and the error bars correspond to one standard deviation~($1\sigma$). We show the results for extensions with the tensor-to-scalar ratio~$r$ and/or the curvature density parameter~$\Omega_{K,0}$. The vertical dotted lines serve as visual references and are the $1\sigma$-boundaries for the \LCDM\ model from \textsc{Planck} data only (top line).}
\end{figure*}

Note that according to current data from Baryon Acoustic Oscillations~(BAOs) these constraints would be pulled towards a flat universe, i.e.\ to $\Omega_{K,0}=0$, and the shift in the spectral index~$n_\mathrm{s}$ would thus be undone. 
However, there has been concern over the combination of BAO with CMB data for curved universes~\cite{Handley2019c, DiValentino2019, DiValentino2020b}, which is why in the present study we restrict ourselves to CMB data only, and leave a more involved analysis including BAOs for future work.
The same applies (albeit to a lesser extent) to CMB lensing.

At first glance it might be surprising that the posterior distribution of the tensor-to-scalar ratio changes so significantly upon including non-zero curvature when computed from P18 data only, but remains essentially unchanged when including BK15 data. 
This phenomenon may be explained by the BK15 data offering an additional observable, the $B$-mode polarisation, which is much more sensitive to changes in the tensor-to-scalar ratio than the temperature data or the $E$-modes. This can be seen in \cref{fig:cl_Or} which shows the CMB temperature~($TT$) and polarisation ($EE$ and $BB$) power spectra respectively for combinations of $r=\{\num{e-3}, \num{e-2}, \num{e-1}\}$ with and without curvature. While in the case of $TT$ and $EE$ spectra different values of the tensor-to-scalar ratio are negligible in comparison to the effects of curvature, this behaviour is reversed in case of the $BB$ spectra, where~$r$ shows a significantly stronger influence. Hence, including the $B$-mode data from BK15 results in essentially the same upper bound on the tensor-to-scalar ratio regardless of whether the universe is assumed to be curved or flat.

\Cref{fig:paramstability_lcdm} summarises how the six cosmological base parameters change across the various extensions as regards their mean and standard deviation. 
Differences owing to the addition of BK15 data are negligible, with the parameter constraints all lying well within one standard deviation of one another across all models. Hence, both P18 and combined P18+BK15 exhibit the same trends when comparing different models with one another.
Extension with the tensor-to-scalar ratio leaves the cosmological base parameters essentially invariant. Adding curvature on the other hand shifts all these parameters by roughly one standard deviation. The biggest shift is in the spectral index~$n_\mathrm{s}$.

\subsubsection{Model comparison of \texorpdfstring{\LCDM}{LCDM} extensions}
\label{sec:stats_extensions}

\begin{figure*}[tb]
    \subfloat{\includegraphics[width=\columnwidth]{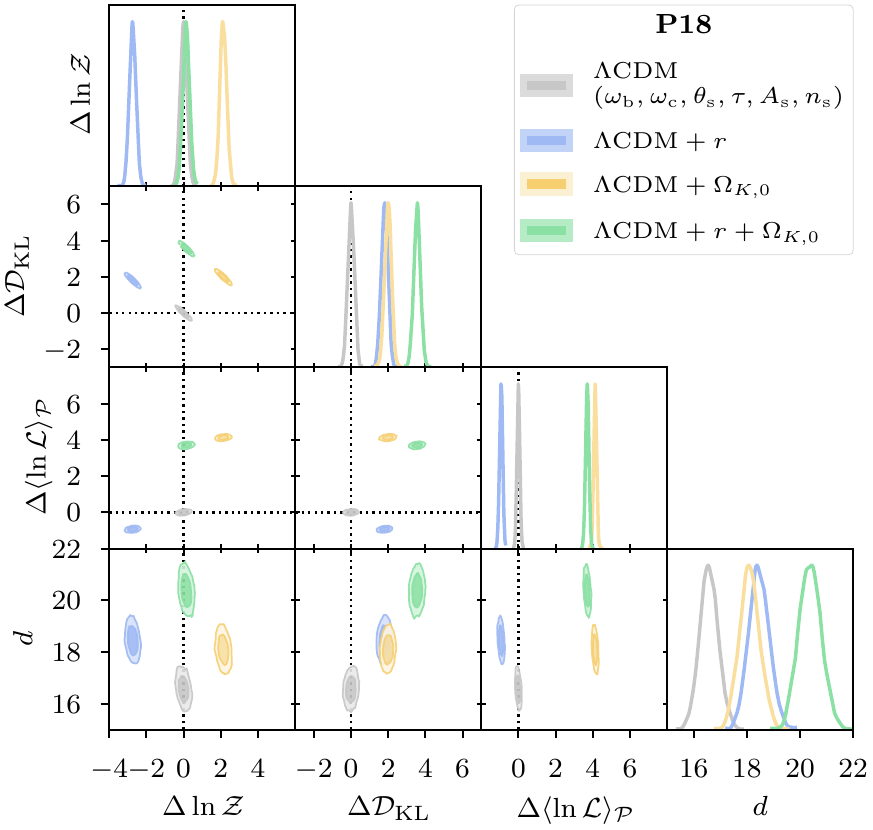}} 
    \hfill 
    \subfloat{\includegraphics[width=\columnwidth]{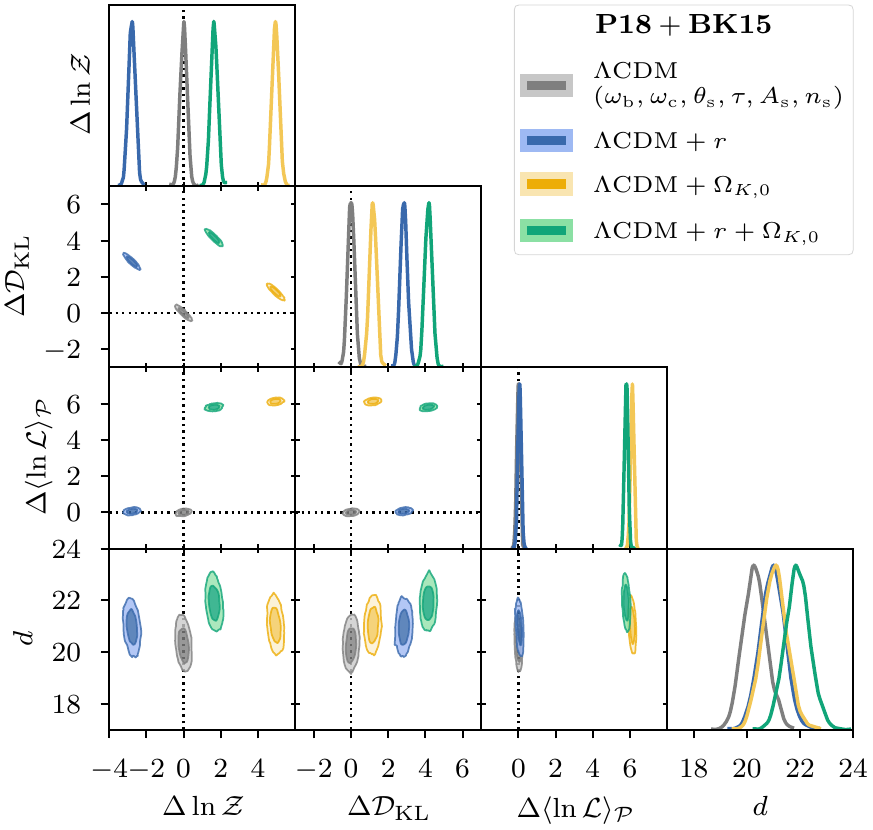}} 
    \caption{\label{fig:stats_extensions} Bayesian model comparison for extensions to the base \LCDM\ model with the tensor-to-scalar ratio~$r$ and/or the curvature density parameter~$\Omega_{K,0}$. 
    On the left side in a light hue we show the results using \textsc{Planck}~2018 $TT,TE,EE+\mathrm{low}E$ data only, whereas on the right side in a darker hue we show the results when additionally including data from the \textsc{Bicep2} and \textsc{Keck Array}.
    We show the log-evidence~$\ln\mathcal{Z}$, Kullback-Leibler divergence~$\mathcal{D}_\mathrm{KL}$ (in \si{\nats}), Bayesian model dimensionality~$d$, and the posterior average of the log-likelihood~$\langle\ln\mathcal{L}\rangle_\mathcal{P} = \ln\mathcal{Z}+\mathcal{D}_\mathrm{KL}$. 
    The~$\Delta$ denotes normalisation with respect to the base \LCDM\ model without extensions (i.e.\ with $r=0$ and $\Omega_{K,0}=0$) for easier comparison.
    The probability distributions represent uncertainty arising from the nested sampling process. In the limit of infinite life points these probability distributions would become point statistics.
    See \cref{tab:stats_extensions} for a full list of the numerical values and uncertainties.
    }
\end{figure*}

For the comparison of the \LCDM\ extensions we investigate the log-evidence~$\ln\mathcal{Z}$, Kullback-Leibler divergence~$\mathcal{D}_\mathrm{KL}$, Bayesian model dimensionality~$d$, and the posterior average of the log-likelihood~$\langle\ln\mathcal{L}\rangle_\mathcal{P} = \ln\mathcal{Z}+\mathcal{D}_\mathrm{KL}$. \Cref{fig:stats_extensions} shows these quantities for \textsc{Planck}~2018 $TT,TE,EE+\mathrm{low}E$ data in the left triangle plot and additionally using data from the \textsc{Bicep2} and \textsc{Keck Array} in the right triangle plot. 
Note that since these different datasets result in fundamentally different likelihood values, their absolute evidence values are not directly comparable. The relative differences of the various models, on the other hand, are comparable. We are using the base \LCDM\ model for any given likelihood combination as our reference point, and denote relative differences to that model with a~$\Delta$ such that for \LCDM\ itself we have $\Delta\ln\mathcal{Z}=\Delta\mathcal{D}_\mathrm{KL}=\Delta\langle\ln\mathcal{L}\rangle_\mathcal{P}=0$.

\paragraph{Tensor modes:} 
Including the tensor-to-scalar ratio (blue) is disfavoured with a log-evidence of $\Delta\ln\mathcal{Z}=-2.8\pm0.2$ which translates to betting odds of about $1:16$ against the $r$-extension. This is mostly driven by the Occam penalty for the additional parameter and because of the lack of any clear $B$-mode signal. Sampling the tensor-to-scalar ratio logarithmically would leave~$\log r$ mostly unconstrained and therefore the Bayesian evidence essentially invariant compared to the base \LCDM\ model. Consequently the KL-divergence, which is effectively a measure of the Occam penalty~\cite{Hergt2021}, would be much smaller, too, such that switching between uniform and logarithmic priors corresponds roughly to moving contours along a $\ln\mathcal{Z}+\mathcal{D}_\mathrm{KL}=C$ degeneracy line for some constant~$C$~\cite{Hergt2021}. 
Adding BK15 data leaves the log-evidence and the betting odds for the $r$-extension unaffected. The posterior average of the log-likelihood is essentially zero, telling us that the BK15 data does not require a non-zero tensor-to-scalar ratio for a sufficiently good fit. The relative entropy, on the other hand, increases to~$\Delta\mathcal{D}_\mathrm{KL}=2.8\pm0.2$, which is a result of the stronger compression from prior to posterior already observed in \cref{fig:posterior_Onr}.

\paragraph{Spatial curvature:} 
Including the present-day curvature density parameter~$\Omega_{K,0}$ (yellow), on the other hand, is favoured with a log-evidence of $\Delta\ln\mathcal{Z}=2.1\pm0.2$ compared to the base \LCDM\ model, which translates to betting odds of about $8:1$ in favour of the curvature extension. Note that this is smaller compared to findings in \cite{Handley2019c,DiValentino2019} which give odds of $50:1$ and a log-evidence of about~3.3 respectively. However, in those cases the upper bound on the flat prior on~$\Omega_{K,0}$ was chosen to be~0.05 or~0, whereas we have chosen our prior range symmetrically around zero as $[-1, +1]$. The preference for the curvature extension is mostly driven by an improved fit as can be seen by the increase in the posterior average of the log-likelihood.
Adding BK15 data further improves the fit compared to the \LCDM\ base model, and thereby significantly increases the odds in favour of the curvature extension to over $100:1$ with a log-evidence of $\Delta\ln\mathcal{Z}=4.9\pm0.2$.

\paragraph{Joint tensor modes and spatial curvature:} 
In a two-parameter extension with tensor modes and spatial curvature, their individual one-parameter effects cancel (for the P18 likelihood) and the log-evidence is essentially equal to that of the base \LCDM\ model. This only holds for the evidence, however, with the KL-divergence effectively adding up to represent the large Occam penalty from two additional parameters. 
It is worth noting how the tensor-to-scalar ratio and the curvature density have an almost orthogonal effect on the \LCDM\ model in the $(\Delta\ln\mathcal{Z}, \Delta\mathcal{D}_\mathrm{KL})$ plane. While the tensor-to-scalar ratio shifts the contour along the $\ln\mathcal{Z}+\mathcal{D}_\mathrm{KL}=C$ line (for some constant~$C$), the curvature density shifts along the $\ln\mathcal{Z}-\mathcal{D}_\mathrm{KL}=C$ line. The former corresponds to a shift caused mostly by an Occam factor (quantified by~$\Delta\mathcal{D}_\mathrm{KL}$). The latter corresponds to a shift mostly driven by a better fit, which can be quantified by the posterior average of the log-likelihood, related to Bayesian evidence and KL-divergence as: $\ln\mathcal{Z} + \mathcal{D}_\mathrm{KL} = \langle\ln\mathcal{L}\rangle_\mathcal{P}$ (cf.\ \cref{eq:evidence_fit_occam}). Hence, this can be seen directly in \cref{fig:stats_extensions} by looking at the joint contours of~$\Delta\mathcal{D}_\mathrm{KL}$ and~$\Delta\langle\ln\mathcal{L}\rangle_\mathcal{P}$.
Adding BK15 data also results in a combination of the one-parameter effects. The change in model preference by adding tensor modes stays mostly unaffected by the addition of BK15 data, regardless of with or without curvature. Similarly, the preference for curved models is further increased by the addition of BK15 data, regardless of with or without tensor modes. 
This means that the BK15 data has a greater effect on model preference involving the curvature parameter~$\Omega_{K,0}$ than involving the tensor-to-scalar ratio~$r$, which is curious in light of our previous discussion in \cref{sec:posterior_extensions} of the changes to the posterior. There the role was inverted, with the BK15 data having a greater effect on the posterior of the tensor-to-scalar ratio~$r$ than on the curvature density parameter~$\Omega_{K,0}$. 

The fourth parameter in \cref{fig:stats_extensions}, the Bayesian model dimensionality~$d$, typically comes with a large sampling uncertainty and is therefore not very specific in its count of the number of constrained parameters. It does show the expected ordering of the models, though, with the base \LCDM\ model having the fewest parameters and therefore the lowest dimensionality~$d$, and with the dimensionality increasing with additional parameters. In addition to the 6 cosmological parameters (plus one or two for extensions), the P18 likelihood comes with 21 nuisance parameters and the BK15 likelihood adds another 7 nuisance parameters. The total number does not match the total number of sampling parameters exactly, because many of the nuisance parameters are completely prior dominated, i.e.\ they show no or only little compression from prior to posterior distribution. We can roughly estimate that $10+3$ of the $21+7$ nuisance parameters are prior dominated (more details in \cref{sec:appendix,fig:posterior_nuisance}), such that we would expect roughly 17 constrained parameters from the P18 likelihood and roughly 21 from the P18+BK15 combination for the \LCDM\ base model, matching the dimensionalities in \cref{fig:stats_extensions}.

\subsection[Inflation models]{Nested sampling results: Inflation models}
\label{sec:result_inflation}

In the following we present the results from our nested sampling runs with fully numerically integrated primordial power spectra for the inflation models considered in \cref{sec:potentials}. For results that are mostly independent of the choice of potential we only show plots from the Starobinsky potential, representative of all potentials. Similarly, we only show results from the combined data of P18 and BK15 for some parameters, when there are no clearly visible differences with or without the BK15 data.

In addition to the prior bounds specified in \cref{tab:params} and to standard constraints from \LCDM\ cosmology, we also enforce the curvature constraint for open universes from \cref{eq:openconstraint}, the horizon constraint from \cref{eq:horizon_constraint} and the reheating constraints from \cref{eq:reheating_constraint_w,eq:reheating_constraint_N} at the prior level.

\subsubsection{Posteriors of primordial sampling parameters}

\begin{figure}[tb]
    \includegraphics[width=\columnwidth]{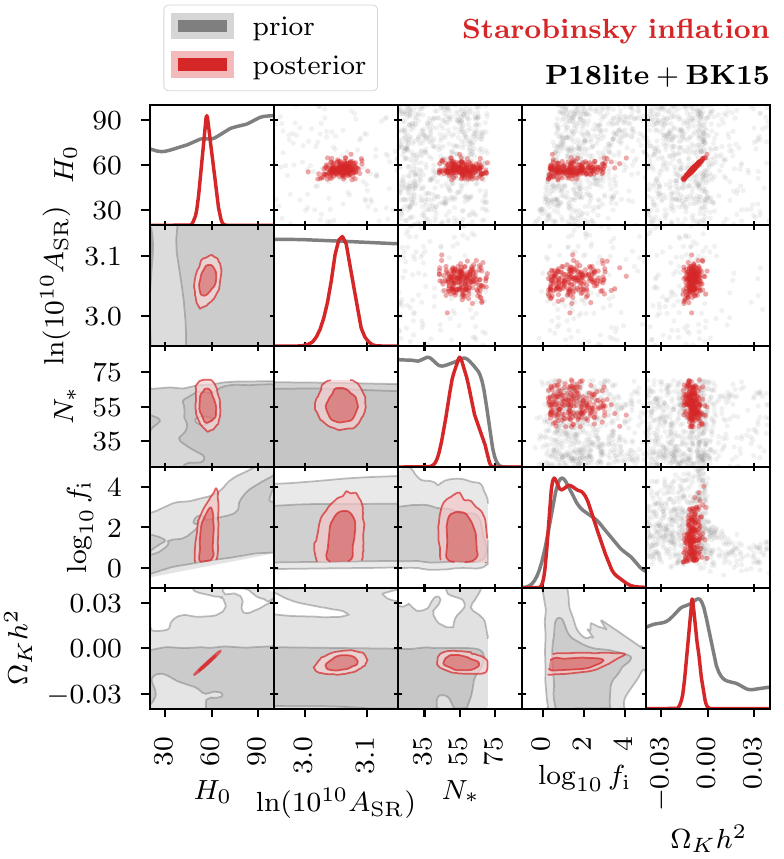}
    \caption{\label{fig:posterior_primordial} Prior (in grey) and posterior (in red) distributions of the parameters used to compute the primordial power spectrum with the Starobinsky potential. Note that all the priors shown here are initially set as uniform priors, but deviate from uniformity owing to additional constraints from curvature, reheating and horizon considerations.
    For the two-dimensional posterior distributions we show the \SIlist{68;95}{\percent} contours.
    }
\end{figure}

In \cref{sec:parametrisation} we introduced the slow-roll approximation of the amplitude of scalar density perturbations~$A_\mathrm{SR}$, the number of e-folds of inflation~$N_\ast$ after horizon crossing of the pivot scale, and the fraction of primordial curvature~$f_\mathrm{i}\equiv\Omega_{K,\mathrm{i}}/\Omega_{K,0}$ as our primordial sampling parameters. We sample these parameters together with the \LCDM\ parameters (using the Hubble parameter~$H_0$), and with the present day curvature density parameter~$\omega_{K,0}\equiv\Omega_{K,0}h^2$. In \cref{fig:posterior_primordial} we show the prior (in grey) and posterior (in red) constraints of the parameters going into the computation of the primordial physics for the Starobinsky model from P18 and BK15 data. 

The picture is very similar across all potentials considered here. The only notable difference between inflation models is in the e-folds parameter~$N_\ast$ which is characteristically linked to the epoch of reheating (see \cref{sec:reheating}) and to features in the primordial power spectrum such as the spectral index~$n_\mathrm{s}$ or the tensor-to-scalar ratio~$r$ (see \cref{sec:pps_sr,fig:nsr_slow_roll}). We will explore the results from these connections in more detail in \cref{sec:posterior_inflation,sec:result_reheating}.

The (approximate) amplitude of the scalar primordial power spectrum~$A_\mathrm{s}$ (or equivalently $A_\mathrm{SR}$, see \cref{fig:As_vs_ASR} in the appendix for a comparison) is by far the best constrained of these parameters. This comes as no surprise, considering that it is also one of the six parameters in the base \LCDM\ cosmology.

\begin{figure}[b]
    \includegraphics[width=\columnwidth]{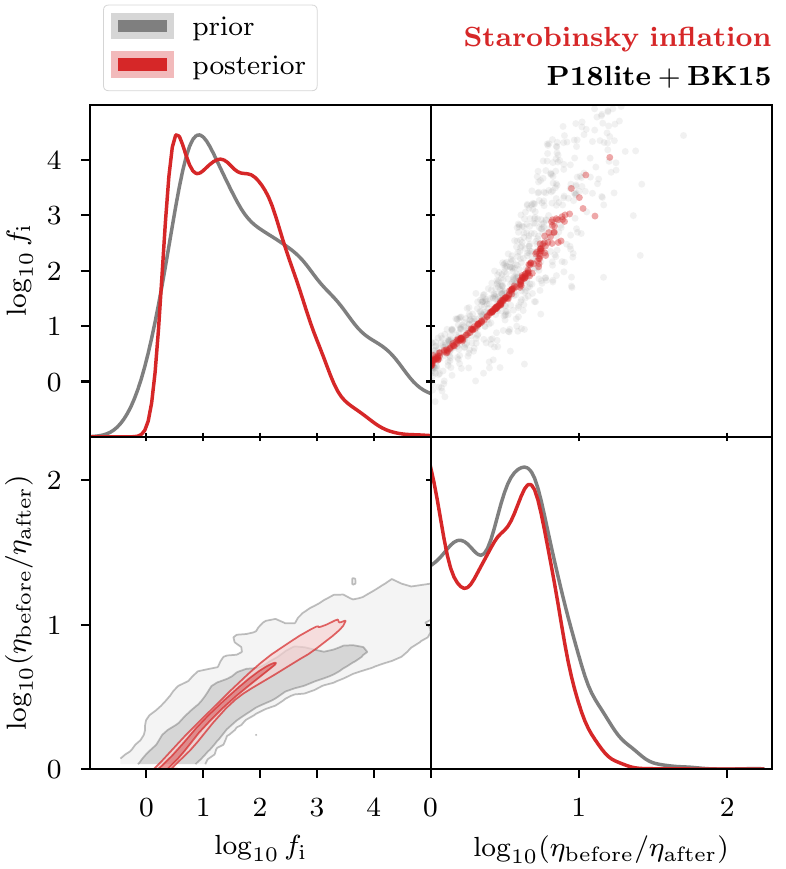}
    \caption{\label{fig:posterior_horizon} Prior (in grey) and posterior (in red) distributions of the primordial curvature fraction~$f_\mathrm{i}$ and the conformal time ratio~$\eta_\mathrm{before}/\eta_\mathrm{after}$.}
\end{figure}

The prior distribution of the primordial curvature fraction~$f_\mathrm{i}$ is the joint result of the curvature constraints and the horizon constraint from \cref{eq:horizon_constraint} specifying that the conformal time that passed before the end of inflation needs to be greater than thereafter in order to solve the horizon problem. 
Towards larger values of~$f_\mathrm{i}$ the prior is reduced owing to curvature constraints. First, the curvature constraint for open universes from \cref{eq:openconstraint} only allows comparatively small values of~$f_\mathrm{i}$. The prior is reduced further for too large curvature in closed universes, as these universes lack a Big Bang in the first place. Also, for large values of~$f_\mathrm{i}$ the maximum (which contributes the most to the conformal time before the end of inflation) of the comoving Hubble horizon (and equivalently the primordial curvature density) becomes very pointy (see \cref{fig:cHH}) and contributes less to the conformal time, which is the integral of the comoving Hubble horizon. 
It is useful in this context to look at the joint distribution with the present-day curvature and also at the corresponding plots from the conformal time analysis in \cref{fig:conformal_time}.
These plots also indicate that a total elapse of conformal time of~$\eta_\mathrm{total}=\pi/2$ from pre-inflationary Big Bang to the future conformal boundary, as proposed in the closed universe theory discussed in~\cite{Lasenby2005}, is consistent with the data.
The sharp drop in the prior towards low values of the primordial curvature fraction~$f_\mathrm{i}$ is driven by the horizon constraint as expected from our analysis in \cref{sec:conformal_time}.
This is confirmed in \cref{fig:posterior_horizon}, where we show the prior (in grey) and posterior (in red) distributions of the primordial curvature fraction~$f_\mathrm{i}$ and the conformal time ratio~$\eta_\mathrm{before}/\eta_\mathrm{after}$. 
Indeed the correlation between~$f_\mathrm{i}$ and the ratio of $\eta_\mathrm{before}/\eta_\mathrm{after}$ together with the cut of $\eta_\mathrm{before}>\eta_\mathrm{after}$ excludes low values of~$f_\mathrm{i}$.
On the posterior level, this correlation reduces almost to a one-to-one correspondence.
The data pushes the conformal time ratio, which already prefers low values below about~10 a priori, further down and thereby towards a scenario where there was just enough inflation to solve the horizon problem.

\begin{figure}[tb]
    \includegraphics[width=\columnwidth]{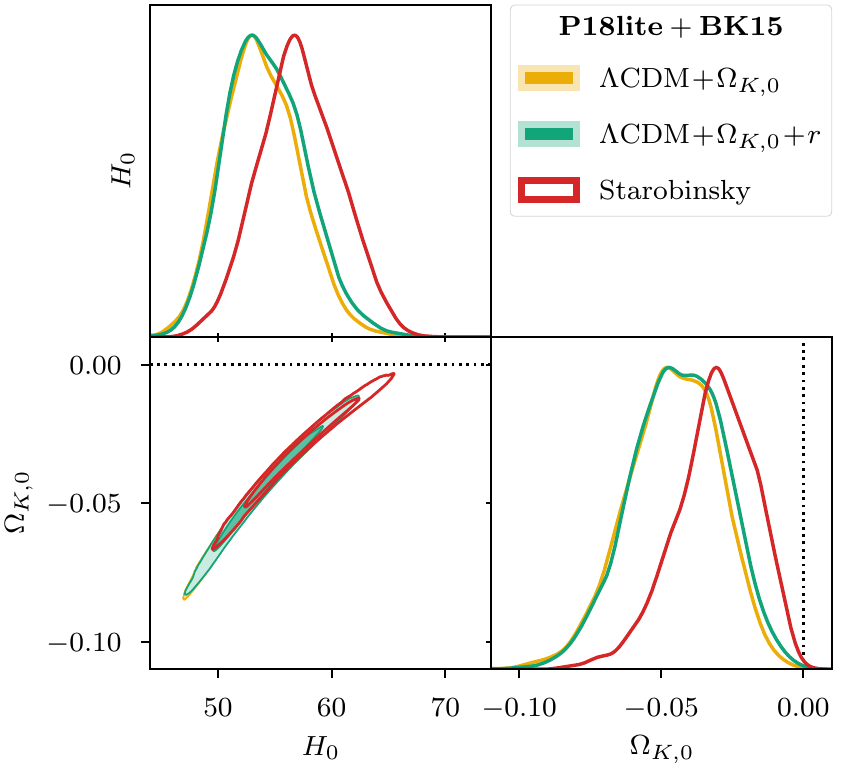}
    \caption{\label{fig:posterior_inflation_curvature} Posterior distributions of the Hubble parameter~$H_0$ and the present-day curvature density parameter~$\Omega_{K,0}$. We compare curvature extensions of the base \LCDM\ model using the standard power-law primordial power spectrum~(PPS) with a PPS computed numerically for Starobinsky inflation. For clarity we omit the posteriors for Quadratic and Natural inflation, which match the Starobinsky one.
    The posterior on the curvature density parameter for the Starobinsky model amounts to~$\Omega_{K,0}=\num{-0.031(14)}$.
    For the two-dimensional posterior distributions we show the \SIlist{68;95}{\percent} contours.}
\end{figure}

The posteriors on Hubble parameter~$H_0$ and present-day curvature density parameter~$\Omega_{K,0}$ match across inflation models. \Cref{fig:posterior_inflation_curvature} representatively shows the posterior of the Starobinsky model. Compared to curvature extensions of the base \LCDM\ cosmology, the bulk of the posterior mass shifts visibly towards flatness. Nevertheless, with a simultaneous narrowing of the posterior width, the probability density still drops to almost zero below $\Omega_{K,0}=0$ with the \SI{95}{\percent} upper bound shifting from about $-0.02$ to $-0.01$.

\subsubsection{Posteriors of derived parameters}
\label{sec:posterior_inflation}

Computing the primordial power spectrum from the inflationary background \cref{eq:background1,eq:background2,eq:eom,eq:background3} and the mode \cref{eq:mukhanov-sasaki_scalar,eq:mukhanov-sasaki_tensor} of scalar and tensor perturbations turns the phenomenological spectral index~$n_\mathrm{s}$ and tensor-to-scalar ratio~$r$ into derived parameters. These parameters mostly depend on the observable number of e-folds of inflation~$N_\ast$ from horizon crossing of the pivot scale~$k_\ast$ until the end of inflation. 
For a fixed start to inflation, e.g.\ through fixing $f_\mathrm{i}$, this dependence is equivalent to a dependence on the total number of e-folds of inflation~$N_\mathrm{tot}$.
As discussed in \cref{sec:reheating}, the equation-of-state parameter of reheating~$w_\mathrm{reh}$ is also mostly driven by the amount of inflation.

\begin{figure}[tb]
    \includegraphics[width=\columnwidth]{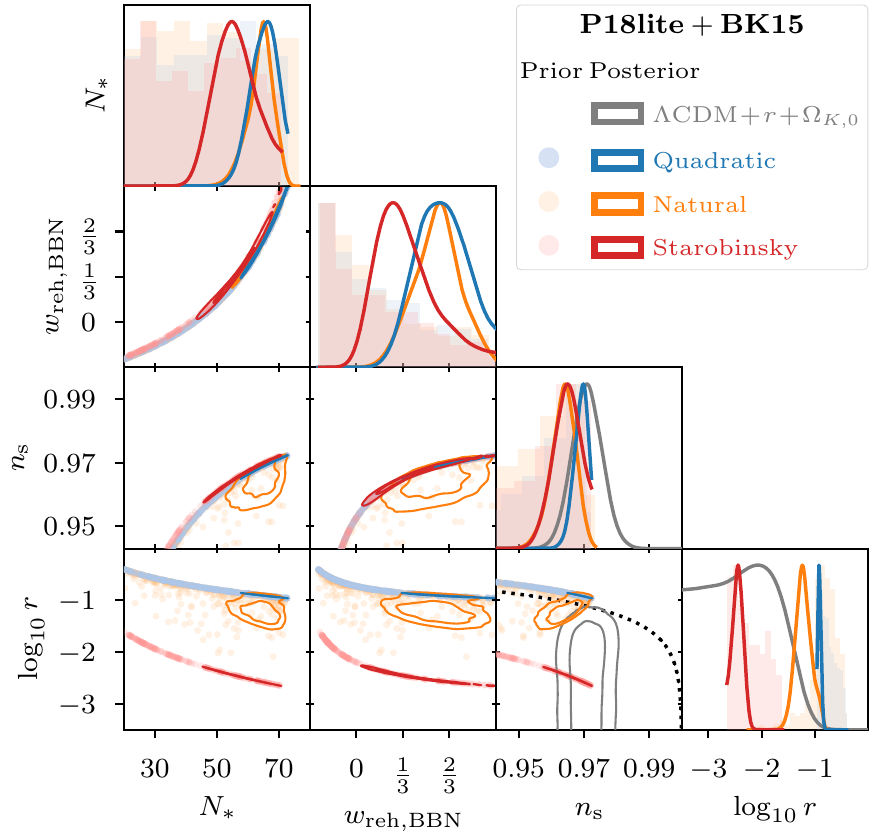}
    \caption{\label{fig:posterior_inflation_nsr} Posterior distributions of the observable number of e-folds of inflation~$N_\ast$, the equation-of-state parameter of reheating until BBN~$w_\mathrm{reh,BBN}$, the spectral index~$n_\mathrm{s}$ and the tensor-to-scalar ratio~$r$ for various inflation models. We show quadratic inflation in blue, natural inflation in orange and Starobinsky inflation in red. In grey we show the contours for a power-law primordial power spectrum following \cref{eq:powerlaw_pps,eq:powerlaw_pps_tensor} with the inflation consistency relation from \cref{eq:inflation_consistency}. The one-dimensional histogram plots and the two-dimensional scatter plots in a lighter hue illustrate the prior distributions of the corresponding parameters, which are non-uniform here since all parameters are derived parameters. The visible cutoffs of both prior and posterior distribution are driven by the (permissive) reheating constraint $-1/3 < w_\mathrm{reh,BBN} < 1$.
    For the two-dimensional posterior distributions we show the \SIlist{68;95}{\percent} contours.
    }
\end{figure}

\Cref{fig:posterior_inflation_nsr} shows all these parameters for three potential models in a triangle plot using the combined P18 with BK15 data. The prior distributions are shown in a lighter hue as histograms for the one-dimensional distributions and as scatter plots for the pairwise joint distributions. Posterior distributions are plotted with a darker hue. For~$n_\mathrm{s}$ and~$r$ we also show the posterior for the \LCDM\ extension with curvature~$\Omega_{K,0}$ and with~$r$ (sampled logarithmically). We show the tensor-to-scalar ratio on a logarithmic scale to better visualise the large difference in the inflation models' predictions. We show results from the quadratic potential in blue, natural potential in orange, and Starobinsky potential in red. 

The two-dimensional distributions (both prior and posterior) show the degeneracy lines between all these parameters for the various inflation models clearly. Only the natural inflation model with the extra inflationary parameter~$\phi_0$ shows a slightly greater spread, which would be more apparent on a linear scale in~$r$.

The reheating parameter~$w_\mathrm{reh,BBN}$ was allowed to vary from~$-1/3$ to~$1$, thereby placing a theoretical upper limit on the spectral index and lower limit on the tensor-to-scalar ratio. This is particularly apparent for the quadratic and natural potential. The data prefer a lower tensor-to-scalar ratio, but the reheating prior limits how far down the posterior contours can be pushed. 
Note that this is a very permissive reheating prior. In \cref{sec:result_reheating} we show the effects of different reheating priors in more detail.

\begin{table*}[tb]
\renewcommand{\arraystretch}{1.5}
    \caption{\label{tab:bestfit} 
    Best-fit parameter values used for generating \cref{fig:bestfit_pps,fig:bestfit_cmb}.
    Although the spectral index~$n_\mathrm{s}$, the tensor-to-scalar ratio~$r$, and the e-folds of inflation before horizon crossing~$N_\dagger$ are all derived parameters in the last row, we include them here for comparison with the \LCDM\ extensions and with the analysis for flat universes in~\cite{Hergt2019a}.
    }
    \begin{ruledtabular}
        \footnotesize
        
\begin{tabular}{ l c D{.}{.}{6} D{.}{.}{5} D{.}{.}{2} D{.}{.}{4} D{.}{.}{4} D{.}{.}{5} c c D{.}{.}{3} D{.}{.}{2} D{.}{.}{3} r }
    Model & Likelihood & \multicolumn{1}{c}{$\omega_\mathrm{b}$} & \multicolumn{1}{c}{$\omega_\mathrm{c}$} & \multicolumn{1}{c}{$H_0$} & \multicolumn{1}{c}{$\tau_\mathrm{reio}$} & \multicolumn{1}{c}{$10^9A_\mathrm{s}$} & \multicolumn{1}{c}{$n_\mathrm{s}$} & \multicolumn{1}{c}{$r$} & \multicolumn{1}{c}{$\Omega_{K,0}$} & \multicolumn{1}{c}{$f_\mathrm{i}$} & \multicolumn{1}{c}{$N_\ast$} & \multicolumn{1}{c}{$N_\dagger$} & \\ \hline
    $\Lambda\mathrm{CDM}$          & P18+BK15     & 0.022377  & 0.1201   & 67.32  & 0.0543  & 2.1004  & 0.96589  & 0        &  0       &        &        &      & \cite{Planck2018Parameters} \\
    $\Lambda\mathrm{CDM}+\Omega_K$ & P18+BK15     & 0.022632  & 0.11792  & 54.09  & 0.0495  & 2.0706  & 0.97235  & 0        & -0.0438  &        &        &      & \cite{Planck2018Parameters} \\
    Starobinsky inflation (flat)   & P15          & 0.022257  & 0.11965  & 67.38  & 0.0790  & 2.2018  & 0.96393  & 0.00331  &  0       &        & 57.1  & 6.09 & \cite{Hergt2019a}            \\
    Starobinsky inflation (closed) & P18lite+BK15 & 0.022552  & 0.11864  & 55.29  & 0.0509  & 2.0849  & 0.95737  & 0.00325  & -0.0376  & 0.128  & 57.6  & 5.93 &                             \\
\end{tabular}
    \end{ruledtabular}
\end{table*}

The results for the quadratic and natural potential are very similar. The preferred number of observable e-folds is roughly $N_\ast\approx60$ when using only P18 data, even larger with BK15 data included. This is somewhat larger than the more commonly quoted~50 to~\SI{60}{\efolds} owing to the larger spectral index when including curvature and because of the pull towards a smaller tensor-to-scalar ratio. The effective reheating parameter~$w_\mathrm{reh,BBN}$ is centred around~$1/3$ for P18 data only. This agrees well with the effective nature of the parameter tending towards~$1/3$ in cases where thermalisation would have happened much earlier than Big Bang Nucleosynthesis as previously discussed in \cref{sec:reheating}. Both the spectral index and tensor-to-scalar ratio are prior constrained. The models maximise their likelihood by pushing~$n_\mathrm{s}$ to its upper and~$r$ to its lower prior bound.

It might seem surprising that the results for the natural potential are so similar to those of the quadratic potential. One could have expected natural inflation's ability to accommodate for a smaller tensor-to-scalar ratio via a smaller potential hill parameter~$\phi_0$ to pull the posterior away from quadratic inflation, which indeed is slightly visible when including BK15 data. However, there is a trade-off between the spectral index and the tensor-to-scalar ratio in natural inflation. The data push simultaneously to smaller~$r$ and larger~$n_\mathrm{s}$, whereas the potential hill parameter~$\phi_0$ gives a smaller~$r$ only for a smaller~$n_\mathrm{s}$. 

The posterior of the Starobinsky model shows a preference for fewer observable e-folds~$N_\ast$ than for the other models. This also yields a lower effective reheating parameter~$w_\mathrm{reh,BBN}$ and spectral index~$n_\mathrm{s}$ and can be attributed to the Starobinsky model's generally lower tensor-to-scalar ratio~$r$. Where the other models push to the limit set by the reheating prior ($w_\mathrm{reh,BBN}<1$) to try and accommodate as small a tensor-to-scalar ratio as possible, the posterior for the Starobinsky model is well within the unconstrained plateau region of the likelihood on~$\log r$, which remains the case even when including BK15 data. Hence, there is no pressure towards smaller~$r$ for the Starobinsky model.

\subsubsection{Best-fit power spectra}

\begin{figure}[b]
    \includegraphics[width=\columnwidth]{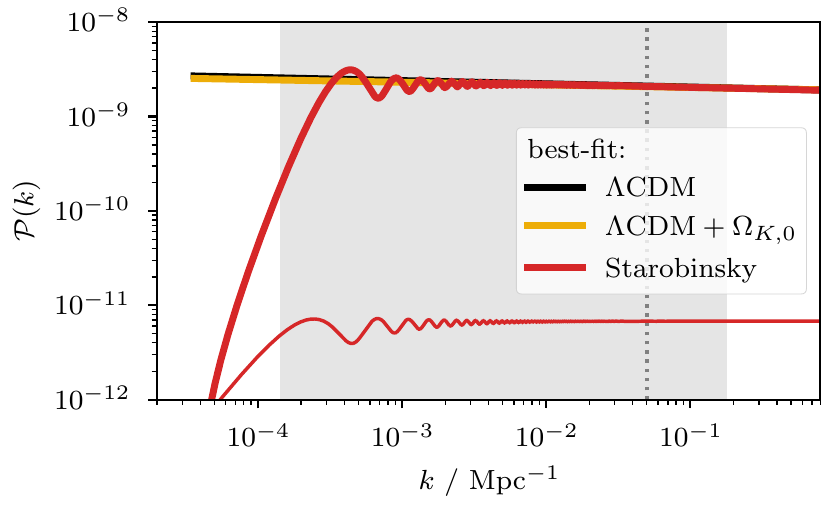}
    \caption{\label{fig:bestfit_pps} Best-fit primordial power spectra for the power-law spectra from \LCDM\ in black and its extension with curvature~$\Omega_{K,0}$ in yellow, and for the fully numerically integrated scalar (heavy upper line) and tensor (thin lower line) spectra from Starobinsky inflation in a closed universe in red. The vertical dotted line corresponds to the pivot scale~$k_\ast=\SI{0.05}{\per\mega\parsec}$, where power amplitude~$A_\mathrm{s}$, spectral index~$n_\mathrm{s}$ and tensor-to-scalar ratio~$r$ are measured. The grey shaded region illustrates roughly the CMB observable window. The corresponding CMB spectra are shown in 
    \cref{fig:bestfit_cmb}.}
\end{figure}

\begin{figure*}[p]
    \subfloat[\label{fig:bestfit_tt} Temperature spectrum]{\includegraphics[scale=1.03]{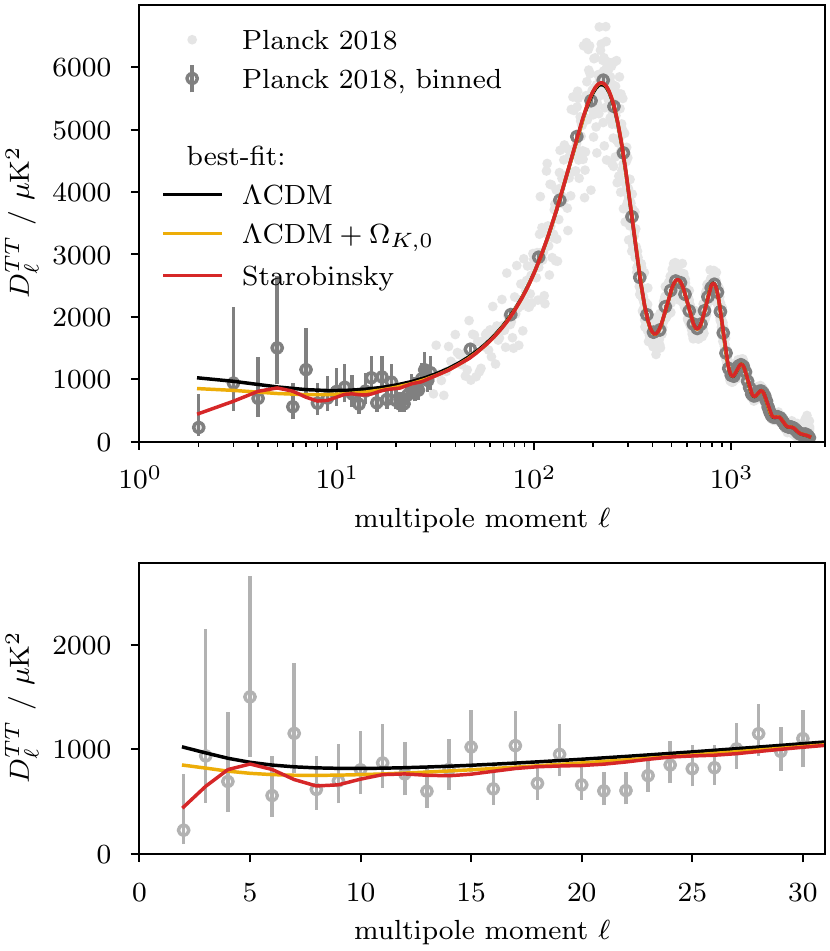}} \hfill
    \subfloat[\label{fig:bestfit_te} $TE$ cross-spectrum]{\includegraphics[scale=1.03]{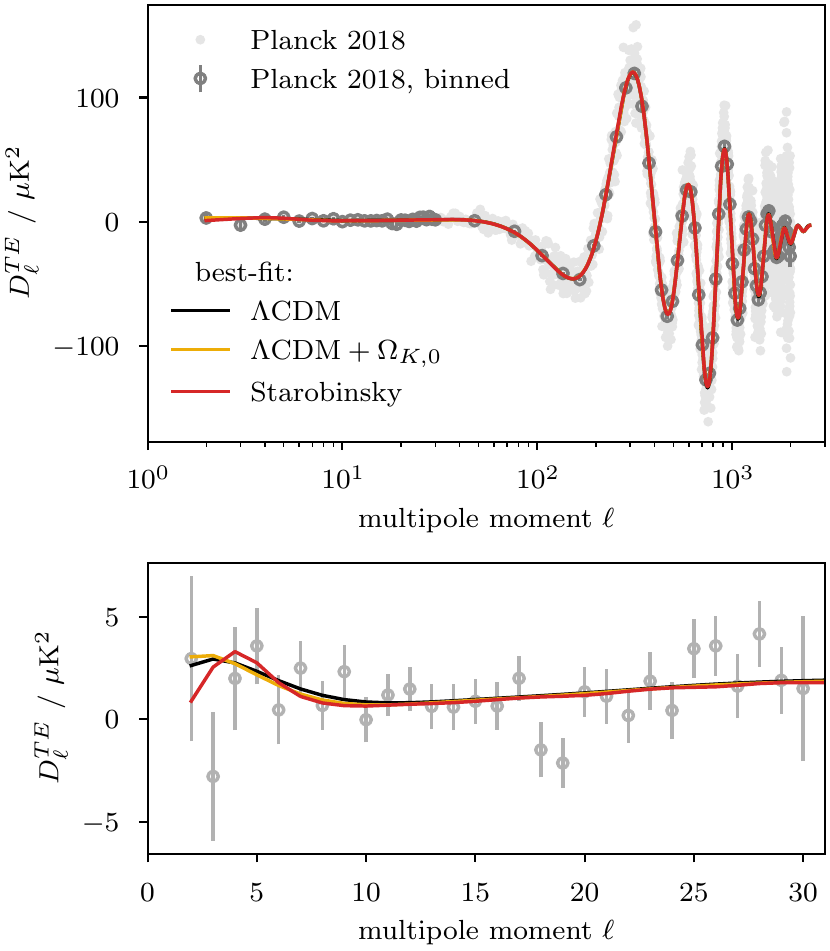}} \\
    \subfloat[\label{fig:bestfit_ee} $EE$ polarisation spectrum]{\includegraphics[scale=1.03]{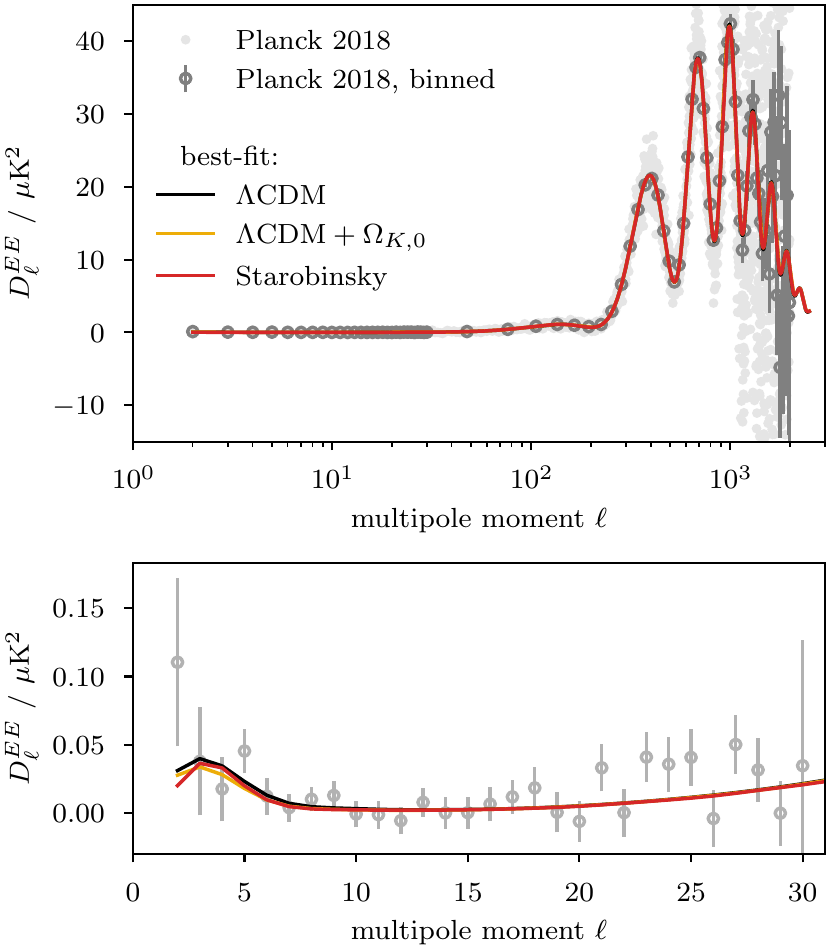}} \hfill
    \subfloat[\label{fig:bestfit_bb} $BB$ polarisation spectrum]{\includegraphics[scale=1.03]{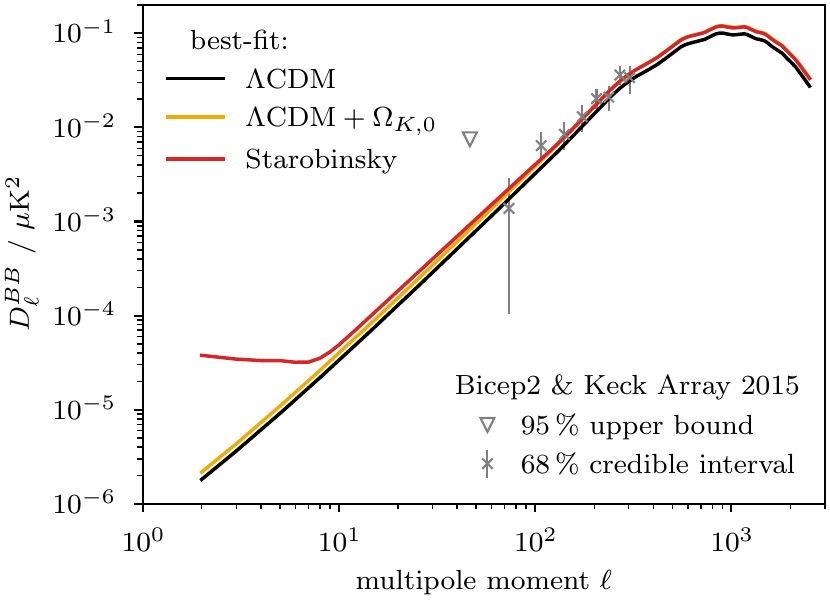}}
    \caption{\label{fig:bestfit_cmb} Comparison of the best-fit angular power spectra to the \textsc{Planck}~2018 data for temperature and $E$\=/mode polarisation, and to \textsc{Bicep2} and \textsc{Keck Array}~2015 data for $B$\=/mode polarisation. The lower plots zoom in on the low-$\ell$ multipole range from the upper plots. The best-fit spectra were computed using the corresponding primordial power spectra from \cref{fig:bestfit_pps} using the values listed in \cref{tab:bestfit}.}
\end{figure*}

In \cref{fig:bestfit_pps} we show the best-fit primordial power spectra~(PPS) that enter the computation of the angular $TT$, $TE$, $EE$, and $BB$ auto- and cross-spectra, plotted on top of the corresponding \textsc{Planck}~2018 data in \cref{fig:bestfit_tt,fig:bestfit_te,fig:bestfit_ee} and on top of the \textsc{Bicep2} and \textsc{Keck Array}~2015 data in \cref{fig:bestfit_bb}. We use the usual normalisation of the angular CMB power spectra according to:
\begin{align}
    D_\ell^{XX} \equiv \frac{\ell(\ell+1)}{2\pi}\ C_\ell^{XX} .
\end{align}
The correponding best-fit parameter values are listed in \cref{tab:bestfit}.

We compare three representative best-fit PPS. Black and yellow correspond to power-law spectra with the best-fit parameters from the base \LCDM\ model and from its extension with the curvature density parameter~$\LCDM+\Omega_{K,0}$ respectively. Red corresponds to the scalar and tensor spectra that were numerically integrated using the Starobinsky potential. 
Except for the amplitude of tensor modes, the power spectra from other inflationary potentials are very similar. Hence, we only show the best-fit result for the Starobinsky model.  

Comparing the \LCDM\ model and its curvature extension shows that the major difference arising from the addition of~$\Omega_{K,0}$ is a slightly larger spectral index~$n_\mathrm{s}$ as previously shown in \cref{fig:paramstability_lcdm,fig:posterior_Onr}. This results in a little less power on large scales, i.e.\ for small wavenumber~$k$ and multipole~$\ell$.

The PPS for the Starobinsky model shows the typical cutoff and oscillations towards large scales (small~$k$) that are associated with kinetic dominance initial conditions. 
The best-fit parameter combination propagates the cutoff and oscillations through to the temperature power spectrum in \cref{fig:bestfit_tt} where they sink into the large-scale lack of power. The effect on the $EE$ polarisation spectrum in \cref{fig:bestfit_ee} is considerably smaller. The $BB$ power spectrum for the Starobinsky model shows the characteristic reionisation bump on the largest scales (smallest multipoles~$\ell$) that comes with a non-zero tensor-to-scalar ratio. The derived best-fit value is $r=0.003$ in this case. However, the BK15 data only probes multipoles $\ell\gtrsim40$ and therefore does not reach to the large scales of the reionisation bump.

Apart from these differences on large scales, all models agree on small scales, driven mainly by the high precision on the power amplitude~$A_\mathrm{s}$ and by the good agreement between a power-law spectrum and the slow-roll predictions from inflation on small scales.

\subsubsection{Model comparison of inflation models}

\begin{figure}[tb]
    \includegraphics[width=\columnwidth]{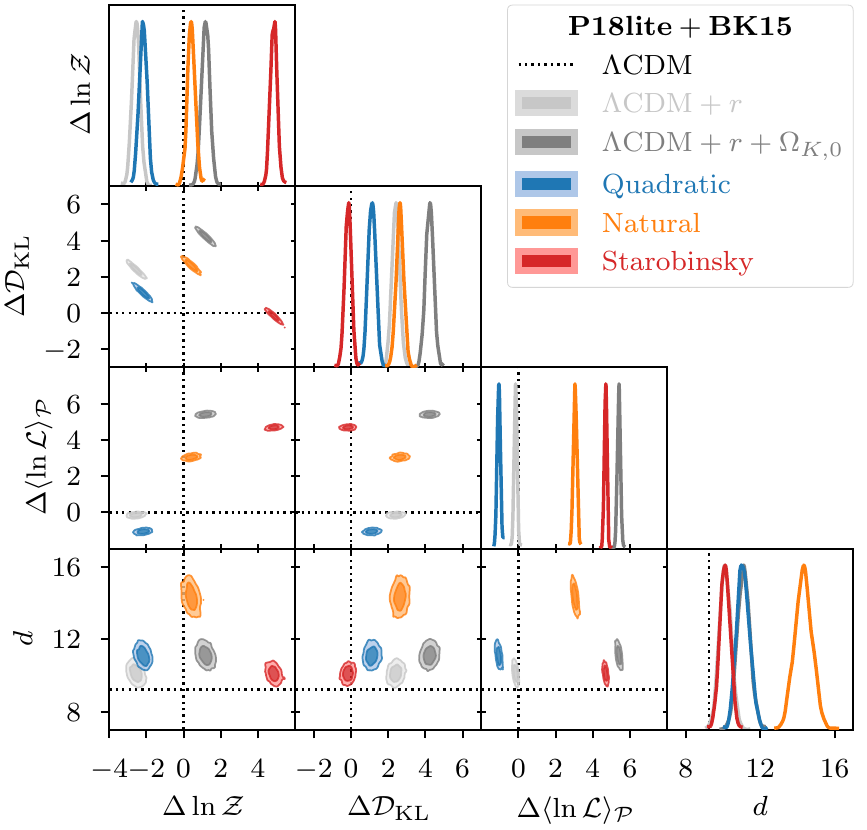}
    \caption{\label{fig:stats_inflation} Bayesian model comparison for various inflation models using the combined (lite) likelihoods of $TT,TE,EE+\mathrm{low}\ell+\mathrm{low}E$ from \textsc{Planck}~2018 with $BB$ from \textsc{Bicep2} and the \textsc{Keck Array}. 
    We show quadratic inflation in blue, natural inflation in orange, and Starobinsky inflation in red.
    We show the log-evidence~$\ln\mathcal{Z}$, the Kullback-Leibler divergence~$\mathcal{D}_\mathrm{KL}$ (in \si{\nats}), the posterior average of the log-likelihood~$\langle\ln\mathcal{L}\rangle_\mathcal{P} = \ln\mathcal{Z}+\mathcal{D}_\mathrm{KL}$, and the Bayesian model dimensionality~$d$.
    The~$\Delta$ denotes normalisation with respect to the base \LCDM\ model without extensions (i.e.\ with $r=0$ and $\Omega_{K,0}=0$) indicated by the vertical and horizontal dotted lines. As an additional reference we also include the results for an extension of \LCDM\ with tensor modes and curvature in grey.
    The probability distributions represent uncertainty arising from the nested sampling process. In the limit of infinite life points these probability distributions would become point statistics.
    See \cref{tab:stats_inflation} for a full list of the numerical values and uncertainties.
    }
\end{figure}

As in \cref{sec:stats_extensions} for extensions to the base \LCDM\ model, we investigate the log-evidence~$\ln\mathcal{Z}$, Kullback-Leibler divergence~$\mathcal{D}_\mathrm{KL}$, Bayesian model dimensionality~$d$, and the posterior average of the log-likelihood~$\langle\ln\mathcal{L}\rangle_\mathcal{P} = \ln\mathcal{Z}+\mathcal{D}_\mathrm{KL}$ for the three inflation models, quadratic, natural, and Starobinsky inflation, in a curved universe.
We show these quantities in \cref{fig:stats_inflation} in a triangle plot, for the combined likelihoods P18 and BK15.

We normalise with respect to the base \LCDM\ model (vertical and horizontal dotted lines marking zero). We also show the results for the $\LCDM+r$ and $\LCDM+r+\Omega_{K,0}$ model.
It should be noted, though, that because of the different sampling parameters and their priors it is difficult to compare the very phenomenological description of the primordial Universe as expressed by power-law parameters~$A_\mathrm{s}$, $n_\mathrm{s}$ and $r$, to the much more specific generation of the PPS from inflation models.
We have tried to mitigate this problem by using the same prior on the power amplitude~$A_\mathrm{s}$ for the inflation models. However, this cannot be done for the spectral index and the tensor-to-scalar ratio. Instead, these become derived parameters, dependent on the e-folds of inflation~$N_\ast$ after horizon crossing of the pivot scale.
As already seen in \cref{fig:posterior_primordial}, the priors on~$N_\ast$ are limited by external constraints from reheating and from needing to solve the horizon problem. This is a feature of curved universes providing an absolute scale for the Universe and thereby a limit on the amount of inflation. 

The Bayesian evidence for different inflation potentials is heavily dependent on the tensor-to-scalar ratio, which is not surprising in light of \cref{fig:nsr_slow_roll} and also apparent in the change of inflationary model comparisons from the Planck~2015 to the Planck~2018 analysis~\cite{Planck2015Inflation,Planck2018Inflation}. As observed in \cref{sec:posterior_extensions,fig:posterior_Onr}, using P18 data on its own, the tensor-to-scalar ratio~$r$ is even less constrained for curved universes than for a flat universe. So from P18 data only we do not expect big differences in the Bayesian evidence between different inflation models. BK15 data is necessary to get good constraints on~$r$ and thereby to properly compare the performance of various inflation models. 

Looking at the relative evidence~$\Delta\ln\mathcal{Z}$ in \cref{fig:stats_inflation}, we can see a clear preference hierarchy between the studied models, with the Starobinsky model clearly preferred, followed by natural inflation, and with quadratic inflation ruled out. This is very similar to previous results from flat universes~\cite{Easther2012,Planck2013Inflation,Planck2015Inflation,Planck2018Inflation,Hergt2019a}.

Quadratic inflation does not manage to provide a sufficiently small tensor amplitude under reheating constraints, and therefore is disfavoured with Bayesian odds of almost $1:10$ compared to \LCDM, and over $1:1000$ compared to the Starobinsky model. Even purely in terms of fit as measured by $\Delta\langle\ln\mathcal{L}\rangle_\mathcal{P}$ it performs poorly compared to \LCDM\ and its extensions.

Natural inflation fares slightly better owing to its ability to provide a smaller tensor-to-scalar ratio and is roughly on par with the \LCDM\ model and with the \LCDM\ extension with tensors and curvature. Compared to the Starobinsky model it is disfavoured with Bayesian odds of about $1:85$.

The Starobinsky model remains a strong competitor. Its fit is similar to that of the tensor and curvature extension of \LCDM, but it achieves this with a much smaller relative entropy~$\Delta\mathcal{D}_\mathrm{KL}$ or Occam penalty (smaller even than \LCDM\ which has fewer sampling parameters), and therefore ends up with a much higher Bayesian evidence. This goes to show that the Starobinsky model naturally manages to accommodate all the phenomenological requirements for the PPS imposed by the data.

\subsubsection{Effect of reheating constraints on evidences}
\label{sec:result_reheating}

In \cref{sec:reheating} we introduced the constraints from reheating on the end of inflation. In this section we will contrast the following two reheating scenarios:
\begin{align}
\label{eq:permissive_reh}
    \mathrm{(permissive)}& & N_\mathrm{reh} &= N_\mathrm{BBN},       & -\tfrac{1}{3} &< w_\mathrm{reh} < 1 , \\
\label{eq:restrictive_reh}
    \mathrm{(restrictive)}& & \rho_\mathrm{reh}^{1/4} &= \SI{e9}{\giga\eV} & -\tfrac{1}{3} &< w_\mathrm{reh} < \tfrac{1}{3} . 
\end{align}
Similar categories can be found in~\cite{Planck2013Inflation}.

\begin{figure}[tb]
    \includegraphics[width=\columnwidth]{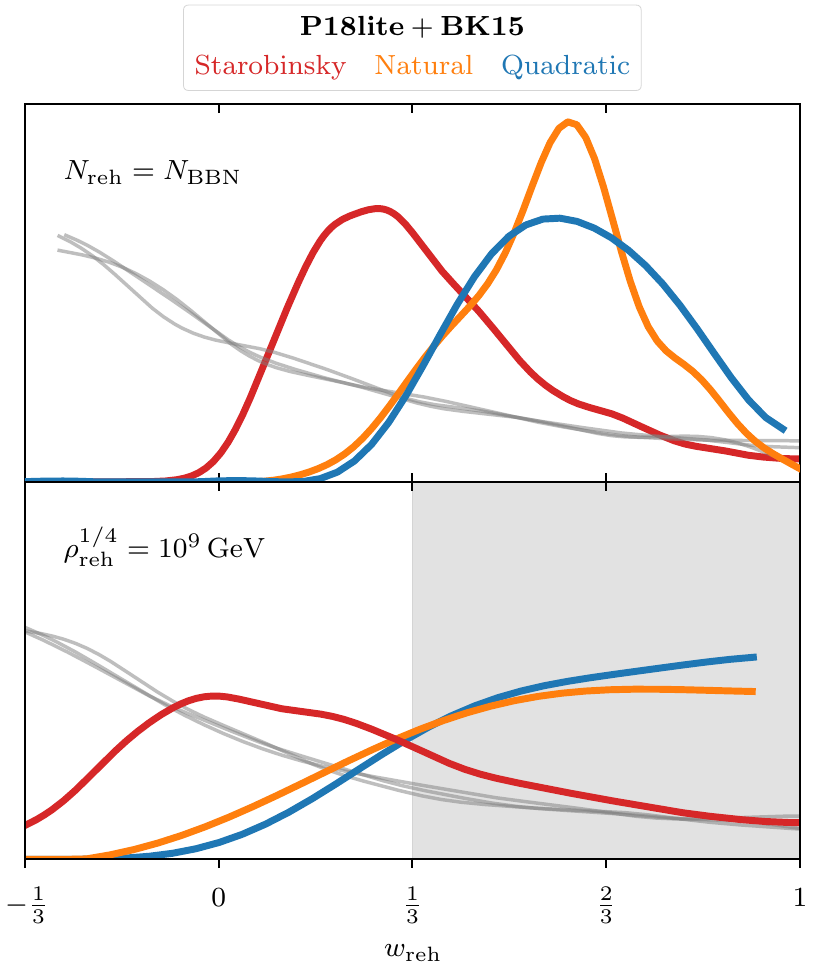}
    \caption{\label{fig:posterior_wreh} Posterior distributions for the derived equation-of-state parameter of reheating~$w_\mathrm{reh}$ for the permissive (upper panel) and restrictive (lower panel) reheating scenario from \cref{eq:permissive_reh,eq:restrictive_reh} respectively. 
    The grey shading in the lower panel highlights that under the restrictive reheating scenario only $-1/3<w_\mathrm{reh}<1/3$ is allowed.
    We show quadratic inflation in blue, natural inflation in orange, results from the quartic double-well potential in green and the Starobinsky model in red.
    The grey lines illustrate the sampled prior distributions.
    Dashed lines correspond to P18 data only, and solid lines to P18 and BK15 data combined.}
\end{figure}

\Cref{fig:posterior_wreh} shows the posterior distributions of the derived equation-of-state parameter~$w_\mathrm{reh}$ for the permissive scenario in the upper panel and the restrictive one in the lower panel. 
We again present the results in blue, orange, and red for quadratic, natural, and Starobinsky inflation respectively. In grey we illustrate the underlying prior distribution, which is derived from the prior distributions listed in \cref{tab:params}. Note how this favours small~$w_\mathrm{reh}$ values a priori, driven by a degeneracy with~$N_\ast$ as seen in \cref{fig:posterior_inflation_nsr}, but is clearly overcome by the data.
The reheating parameter is significantly larger for quadratic and natural inflation compared to the Starobinsky model.
This is driven by the smaller tensor-to-scalar ratio required by the BK15 data, which is mostly independent of the reheating scenario used. 

Comparing both reheating scenarios overall shows how the posterior for~$w_\mathrm{reh}$ is diluted away from instant reheating at $w_\mathrm{reh}=1/3$ the shorter the duration of reheating, i.e.\ for an earlier (stricter) end to reheating at a higher energy density. Phrased the other way round, the longer reheating is allowed to last, the more the posterior on the \emph{effective} equation-of-state parameter gets concentrated around $w_\mathrm{reh}=1/3$, which is equal to the equation-of-state parameter during the subsequent epoch of radiation domination (see \cref{fig:univolution_curvature,fig:reheating} for a visual aid).
At a first glance it might appear counter-intuitive that a permissive reheating scenario should result in tighter constraints on~$w_\mathrm{reh}$. However, the way to read this is that for the permissive reheating scenario essentially all posterior samples fall into the acceptable range of~$w_\mathrm{reh}$, which is not the case for the restrictive reheating scenario.

The posteriors for quadratic and natural inflation both peak at values $w_\mathrm{reh}>1/3$, meaning that the comoving Hubble horizon needs to grow faster during reheating than during radiation domination to catch up with the standard Big Bang evolution. Note that this result is in stark contrast to the analytic prediction of matter dominated reheating, i.e.\ $w_\mathrm{reh}\approx0$, from the time averaged oscillations of the inflaton field around its potential minimum (see also \cref{sec:reheating}). The dilution of the posterior with strict reheating somewhat reconciles these models with matter dominated reheating, but this shows that any such oscillations can only last for a short time in case of quadratic or natural inflation.

The Starobinsky model peaks in-between~$0$ and~$1/3$ in case of permissive reheating and roughly at~$0$ for restrictive reheating. Thus, for Starobinsky inflation, matter dominated oscillations around the potential minimum agree very well with the data, further adding to the success of the model, which it already accumulated on the level of the spectral index and the tensor-to-scalar ratio (although we recognise that these are all connected). 

For the nested sampling runs presented in the previous sections and in \cref{fig:posterior_inflation_nsr,fig:stats_inflation} in particular, we only used the permissive reheating scenario from \cref{eq:permissive_reh} as a prior constraint. In order to infer the evidence~$\mathcal{Z}$ and Kullback--Leibler divergence~$\mathcal{D}_\mathrm{KL}$ with the restrictive scenario as prior constraint, we use \texttt{anesthetic}'s~\cite{Anesthetic} importance sampling feature for nested samples. This frees us from the need to recompute entire nested sampling runs. However, as with importance sampling of MCMC chains, it only works well provided sufficient coverage of the importance sampled subspace of the original parameter space. Hence, the uncertainties tend to increase, which is especially the case for quadratic and natural inflation, for which most sample points belong to the excluded region of parameter space with $w_\mathrm{reh}>1/3$, as is clearly visible in \cref{fig:posterior_wreh}.

\begin{figure}[tb]
    \includegraphics[width=\columnwidth]{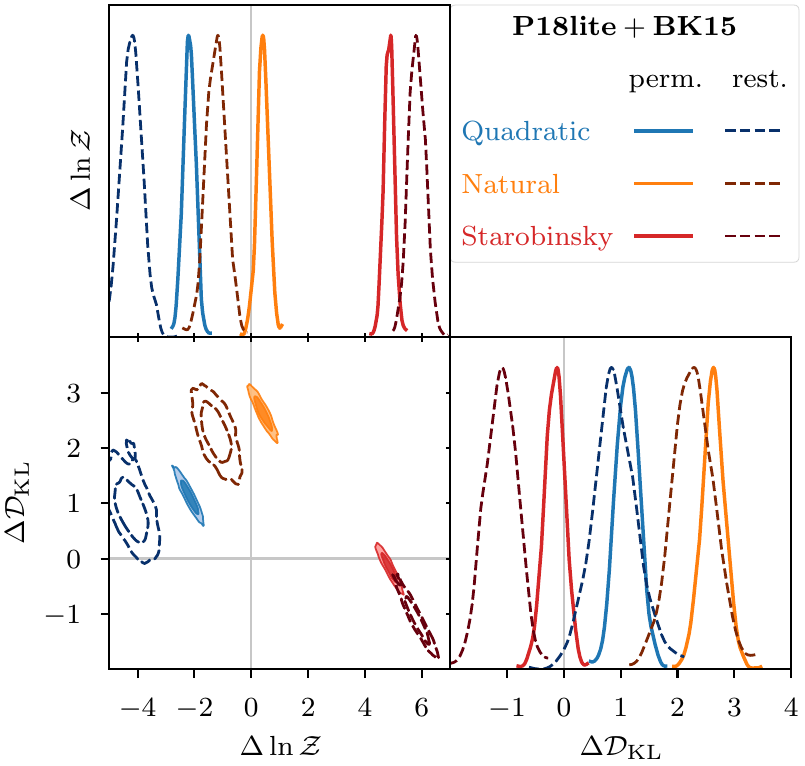}
    \caption{\label{fig:stats_reheating} Bayesian model comparison of permissive (solid lines, cf.\ \cref{eq:permissive_reh}) and restrictive (dashed lines, cf.\ \cref{eq:restrictive_reh}) reheating scenarios for various inflation models using P18 and BK15 data combined. 
    We show quadratic inflation in blue, natural inflation in orange, results from the quartic double-well potential in green and the Starobinsky model in red.
    We show the log-evidence~$\ln\mathcal{Z}$ and Kullback-Leibler divergence~$\mathcal{D}_\mathrm{KL}$ (in \si{\nats}). 
    The~$\Delta$ denote normalisation with respect to the base \LCDM\ model indicated by the vertical and horizontal grey lines.}
\end{figure}

In \cref{fig:stats_reheating} we compare the log-evidence~$\ln\mathcal{Z}$ and KL-divergence~$\mathcal{D}_\mathrm{KL}$ of the permissive with those of the restrictive reheating scenario for combined P18 and BK15 data. 
The filled contours in \cref{fig:stats_reheating} for the permissive reheating case from \cref{eq:permissive_reh} are the same as in \cref{fig:stats_inflation}. The unfilled contours with dashed lines come from the importance sampling with the restrictive reheating case from \cref{eq:restrictive_reh}.

For quadratic and natural inflation, the Bayesian evidence~$\mathcal{Z}$ shrinks further by two to three log-units, while the KL-divergence~$\mathcal{D}_\mathrm{KL}$ remains relatively unchanged, indicating a poorer fit of these models in a restrictive reheating scenario.
The fit of Starobinsky inflation, on the other hand, remains unchanged, but the KL-divergence and hence the Occam penalty decreases, leading to a larger evidence.
This increases the gap between quadratic and Starobinsky inflation beyond Bayesian odds of overwhelming $1:\num{20000}$, and for natural to Starobinsky inflation beyond $1:1000$.

\section{Conclusion}
\label{sec:conclusion}

Despite the success of flat \LCDM\ there has been a persistent tendency towards positive curvature (closed universes) in CMB temperature and polarisation data. The 2018 data release from \textsc{Planck} in particular has sparked some discussion of possible evidence for spatial curvature in the CMB.
In this paper we have investigated what the presence of such non-zero spatial curvature means for inflation.

We have reviewed how curvature links primordial and late-time scales, and how the detection of non-zero late-time curvature limits the total amount of inflation, thereby placing a bound on the comoving Hubble horizon, which becomes maximal at the onset of inflation. This sets tight constraints on initial conditions for inflation in order to solve the horizon and flatness problems, which we have folded into a Bayesian comparison of various inflationary potentials.

We have computed the primordial power spectra from these inflation models numerically, revealing oscillations and a cutoff towards large scales, which are common features of finite inflation. Additionally we have shown how curvature leads to an additional suppression or to an amplification of power on large scales for closed and open universes respectively, which holds for both scalar and tensor perturbations.

In our Bayesian analysis we have used CMB data from the \textsc{Planck}~2018 legacy archive and from the 2015 observing season of \textsc{Bicep2} and the \textsc{Keck Array}.
We chose this approach of purely using CMB data and not including data from lensing or BAOs in order to test how far cosmic inflation, which drives the primordial universe towards flatness, affects the preference for closed universes observed in CMB data.

Nested sampling runs of the base \LCDM\ cosmology and its extensions with the present-day curvature density parameter~$\Omega_{K,0}$ and/or the tensor-to-scalar ratio~$r$, presented in \cref{fig:posterior_Onr}, have confirmed that the inclusion of curvature significantly weakens the bounds on the tensor-to-scalar ratio when only taking temperature and $E$-mode polarisation into account. This fails to hold, however, when including $B$-mode data, in which case the bounds on~$r$ match those of a flat cosmology. Note that we have adopted a nominally uniform prior on the curvature density parameter~$\Omega_{K,0}$. The CMB constraints on the spectral index~$n_\mathrm{s}$, on the other hand, point to a persistently larger value, roughly one standard deviation greater with than without curvature, albeit with roughly~$\SI{6}{\sigma}$ still clearly below scale invariance. Nevertheless, this changes the picture of slow-roll predictions from various inflation models, as shown in \cref{fig:nsr_slow_roll}. 
We have computed the Bayesian evidence and Kullback--Leibler divergence for the various extensions and confirmed previous findings of the CMB having a preference for closed cosmologies. This preference is reduced when the tensor-to-scalar ratio is included, which comes with a significant Occam penalty (same as for flat universes). Interestingly, the addition of $B$-mode data further strengthens the preference for closed universes. The details of this model comparison are presented in \cref{fig:stats_extensions}.

Using the aforementioned numerically integrated primordial power spectra, we have also computed the Bayesian posteriors and evidence from three single-field inflationary potentials: the quadratic, natural, and Starobinsky potential.
We have found prior constraints on the primordial curvature, giving a lower bound from horizon considerations and upper bounds from considerations of an open or closed global geometry and from reheating.
Similarly, there are prior constraints on the amount of inflation. These are the combined effect of curvature linking the primordial to the late-time universe and of possible reheating scenarios. 

As in previous findings considering curvature or finite inflation, an improved fit to CMB data is achieved via a suppression of power and smoothing of peaks on the largest scales (analogous to effects of the artificial lensing parameter~$A_\mathrm{lens}$, see e.g.~\cite{Planck2018Parameters}).
In the absence of $B$-mode data all inflation models considered perform similarly well, a result of the weaker bound on the tensor-to-scalar ratio. However, with $B$-mode data taken into account, we obtain similar results to the flat case, with the Starobinsky model significantly outperforming the other inflation models, as seen in \cref{fig:stats_inflation}.

Quadratic and natural inflation are reheating constrained, which becomes very clear when looking at the spectral index and the tensor-to-scalar ratio as shown in \cref{fig:posterior_inflation_nsr}. Both the high spectral index from the inclusion of non-zero curvature and the low tensor-to-scalar ratio from $B$-mode data push those inflation models to the edges of their prior constraints, with the limits set by the equation-of-state parameter of reheating. 
In the first instance we have only used very permissive reheating constraints. We have then used importance sampling to explore stricter reheating constraints, presented in \cref{fig:posterior_wreh,fig:stats_reheating}, which has significantly penalised quadratic and natural inflation while strengthening the Starobinsky model.

\begin{acknowledgments}
LTH was supported by the Isaac Newton Trust, the STFC, and the Cavendish Laboratory as well as a UBC Killam Postdoctoral Research Fellowship. FJA thanks the STFC for their support. WJH was supported by a Gonville \& Caius Research Fellowship and a Royal Society University Research Fellowship.  

This work was performed using the resources provided by the Cambridge Service for Data Driven Discovery~(CSD3) operated by the University of Cambridge Research Computing Service (www.csd3.cam.ac.uk), provided by Dell EMC and Intel using Tier-2 funding from the Engineering and Physical Sciences Research Council (capital grant EP/P020259/1), and DiRAC (www.dirac.ac.uk) funding from the Science and Technology Facilities Council~(STFC) (capital grants ST/P002307/1 and ST/R002452/1 and operations grant ST/R00689X/1). 
DiRAC is part of the National e-Infrastructure.

This work was also performed using the DiRAC Data Intensive service at Leicester~(DiaL), operated by the University of Leicester IT Services, which forms part of the STFC DiRAC HPC Facility (www.dirac.ac.uk). 
The equipment was funded by BEIS capital funding via STFC (capital grants ST/K000373/1 and ST/R002363/1 and operations grant ST/R001014/1). 
DiRAC is part of the National e-Infrastructure.
\end{acknowledgments}

\appendix
\section{Robustness checks}
\label{sec:appendix}

We use this appendix to expand on some of the tests and checks we have performed.

\begin{figure*}[b]
    \includegraphics[width=\textwidth]{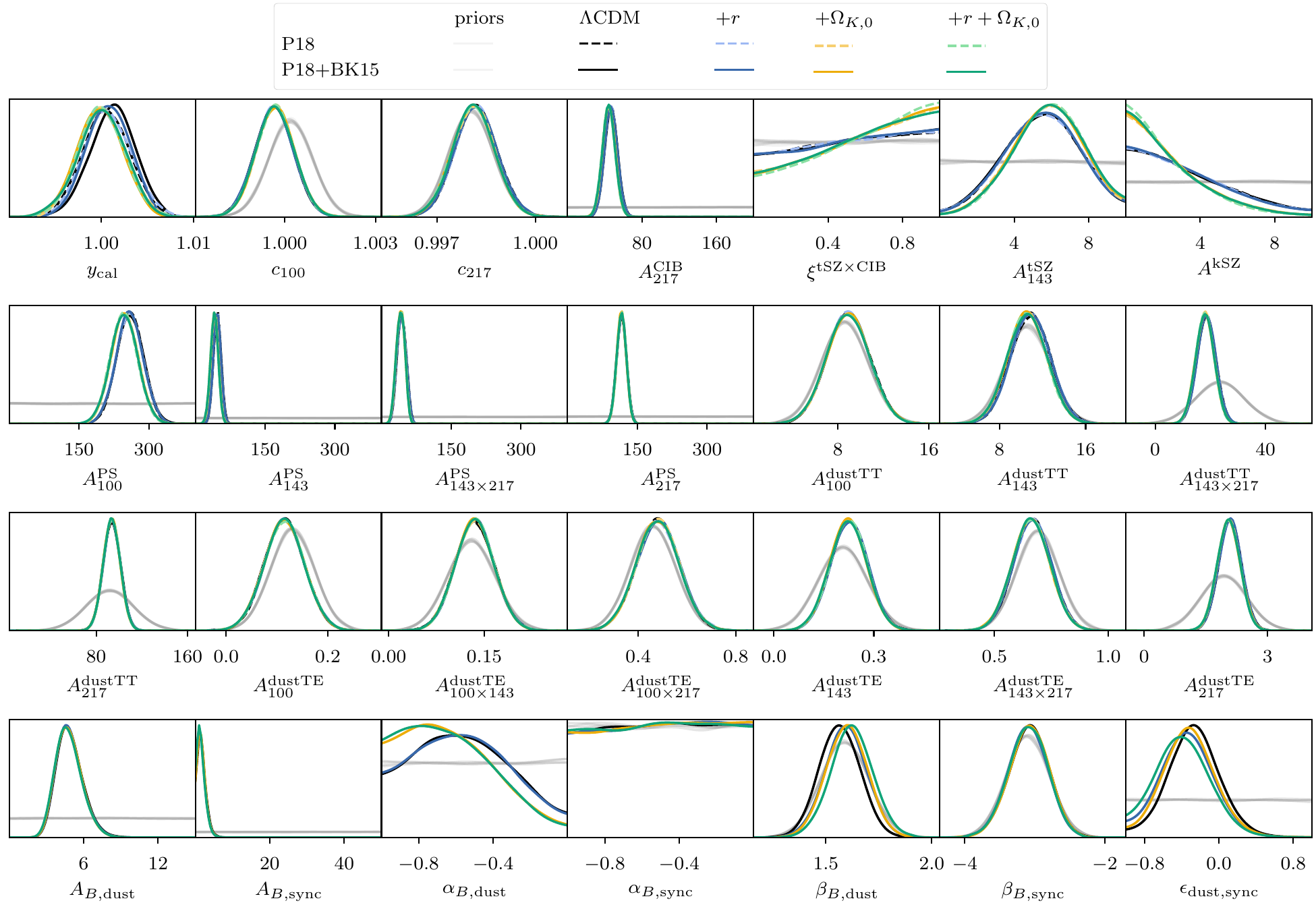}
    \caption{\label{fig:posterior_nuisance} Posteriors of P18 and BK15 nuisance parameters for the base \LCDM\ cosmology and one- or two-parameter extensions with the tensor-to-scalar ratio~$r$ and/or the present-day curvature density parameter~$\Omega_{K,0}$.}
\end{figure*}

\Cref{fig:posterior_nuisance} shows the one-dimensional posterior distributions for the 21 nuisance parameters of the \textsc{Planck}~2018 likelihoods and the extra 7 nuisance parameters of the likelihood from \textsc{Bicep2} and the \textsc{Keck Array}. We show the distributions for the base \LCDM\ model and three of its parameter extensions, demonstrating how the posteriors of the nuisance parameters are mostly unaffected by the choice of cosmological model.

\Cref{tab:stats_extensions,tab:stats_inflation} list the nested sampling results for log-evidence~$\ln\mathcal{Z}$, Kullback--Leibler divergence~$\mathcal{D}_\mathrm{KL}$, posterior average of the log-likelihood~$\langle\ln\mathcal{L}\rangle_\mathcal{P}$, and Bayesian model dimensionality~$d$. \Cref{tab:stats_extensions} shows results for the base \LCDM\ model and three of its parameter extensions, each for various likelihood runs. \Cref{tab:stats_inflation} shows results from three inflation models, each for two reheating constraints.

\begin{figure}[tbb]
    \includegraphics[width=\columnwidth]{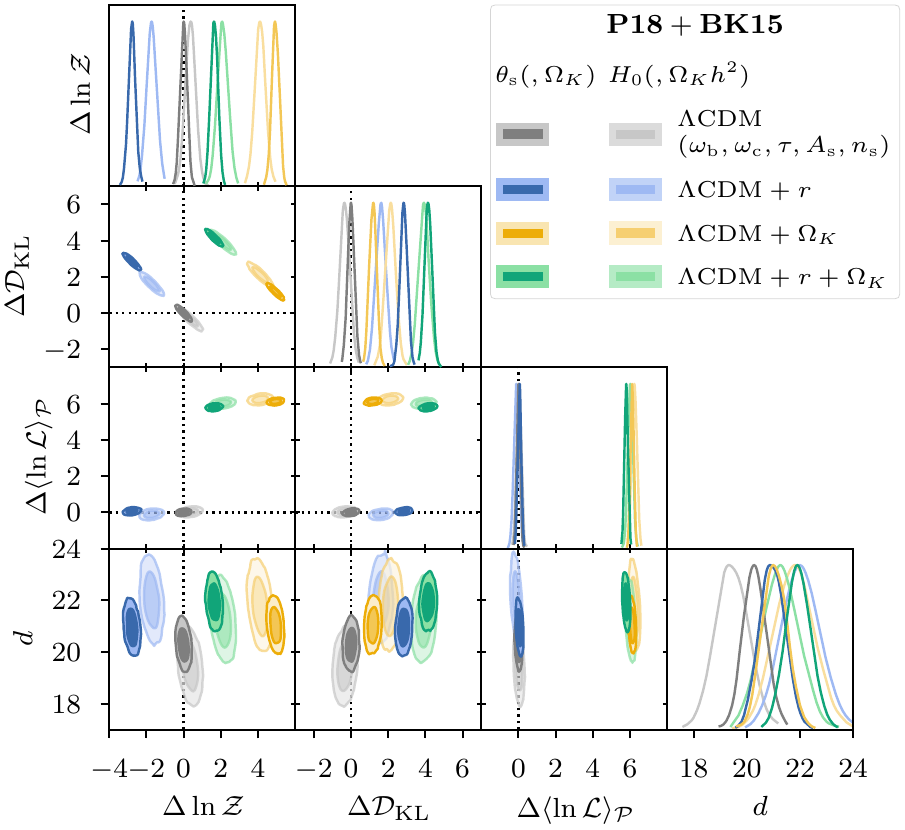}
    \caption{\label{fig:stats_theta_vs_H0} Robustness check of Bayesian model comparison: $\theta_\mathrm{s}$ vs $H_0$. Sampling the six \LCDM\ parameters using the angular distance to the sound horizon $\theta_\mathrm{s}$ (dark shade) or the Hubble parameter $H_0$ (light shade) with flat priors respectively. For curvature extensions we sample $\Omega_{K,0}$ alongside $\theta_\mathrm{s}$, and $\omega_{K,0}\equiv\Omega_{K,0}h^2$ alongside $H_0$. We do the latter chage.}
\end{figure}

\Cref{fig:stats_theta_vs_H0} summarises the model comparison results for \LCDM\ extensions using either~$\theta_\mathrm{s}$ or~$H_0$ as one of the cosmological sampling parameters. The qualitative picture remains the same between the two cases, but quantitatively there can be shifts of up to \SIrange[range-units=single]{2}{3}{\sigma} with respect to the sampling uncertainty.

\begin{figure}[tbp]
    \includegraphics[width=\columnwidth]{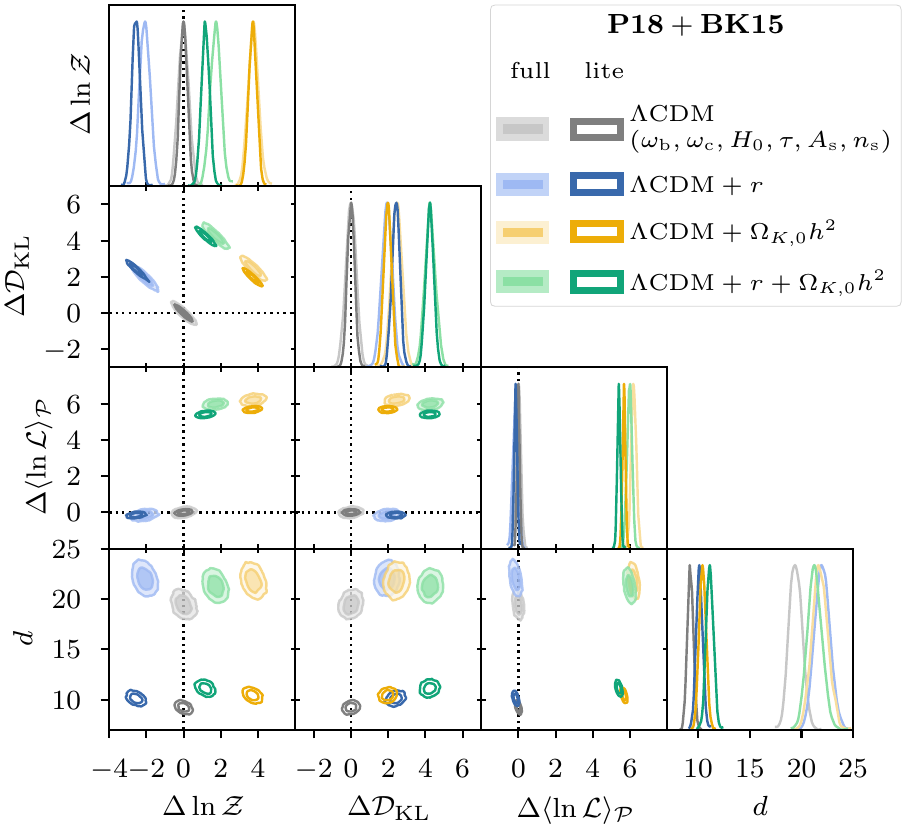}
    \caption{\label{fig:stats_full_vs_lite} Robustness check of Bayesian model comparison: \emph{full} vs \emph{lite} \textsc{Planck}~2018 $TT,TE,EE$ likelihood (also with \textsc{Planck}~2018 low-$\ell$ temperature and $E$-mode data and with $B$-mode data from the 2015 \textsc{Bicep2} and \textsc{Keck Array}). The \emph{full} P18 likelihood uses 21 nuisance parameters, whereas the \emph{lite} P18 likelihood has only a single nuisance parameter. The BK15 likelihood comes with an additional 7 nuisance parameters in both cases.}
\end{figure}

\Cref{fig:stats_full_vs_lite} summarises the model comparison results for \LCDM\ extensions run with either the full or only the lite version of the P18 likelihood. The lite version marginalises over all Planck nuisance parameters except for~$y_\mathrm{cal}$ (see \cref{fig:posterior_nuisance} for the posterior distributions of those nuisance parameters). This is reflected in the lower Bayesian model dimensionality~$d$ for the lite case, the difference of 10 between full and lite matches roughly the number of constrained nuisance parameters (unconstrained nuisance parameters do not contribute to~$d$). Since these are nested sampling runs with different likelihoods, we cannot directly compare Bayesian evidence and KL-divergence. However, we can use the \LCDM\ model as normalisation and then compare the relative change for the various extensions, giving roughly the same results for full and lite likelihood.

\begin{table*}[tbp]
\renewcommand{\arraystretch}{1.5}
    \caption{\label{tab:stats_extensions} Block~1 was run with only the P18 likelihood. Blocks~2 and~3 were run with the P18+BK15 likelihoods. Block~4 was run with the P18lite+BK15 likelihoods. Blocks~1 and~2 used the angular distance~$\theta_\mathrm{s}$ to the sound horizon as one of the cosmological sampling parameters, whereas blocks~3 and~4 changed this to the Hubble parameter~$H_0$.}
    \begin{ruledtabular}
        {\scriptsize
        
\begin{tabular}{ l c c c c c c c c } 
    Model parameters      & $\ln\mathcal{Z}$ & $\mathcal{D}_\mathrm{KL}$ & $\langle\ln\mathcal{L}\rangle_\mathcal{P}$ & $d$ & $\Delta\ln\mathcal{Z}$ & $\Delta\mathcal{D}_\mathrm{KL}$ & $\Delta\langle\ln\mathcal{L}\rangle_\mathcal{P}$ & $\Delta d$ \\ \hline 
    P18 \\
    $\omega_\mathrm{b}, \omega_\mathrm{c}, \theta_\mathrm{s}, \tau, A_\mathrm{s}, n_\mathrm{s}$              & $-1431.94\pm0.17$ & $39.48\pm0.17$ & $-1392.46\pm0.07$ & $+16.6\pm0.4$ & $-0.0\pm0.2$ & $+0.0\pm0.2$ & $-0.0\pm0.1$ & $+0.0\pm0.4$ \\  
    $\omega_\mathrm{b}, \omega_\mathrm{c}, \theta_\mathrm{s}, \tau, A_\mathrm{s}, n_\mathrm{s}, r$           & $-1434.69\pm0.18$ & $41.31\pm0.18$ & $-1393.39\pm0.08$ & $+18.5\pm0.4$ & $-2.8\pm0.2$ & $+1.8\pm0.2$ & $-0.9\pm0.1$ & $+1.9\pm0.4$ \\  
    $\omega_\mathrm{b}, \omega_\mathrm{c}, \theta_\mathrm{s}, \tau, A_\mathrm{s}, n_\mathrm{s}, \Omega_K$    & $-1429.81\pm0.18$ & $41.48\pm0.18$ & $-1388.33\pm0.08$ & $+18.1\pm0.4$ & $+2.1\pm0.2$ & $+2.0\pm0.2$ & $+4.1\pm0.1$ & $+1.5\pm0.4$ \\  
    $\omega_\mathrm{b}, \omega_\mathrm{c}, \theta_\mathrm{s}, \tau, A_\mathrm{s}, n_\mathrm{s}, r, \Omega_K$ & $-1431.81\pm0.19$ & $43.05\pm0.18$ & $-1388.76\pm0.09$ & $+20.4\pm0.4$ & $+0.1\pm0.2$ & $+3.6\pm0.2$ & $+3.7\pm0.1$ & $+3.8\pm0.4$ \\  \hline 
    P18 + BK15 \\
    $\omega_\mathrm{b}, \omega_\mathrm{c}, \theta_\mathrm{s}, \tau, A_\mathrm{s}, n_\mathrm{s}$              & $-1807.26\pm0.19$ & $44.53\pm0.19$ & $-1762.73\pm0.09$ & $+20.3\pm0.4$ & $-0.0\pm0.2$ & $-0.0\pm0.2$ & $-0.0\pm0.1$ & $-0.0\pm0.4$ \\  
    $\omega_\mathrm{b}, \omega_\mathrm{c}, \theta_\mathrm{s}, \tau, A_\mathrm{s}, n_\mathrm{s}, r$           & $-1810.04\pm0.19$ & $47.37\pm0.19$ & $-1762.66\pm0.09$ & $+21.0\pm0.5$ & $-2.8\pm0.2$ & $+2.8\pm0.2$ & $+0.1\pm0.1$ & $+0.7\pm0.5$ \\  
    $\omega_\mathrm{b}, \omega_\mathrm{c}, \theta_\mathrm{s}, \tau, A_\mathrm{s}, n_\mathrm{s}, \Omega_K$    & $-1802.32\pm0.19$ & $45.72\pm0.19$ & $-1756.61\pm0.09$ & $+21.0\pm0.5$ & $+4.9\pm0.2$ & $+1.2\pm0.2$ & $+6.1\pm0.1$ & $+0.7\pm0.5$ \\  
    $\omega_\mathrm{b}, \omega_\mathrm{c}, \theta_\mathrm{s}, \tau, A_\mathrm{s}, n_\mathrm{s}, r, \Omega_K$ & $-1805.62\pm0.19$ & $48.69\pm0.19$ & $-1756.92\pm0.09$ & $+21.9\pm0.5$ & $+1.6\pm0.2$ & $+4.2\pm0.2$ & $+5.8\pm0.1$ & $+1.6\pm0.5$ \\  \hline 
    P18 + BK15 \\
    $\omega_\mathrm{b}, \omega_\mathrm{c}, H_0, \tau, A_\mathrm{s}, n_\mathrm{s}$                            & $-1806.91\pm0.27$ & $44.22\pm0.27$ & $-1762.69\pm0.13$ & $+19.4\pm0.6$ & $+0.0\pm0.3$ & $-0.0\pm0.3$ & $-0.0\pm0.1$ & $+0.0\pm0.6$ \\  
    $\omega_\mathrm{b}, \omega_\mathrm{c}, H_0, \tau, A_\mathrm{s}, n_\mathrm{s}, r$                         & $-1808.98\pm0.27$ & $46.15\pm0.28$ & $-1762.83\pm0.14$ & $+22.0\pm0.7$ & $-2.1\pm0.3$ & $+1.9\pm0.3$ & $-0.1\pm0.1$ & $+2.6\pm0.7$ \\  
    $\omega_\mathrm{b}, \omega_\mathrm{c}, H_0, \tau, A_\mathrm{s}, n_\mathrm{s}, \omega_K$                  & $-1803.14\pm0.28$ & $46.65\pm0.28$ & $-1756.48\pm0.14$ & $+21.8\pm0.7$ & $+3.8\pm0.3$ & $+2.4\pm0.3$ & $+6.2\pm0.1$ & $+2.4\pm0.7$ \\  
    $\omega_\mathrm{b}, \omega_\mathrm{c}, H_0, \tau, A_\mathrm{s}, n_\mathrm{s}, r, \omega_K$               & $-1805.16\pm0.29$ & $48.45\pm0.28$ & $-1756.71\pm0.13$ & $+21.3\pm0.7$ & $+1.7\pm0.3$ & $+4.2\pm0.3$ & $+6.0\pm0.1$ & $+1.9\pm0.7$ \\  \hline 
    P18lite + BK15 \\
    $\omega_\mathrm{b}, \omega_\mathrm{c}, H_0, \tau, A_\mathrm{s}, n_\mathrm{s}$                            &  $-899.06\pm0.20$ & $24.56\pm0.19$ &  $-874.50\pm0.07$ &  $+9.2\pm0.3$ & $-0.0\pm0.2$ & $-0.0\pm0.2$ & $-0.0\pm0.1$ & $-0.0\pm0.3$ \\  
    $\omega_\mathrm{b}, \omega_\mathrm{c}, H_0, \tau, A_\mathrm{s}, n_\mathrm{s}, r$                         &  $-901.61\pm0.21$ & $26.97\pm0.20$ &  $-874.64\pm0.08$ & $+10.1\pm0.3$ & $-2.6\pm0.2$ & $+2.4\pm0.2$ & $-0.1\pm0.1$ & $+0.9\pm0.3$ \\  
    $\omega_\mathrm{b}, \omega_\mathrm{c}, H_0, \tau, A_\mathrm{s}, n_\mathrm{s}, \omega_K$                  &  $-895.36\pm0.21$ & $26.54\pm0.20$ &  $-868.82\pm0.08$ & $+10.4\pm0.3$ & $+3.7\pm0.2$ & $+2.0\pm0.2$ & $+5.7\pm0.1$ & $+1.2\pm0.3$ \\  
    $\omega_\mathrm{b}, \omega_\mathrm{c}, H_0, \tau, A_\mathrm{s}, n_\mathrm{s}, r, \omega_K$               &  $-897.89\pm0.22$ & $28.79\pm0.21$ &  $-869.09\pm0.08$ & $+11.1\pm0.3$ & $+1.2\pm0.2$ & $+4.2\pm0.2$ & $+5.4\pm0.1$ & $+1.9\pm0.3$ \\  
\end{tabular}

        }
    \end{ruledtabular}
\end{table*}

\begin{table*}[tbp]
\renewcommand{\arraystretch}{1.5}
    \caption{\label{tab:stats_inflation} Inflationary model comparison using P18lite+BK15 likelihoods. Normalisation with respect to \LCDM\ from \cref{tab:stats_extensions}. 
    The first block uses the permissive reheating scenario with reheating ending at $N_\mathrm{reh}=N_\mathrm{BBN}$ and with an allowed range of the effective reheating parameter of $-\frac{1}{3}<w_\mathrm{reh}<1$.
    The second block uses the restrictive reheating scenario with reheating ending at an energy density of $\rho_\mathrm{reh}^{1/4}=\SI{e9}{\giga\eV}$ and with the allowed range of the effective reheating parameter reduced to $-\frac{1}{3}<w_\mathrm{reh}<\frac{1}{3}$.
    }
    \begin{ruledtabular}
        {\scriptsize
        
\begin{tabular}{ l c c c c c c c c } 
    Inflation Model & $\ln\mathcal{Z}$ & $\mathcal{D}_\mathrm{KL}$ & $\langle\ln\mathcal{L}\rangle_\mathcal{P}$ & $d$ & $\Delta\ln\mathcal{Z}$ & $\Delta\mathcal{D}_\mathrm{KL}$ & $\Delta\langle\ln\mathcal{L}\rangle_\mathcal{P}$ & $\Delta d$ \\ \hline 
    \multicolumn{3}{l}{(permissive reheating)} \\
    Quadratic   & $-901.24\pm0.21$ & $+25.68\pm0.20$ & $-875.55\pm0.08$ & $+11.1\pm0.3$ & $-2.18\pm0.21$ & $+1.12\pm0.20$ & $-1.06\pm0.08$ & $+1.9\pm0.3$ \\ 
    Natural     & $-898.65\pm0.21$ & $+27.19\pm0.20$ & $-871.46\pm0.09$ & $+14.3\pm0.5$ & $+0.41\pm0.21$ & $+2.63\pm0.20$ & $+3.04\pm0.09$ & $+5.1\pm0.5$ \\ 
    Starobinsky & $-894.21\pm0.18$ & $+24.40\pm0.18$ & $-869.81\pm0.07$ & $+10.1\pm0.3$ & $+4.85\pm0.18$ & $-0.16\pm0.18$ & $+4.69\pm0.07$ & $+0.9\pm0.3$ \\ \hline 
    \multicolumn{3}{l}{(restrictive reheating)} \\
    Quadratic   & $-903.27\pm0.37$ & $+25.47\pm0.39$ & $-877.80\pm0.37$ & $+12.3\pm1.7$ & $-4.21\pm0.37$ & $+0.90\pm0.39$ & $-3.30\pm0.37$ & $+3.1\pm1.7$ \\ 
    Natural     & $-900.30\pm0.35$ & $+26.84\pm0.36$ & $-873.46\pm0.31$ & $+14.5\pm1.3$ & $-1.24\pm0.35$ & $+2.28\pm0.36$ & $+1.03\pm0.31$ & $+5.2\pm1.3$ \\ 
    Starobinsky & $-893.23\pm0.29$ & $+23.46\pm0.28$ & $-869.77\pm0.10$ & $+9.7\pm0.4$ & $+5.83\pm0.29$ & $-1.11\pm0.28$ & $+4.73\pm0.10$ & $+0.5\pm0.4$ \\ 
\end{tabular}
        }
    \end{ruledtabular}
\end{table*}

\begin{figure}[tbp]
    \includegraphics[width=\columnwidth]{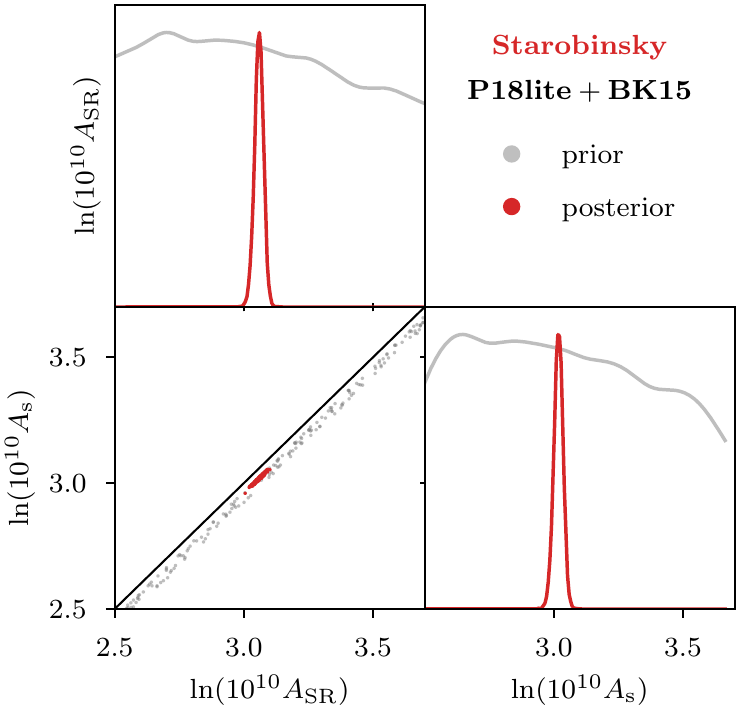}
    \caption{\label{fig:As_vs_ASR} Parameter comparison of the approximate input parameter~$A_\mathrm{SR}$ (using the slow-roll approximation, see \cref{eq:A_SR}) and the derived output parameter~$A_\mathrm{s}$ for the amplitude of the primordial power spectrum at the pivot scale~$k_\ast=\SI{0.05}{\per\mega\parsec}$. In a perfect scenario, these two parameters would be identical. The slight shift is negligible for sampling purposes.}
\end{figure}

\clearpage

\bibliography{ref}

\end{document}